\pdfoutput=1
\documentclass[twocolumn]{aastex63}

\newcommand{\HII}{H\,{\scriptsize II}}

\newcommand{\CII}{[C\,{\scriptsize II}]}
\newcommand{\dCII}{[$^{\rm 13}$C\,{\scriptsize II}]}

\newcommand{\OI}{[O\,{\scriptsize I}]}

\newcommand{\two}{2$\to$1}

\newcommand{\XSPEC}  {{\em XSPEC}}

\received{-}
\revised{-}
\accepted{-}
\submitjournal{ApJ}

\shorttitle{RCW 36}
\shortauthors{Bonne et al.}

\begin{document}
\shorttitle{Mass ejection in RCW 36}
\shortauthors{Bonne et al.}
\title{The SOFIA FEEDBACK Legacy Survey\\ Dynamics and mass ejection in the bipolar \HII\ region RCW 36\footnote{Released on --}}

\correspondingauthor{Lars Bonne}
\email{lbonne@usra.edu}

\author{L. Bonne}
\affiliation{SOFIA Science Center, NASA Ames Research Center, Moffett Field, CA 94 045, USA}
\author{N. Schneider}
\affiliation{I. Physik. Institut, University of Cologne, Z\"ulpicher Str. 77, 50937 Cologne, Germany}
\author{P. Garc\'ia}
\affiliation{Chinese Academy of Sciences South America Center for Astronomy, National Astronomical Observatories, CAS, Beijing 100101, China}
\affiliation{Instituto de Astronom\'ia, Universidad Cat\'olica del Norte, Av. Angamos 0610, Antofagasta, 1270709, Chile}
\author{A. Bij}
\affiliation{Engineering Physics and Astronomy, Queen's University, Kingston, ON K7L 3N6, Canada}
\author{P. Broos}
\affiliation{Department of Astronomy \& Astrophysics, 525 Davey Laboratory, Pennsylvania State University, University Park, PA, 16802, USA}
\author{L. Fissel}
\affiliation{Engineering Physics and Astronomy, Queen's University, Kingston, ON K7L 3N6, Canada}
\author{R. Guesten}
\affiliation{Max-Planck Institut f\"ur Radioastronomie, Auf dem H\"ugel 69, 53121 Bonn, Germany}
\author{J. Jackson}
\affiliation{SOFIA Science Center, NASA Ames Research Center, Moffett Field, CA 94 045, USA}
\author{R. Simon}
\affiliation{I. Physik. Institut, University of Cologne, Z\"ulpicher Str. 77, 50937 Cologne, Germany}
\author{L. Townsley}
\affiliation{Department of Astronomy \& Astrophysics, 525 Davey Laboratory, Pennsylvania State University, University Park, PA, 16802, USA}
\author{A. Zavagno}
\affiliation{Aix Marseille Univ, CNRS, CNES, LAM, Marseille, France}
\author{R. Aladro}
\affiliation{Max-Planck Institut f\"ur Radioastronomie, Auf dem H\"ugel 69, 53121 Bonn, Germany}
\author{C. Buchbender}
\affiliation{I. Physik. Institut, University of Cologne, Z\"ulpicher Str. 77, 50937 Cologne, Germany}
\author{C. Guevara}
\affiliation{I. Physik. Institut, University of Cologne, Z\"ulpicher Str. 77, 50937 Cologne, Germany}
\author{R. Higgins}
\affiliation{I. Physik. Institut, University of Cologne, Z\"ulpicher Str. 77, 50937 Cologne, Germany}
\author{A. M. Jacob}
\affiliation{Max-Planck Institut f\"ur Radioastronomie, Auf dem H\"ugel 69, 53121 Bonn, Germany}
\affiliation{Department of Physics \& Astronomy, Johns Hopkins University, Baltimore, MD 21218, USA}
\author{S. Kabanovic}
\affiliation{I. Physik. Institut, University of Cologne, Z\"ulpicher Str. 77, 50937 Cologne, Germany}
\author{R. Karim}
\affiliation{University of Maryland, Department of Astronomy, College Park, MD 20742-2421, USA}
\author{A. Soam}
\affiliation{SOFIA Science Center, NASA Ames Research Center, Moffett Field, CA 94 045, USA}
\author{J. Stutzki}
\affiliation{I. Physik. Institut, University of Cologne, Z\"ulpicher Str. 77, 50937 Cologne, Germany}
\author{M. Tiwari}
\affiliation{University of Maryland, Department of Astronomy, College Park, MD 20742-2421, USA}
\author{F. Wyrowski}
\affiliation{Max-Planck Institut f\"ur Radioastronomie, Auf dem H\"ugel 69, 53121 Bonn, Germany}
\author{A. G. G. M. Tielens}
\affiliation{University of Maryland, Department of Astronomy, College Park, MD 20742-2421, USA}
\affiliation{Leiden Observatory, PO Box 9513, 2300 RA Leiden, The Netherlands}

\begin{abstract}

We present \CII\ 158 $\mu$m and \OI\ 63 $\mu$m observations of the bipolar \HII\ region RCW 36 in the Vela C molecular cloud, obtained within the SOFIA legacy project FEEDBACK, which is complemented with APEX $^{12/13}$CO(3-2) and Chandra X-ray (0.5-7 keV) data. This shows that the molecular ring, forming the waist of the bipolar nebula, expands with a velocity of 1 - 1.9 km s$^{-1}$. We also observe an increased linewidth in the ring indicating that turbulence is driven by energy injection from the stellar feedback. The bipolar cavity hosts blue-shifted expanding \CII\ shells at 5.2$\pm$0.5$\pm$0.5 km s$^{-1}$ (statistical and systematic uncertainty) which indicates that expansion out of the dense gas happens non-uniformly and that the observed bipolar phase might be relatively short ($\sim$0.2 Myr). The X-ray observations show diffuse emission that traces a hot plasma, created by stellar winds, in and around RCW 36. At least 50 \% of the stellar wind energy is missing in RCW 36. This is likely due to leakage which is clearing even larger cavities around the bipolar RCW 36 region. Lastly, the cavities host high-velocity wings in \CII\ which indicates relatively high mass ejection rates ($\sim$5$\times$10$^{-4}$ M$_{\odot}$ yr$^{-1}$). This could be driven by stellar winds and/or radiation pressure, but remains difficult to constrain. This local mass ejection, which can remove all mass within 1 pc of RCW 36 in 1-2 Myr, and the large-scale clearing of ambient gas in the Vela C cloud indicates that stellar feedback plays a significant role in suppressing the star formation efficiency (SFE).

\end{abstract}

\keywords{HII regions --- ISM: kinematics and dynamics --- X-rays: ISM --- ISM: bubbles --- ISM: magnetic fields --- photon-dominated region (PDR)}


\section{Introduction} \label{sec:intro}

\begin{figure*}
    \centering
    \includegraphics[width=0.475\hsize]{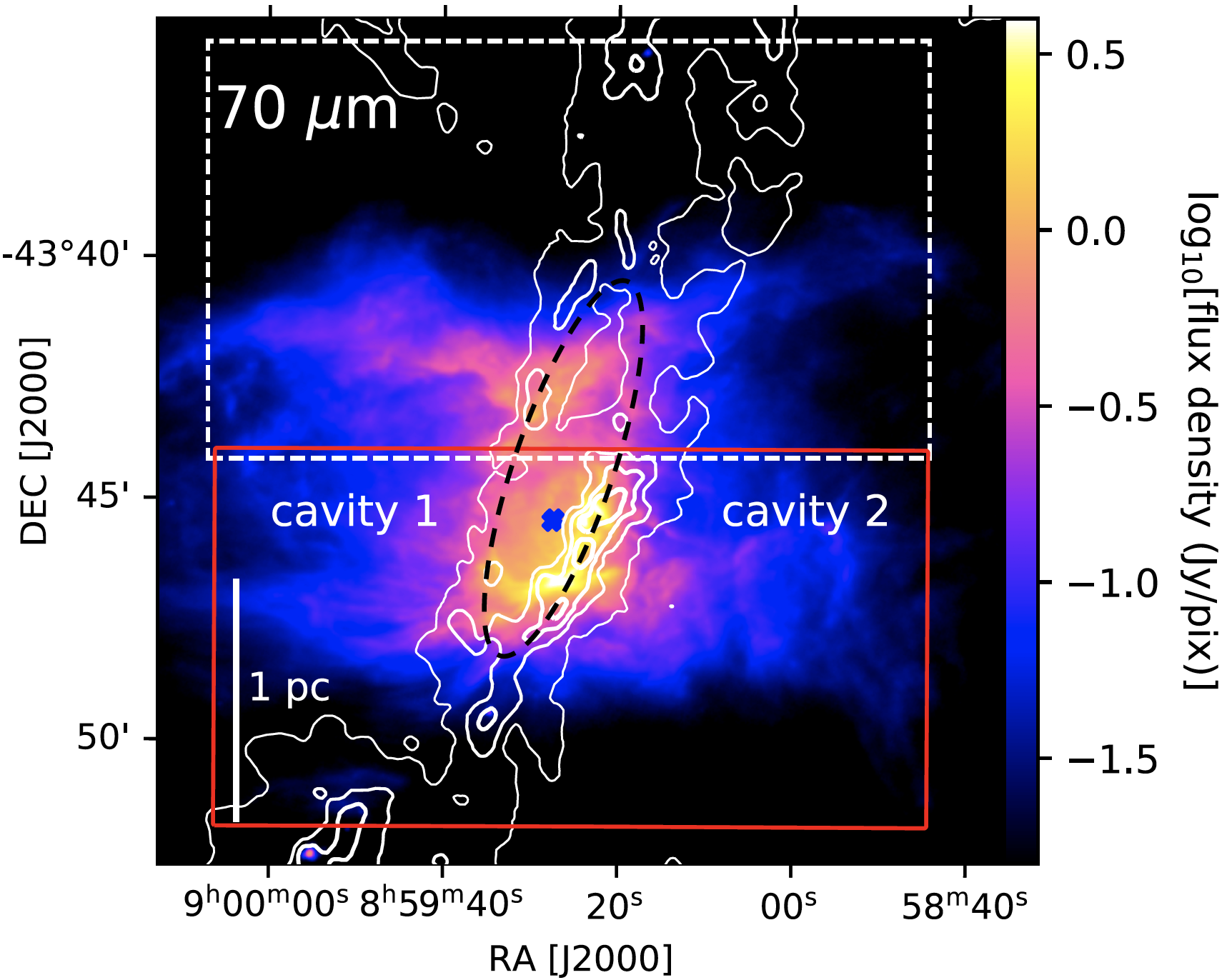}
    \includegraphics[width=0.515\hsize]{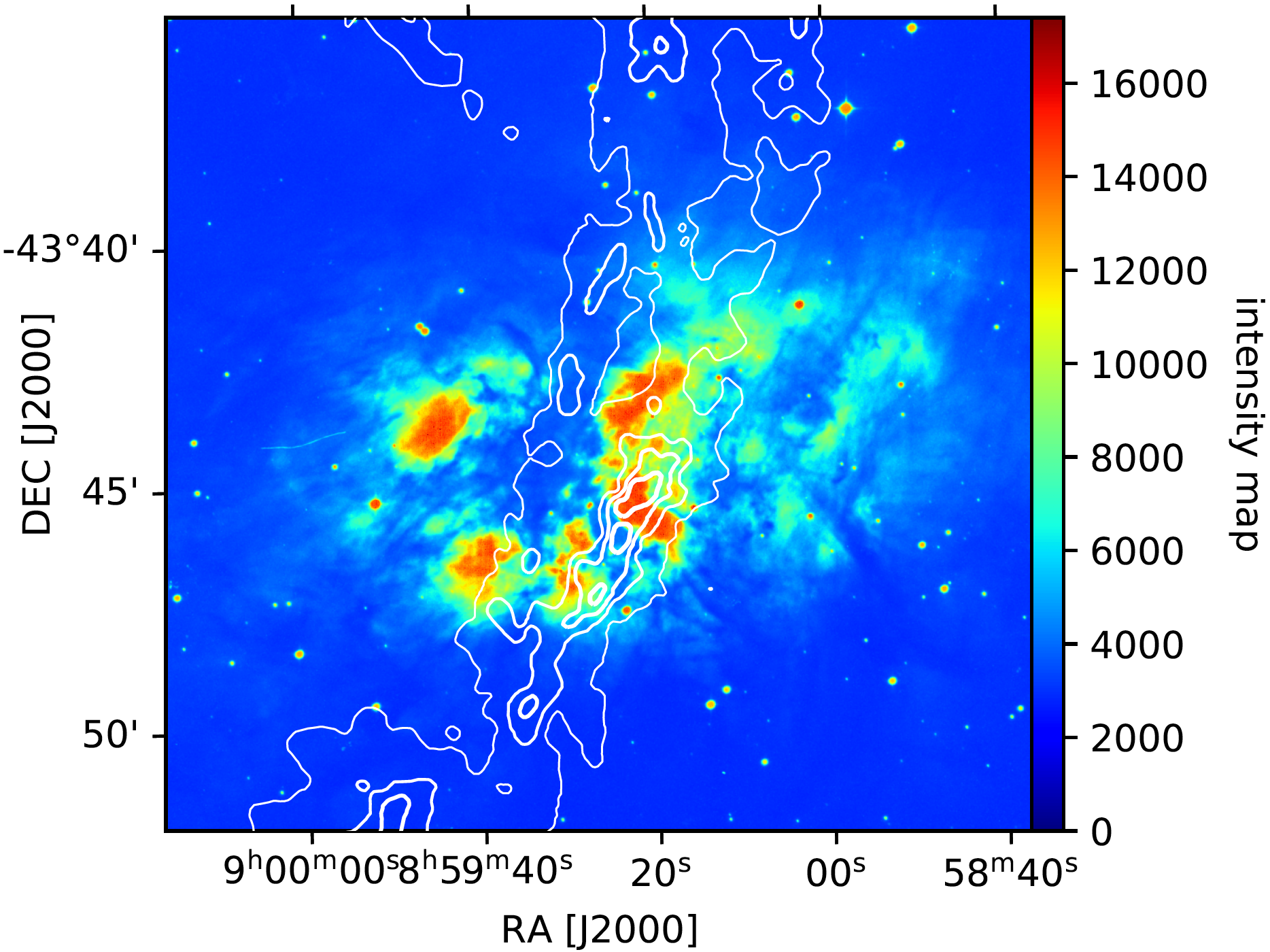}
    \caption{{\bf Left}: The \textit{Herschel} 70 $\mu$m map of RCW 36 \citep{Minier2013}. The red box outlines the region that was already observed for the FEEDBACK program and the dashed white box shows the area that will be mapped.
    The blue crosses mark the location of the two O star candidates. Note that the close proximity of the two O star candidates does not allow to distinguish them clearly in this figure. The black dashed ellipse indicates the location of the central molecular ring, first reported by \citet{Minier2013}. The \textit{Herschel} 70 $\mu$m map also traces the limb-brightened shells at the cavity walls. The white contours indicate the \textit{Herschel} column density at N$_{\rm H_{2}}$ = 1, 2, 4 and 8$\times$10$^{22}$ cm$^{-2}$ determined using the method presented in \citet{Palmeirim2013}.  
    {\bf Right}: H$\alpha$ intensity map towards RCW 36 from the SuperCOSMOS H$\alpha$ Survey. The intensity units of the map are not calibrated to a specific unit and are only relative \citep{Hambly2001,Parker2005}. The white contours again indicate the \textit{Herschel} column density. 
    }
    \label{fig:observedMap}
\end{figure*}

The interstellar medium (ISM) is a complex mixture of multiple phases \citep[][]{McKee1977,Wolfire1995,Wolfire2003,Walch2015b} where stars form in the densest regions \citep{Bergin2007} which are mostly organised in dense filaments \citep{Andre2010,Molinari2010,Schneider2012}. Feedback from the star formation process in the form of ionizing radiation and stellar winds provides substantial energy injection into the ISM which can drive its chemical and dynamical evolution. Feedback might also play a significant role in suppressing the star formation rate (SFR) in molecular clouds. It has been proposed that massive stars form in gravitationally collapsing hubs and ridges \citep{Motte2018} which results in the rapid accumulation of mass at the center of collapse (i.e. in a few free-fall timescales). This would result in high SFRs, which are not observed, if not counteracted by feedback processes \citep[e.g.][]{MacLow2017,VazquezSemadeni2019}. Additionally, with astrophysical simulations it remains extremely challenging to predict the impact and role of the different stellar feedback processes on the molecular cloud evolution because of the large computational requirements and complex subgrid physics  \citep[e.g.][]{Dale2014,Geen2020,Grudic2021,Kim2021}.\\
Stellar feedback leads to the formation of \HII\ regions which come in the form of \HII\ bubbles \citep{Churchwell2006,Deharveng2010,Beaumont2010,Kendrew2012}, bipolar \HII\ regions \citep{Bally1982,Deharveng2015,Schneider2018,Samal2018} and more irregular features \citep{Anderson2011,Anderson2014}. Recently, the kinematics of a couple of \HII\ bubbles have been studied with the \CII\ 158 $\mu$m fine-structure line  \citep[][]{Anderson2019,Pabst2019,Pabst2020,Luisi2021,Tiwari2021,Beuther2022} because it traces the photodissociation regions (PDRs) at the interfaces between \HII\ regions and molecular clouds, where mostly far-ultraviolet (FUV) photons in the energy range of 6 - 13.6 eV regulate the physical conditions of the ISM \citep[][]{Hollenbach1999}. The \CII\ kinematics in some of these regions demonstrated the presence of expanding shells with velocities up to 15 km s$^{-1}$  \citep[][]{Pabst2019,Pabst2020,Luisi2021,Tiwari2021,Beuther2022}. 
However, they often show expansion in only one direction, mostly blue-shifted, which suggests that the bubbles might not be fully spherical.  
This more complex geometry is important for the interpretation of the various emission distributions. Observations of hot X-ray emitting plasmas, produced by stellar winds from OB stars, showed that these plasmas have 
an unbalanced energy budget, i.e. the energy of the hot plasmas is significantly lower than the energy injected by stellar winds  \citep[e.g.][]{Townsley2003,Townsley2006,Townsley2011,HarperClark2009,Lopez2011,Lopez2014,Rosen2014,Tiwari2021}. The more complex geometry of expanding \HII\ regions could thus allow the hot plasma to leak out of the region, which might explain the mismatch in the energy budget. The lack of an expanding shell at one side of a region leading to rapid hot plasma leakage was recently demonstrated for RCW 49 by \citet{Tiwari2021}. Studying the \CII\ emission dynamics can also help to constrain the role of stellar feedback in suppressing the star formation of a molecular cloud as well as triggering a second generation of star formation, which is still highly debated \citep[e.g.][]{Zavagno2010,Zavagno2020,Dale2013,Walch2015a}. \citet{Luisi2021} proposed, based on \CII\ data taken within the SOFIA (Stratospheric Observatory for Infrared Astronomy) legacy program FEEDBACK\footnote{\href{https://feedback.astro.umd.edu}{https://feedback.astro.umd.edu}} \citep{Schneider2020}, that star-formation is triggered in RCW120. They propose a model in which a wind-driven expanding \CII\ shell sweeps up a gas shell that  fragments and forms stars on very short time scales ($\sim$ 0.1 Myr). \\ 
The formation of bipolar \HII\ regions has received less attention than single bubbles, which might be related to the fact that bipolar \HII\ regions are not as common \citep[e.g.][]{Samal2018}. Early work by \citet{Bodenheimer1979} demonstrated that a bipolar \HII\ region can form by expansion in a flattened molecular cloud, consistent with simulations by \citet{Wareing2017}. \citet{Whitworth2018} proposed that bipolar \HII\ regions emerge from a sheet that was formed by a cloud-cloud collision, whereas \citet{Kumar2020} propose that bipolar \HII\ regions form from a hub-filament cloud that was flattened by rotation. Numerical simulations predict that a significant fraction of the dense molecular gas is organized in filaments and sheets \citep[e.g.][]{Padoan2001,VazquezSemadeni2006,Banerjee2009,Wareing2019}, allowing for the formation of bipolar \HII\ regions. This view of the molecular ISM organized in filaments and sheets is now increasingly supported by observations \citep[e.g.][]{Andre2014,Qian2015,Tritsis2018,Shimajiri2019,Bonne2020b}.\\\\
Here, we focus on the bipolar \HII\ region RCW 36 which is located in the dense ridge of the Vela C molecular cloud \citep{Hill2011,Minier2013,Fissel2016}. The distance of the Vela C molecular cloud has recently been updated to 900 $\pm$ 50 pc with the GAIA data \citep{Zucker2020,GAIAcollaboration2018}. Earlier distance estimates \citep{Liseau1992,Yamaguchi1999,Massi2019} 
derived values between 700 and 1000 pc. 
RCW 36 is probably the first developing \HII\ region in Vela C, which is well illustrated by Fig. 1 from \citet{Hill2011}, suggesting Vela C is still in a relatively early stage of star formation. The young cluster at the origin of the bipolar cavities has more than 350 stars located within a radius of 0.5 pc \citep{Baba2004}, an estimated age of 1.1 $\pm$ 0.6 Myr \citep{Ellerbroek2013a} and is proposed to host two O stars of type O9.5 V \& O9 V \citep{Massi2003,Bik2005,Ellerbroek2013a}. This  is consistent with the estimated total luminosity of 1.3$\times$10$^{5}$ L$_{\odot}$ for the cluster (scaled to the adopted distance in this work: d = 900 pc) \citep{Verma1994}. \textit{Herschel} observations indicated that the young cluster is surrounded by a dense molecular ring of swept-up gas with a radius of $\sim$ 1 pc which is proposed to expand with a velocity of 1-2 km s$^{-1}$ \citep{Minier2013}. In Fig. \ref{fig:observedMap}, the H$\alpha$ absorption in the eastern part of the ring would then suggest that this part of the ring is expanding towards us and that the western part is expanding away from us \citep{Minier2013}. Multiwavelength observations of RCW 36 have also established the presence of several protostar and young stellar object (YSO) candidates \citep{Bik2006,Hill2011,Ellerbroek2011,Giannini2012,Massi2019}, indicating relatively low-mass ongoing star formation in the region, as well as the presence of two Herbig-haro (HH) jets \citep{Ellerbroek2013b}.\\\\ 
In this paper, we present the first \CII\ and \OI\ observations of RCW 36 obtained by the SOFIA FEEDBACK Legacy survey
in combination with complementary data to acquire a global understanding of the evolution of this region. Both the \CII\ and \OI\ line are PDR tracers where \OI, with a critical density of 5$\times$10$^{5}$ cm$^{-3}$, is particularly used to trace dense PDRs \citep{Roellig2006}. These FEEDBACK observations unveil new dynamical information that was not available before. 
 Sect.~\ref{obs} describes our new [CII], [OI], and CO(3-2) observations, as well as archival X-ray and submillimeter continuum  datasets. The results of these observations are presented in Sect.~\ref{sec:results}. In Sect.~\ref{sec:analysis}, the expansion energetics and mass ejection parameters are calculated, and in Sect.~\ref{sec:chandra} the \CII\ data is compared with Chandra X-ray observations of RCW 36. In Sect.~\ref{sec:discussion}, these results are combined to propose a scenario for the evolution of RCW 36 and discuss the results in the global context of stellar feedback on molecular clouds.

\section{Observations} \label{obs}
\subsection{SOFIA}
The FEEDBACK C$^{+}$ Legacy survey observed the \CII\ $^{3}$P$_{3/2}$ $\to$ $^{3}$P$_{1/2}$ transition, at 158 $\mu$m, and \OI\ $^{3}$P$_{1}$ $\to$ $^{3}$P$_{2}$ transition, at 63 $\mu$m, simultaneously with the upGREAT instrument \citep{Risacher2018} onboard of the airborne Stratospheric Observatory for Infrared Astronomy 
(SOFIA) \citep{Young2012} during a flight on June 6th 2019 from Christchurch, New Zealand. 
The observation strategy was to carry out \CII\ and \OI\ on-the-fly (OTF) maps and is described in detail in \citet{Schneider2020}. 
For RCW 36, the planned map within the legacy program has a size of 14.4$^{\prime}\times$14.4$^{\prime}$, indicated in Fig. \ref{fig:observedMap}. Here, we present the lower part of the map (14.4$^{\prime}\times$7.2$^{\prime}$) that has been observed so far. 
The center of the full map is located at $\alpha_{(J2000)}$ = 08$^h$59$^m$26.8$^s$ and $\delta_{(J2000)}$ = -43$^\circ$44$'$14.1$''$, and the emission-free OFF position is at $\alpha_{(J2000)}$ = 09$^h$01$^m$31.7$^s$ and $\delta_{(J2000)}$ = -43$^\circ$22$'$50.0$''$.\\ 
 Both, the reduced \CII\ (smoothed to 20$^{\prime\prime}$) and \OI\ (smoothed to 30$^{\prime\prime}$) map, used in this paper, have a spectral binning of 0.2 km s$^{-1}$. The typical noise rms of this smoothed data within 0.2 km s$^{-1}$ is 0.8-1.0 K for \CII\ and $\sim$ 0.8-1.5 K for \OI. 
The calibration was done using the GREAT pipeline \citep{Guan2012}. To convert the antenna to main beam temperatures, a forward efficiency $\eta_{f}$ = 0.97 and main beam efficiency $\eta_{mb}$ = 0.65 for \CII\ and 0.69 for \OI\ were applied. 
To improve the baseline removal of the data, a new method based on Principal Component Analysis (PCA) was used 
(Buchbender et al. in prep.). This technique was also employed for the data published in \citet{Luisi2021}, \citet{Tiwari2021} and \citet{Kabanovic2022}. Specifically, PCA identifies the systematic components of baseline variation in different spectra which allows to accurately remove them and recover the actual spectrum. For a more general overview of PCA we refer the reader to \citet{Jolliffe2011}. 


\begin{figure*}
    \centering
    \includegraphics[width=\hsize]{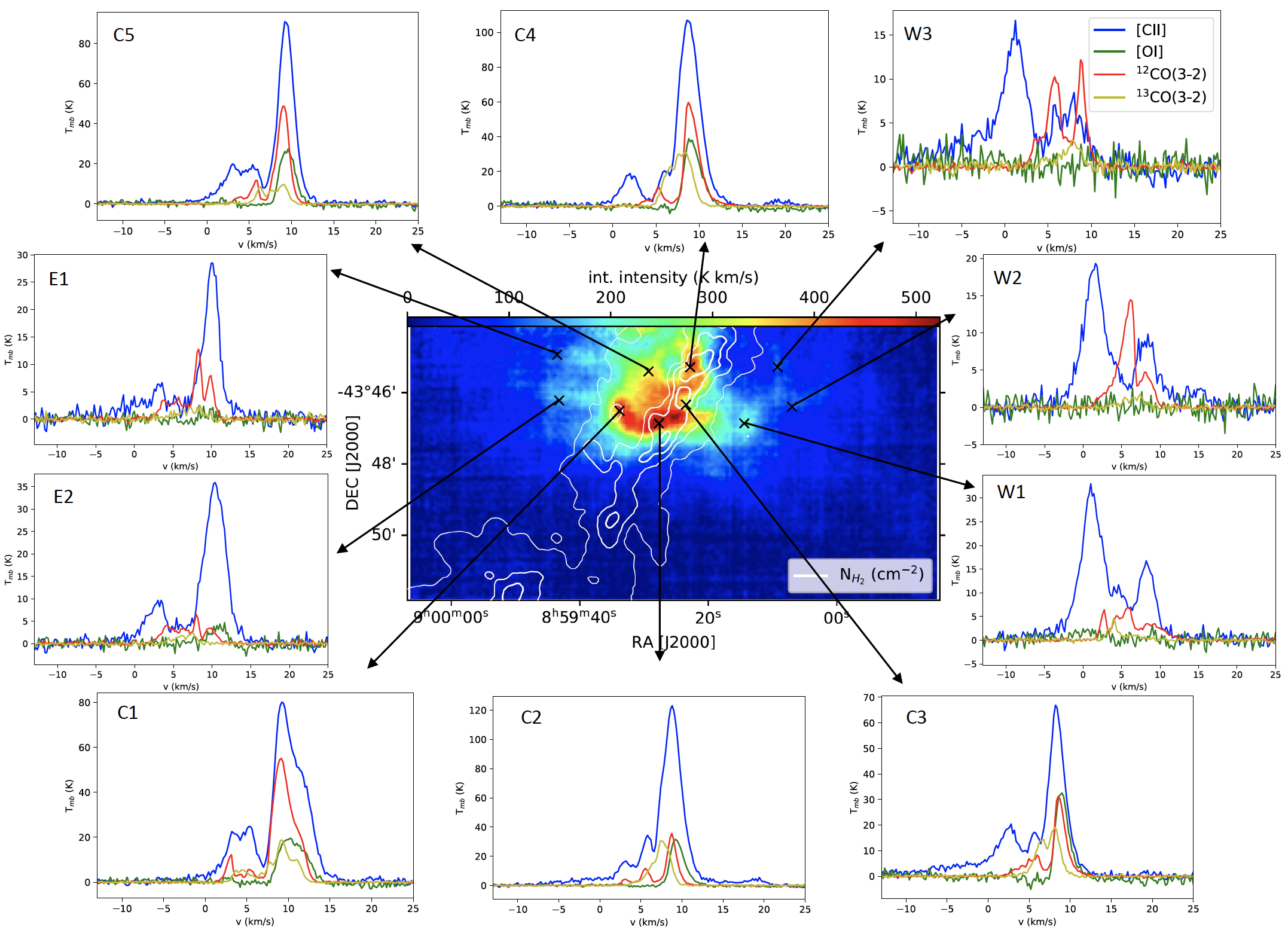}
    \caption{The middle panel shows the line integrated intensity map of \CII\ from -20 to 20 km s$^{-1}$ with the white contours indicating the \textit{Herschel} column density N$_{H_{2}}$ at 1, 2, 4 and 8$\times$10$^{22}$ cm$^{-2}$. The \CII, \OI, $^{12}$CO(3-2) and $^{13}$CO(3-2) spectra at the indicated positions are shown around the map. These spectra are named after their position in the map. The spectra from the central molecular ring are indicated with a `C', the spectra to the east with `E' and the spectra to the west with `W'. The spectra indicate the presence of multiple velocity components and show signatures of self-absorption towards the ring (C). In several spectra (C1-C3), high-velocity blue- and red-shifted wings are detected. The spectrum `W3' at the top right highlights the color legend for the spectra. When inspecting the spectra, note that the y-scale is different in each sub-panel.}
    \label{fig:integratedIntensities}
\end{figure*}

\begin{figure}
    \centering
    \includegraphics[width=\hsize]{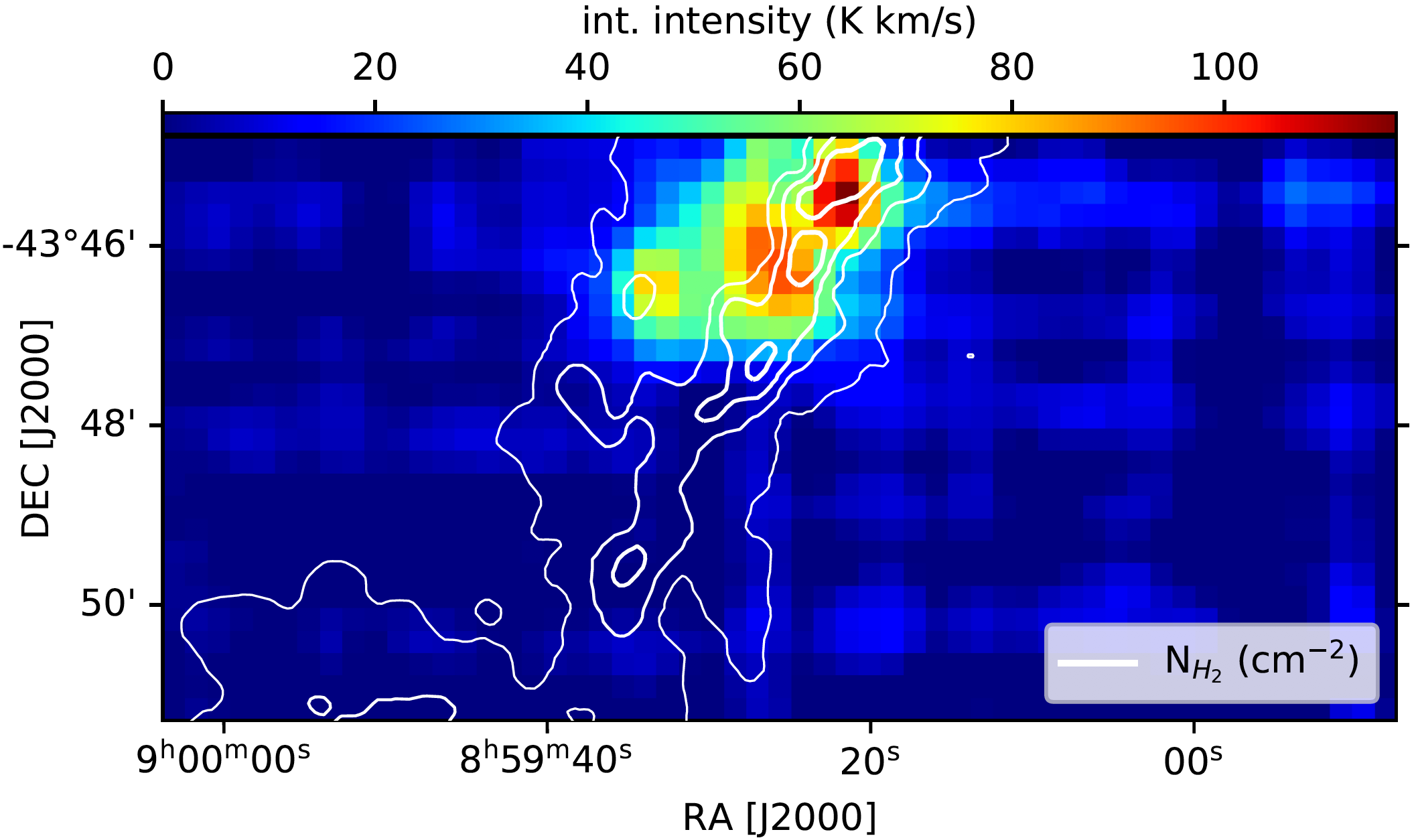}
    \caption{The integrated intensity map of \OI\ at 63 $\mu$m from -20 to 20 km s$^{-1}$ with the white contours indicating the column density 
    N$_{H_{2}}$ at 1, 2, 4, and 8$\times$10$^{22}$ cm$^{-2}$.}
    \label{fig:integratedOI}
\end{figure}

\subsection{APEX}
Complementary $^{12}$CO(3-2) and $^{13}$CO(3-2) data were simultaneously obtained with the LAsMA 7-pixel receiver on September 27, 2019 and July 21, 2021 in good weather conditions (precipitable water vapor, pwv $\approx$ 1 mm) using the APEX telescope\footnote{APEX, the Atacama Pathfinder EXperiment is a collaboration between the Max-Planck-Institut f\"ur Radioastronomie, Onsala Space Observatory (OSO), and the European Southern Observatory (ESO).} \citep{Guesten2006}. This heterodyne spectrometer allows simultaneous observations of the two CO isotopomers in the upper ($^{12}$CO(3-2) at 345.7959899 GHz) and lower ($^{13}$CO(3-2), 330.5879653 GHz) sideband of the receiver. The LAsMA backends are fast Fourier transform spectrometers (FFTS) with 2$\times$4 GHz bandwidth \citep{Klein2012} and a native spectral resolution of 61 kHz. The main beam size of the APEX observations ($\theta_{mb}$ = 18.2$^{\prime\prime}$ at 345.8 GHz) is similar to the resolution of the \CII\ data. The mapped field with a size of 20$^{\prime}\times$15$^{\prime}$ was scanned in total power on-the-fly mode, with a spacing of 9$^{\prime\prime}$ between rows (while oversampling to 6$^{\prime\prime}$ in scanning direction, R.A.). The map data were calibrated against a sky reference position at $\alpha_{(J2000)}$ = 09$^h$01$^m$31.7$^s$ and $\delta_{(J2000)}$ = -43$^\circ$22$'$50.0$''$. 
For the APEX observations, a first order baseline was removed, and a main beam efficiency $\eta_{mb}$ = 0.68 was used to convert the antenna temperatures to main beam temperatures. The baseline-subtracted spectra have been binned to a spectral resolution of 0.2 km s$^{-1}$. The final data cubes (with a pixel size of 9.1$^{\prime\prime}$) have an rms of 0.45 K.

\subsection{Ancillary data}
The Chandra X-ray Observatory ACIS camera \citep{Garmire2003} observed a 17$'$x17$'$ field centred on RCW~36 for $\sim$70~ksec in 2006.  The MYStIX project \citep{Kuhn2013} found 502 X-ray point sources in this field.  These were excised, and the remaining data smoothed, to obtain an 0.5--7~keV diffuse X-ray image of the field \citep{Townsley2014}.  ACIS is an intrinsic spectrometer, tagging every X-ray event with its energy as well as its position.  We will use the Chandra/ACIS data to study the X-ray spectra in a variety of regions in and around RCW~36.\\
The RCW 36 region, being part of the Vela C molecular cloud, was observed with the PACS \citep{Poglitsch2010} and SPIRE \citep{Griffin2010} bolometers on the \textit{Herschel} telescope as part of the {\sl Herschel} imaging survey of OB young stellar objects (HOBYS) program \citep{Motte2010}.  In this work, we use the flux map at 70 $\mu$m with an angular resolution of $\sim$6$''$ as it clearly outlines the bipolar \HII\ region.  Further, we use the dust temperature map at 36$''$ resolution and a dust column density map at 18.2$^{\prime\prime}$ resolution. 
This dust temperature and column density map was obtained by fitting the spectral energy distribution (SED) to the 160 to 500 $\mu$m data, using the method presented in \citet{Palmeirim2013}.\\
We further make use of data from the WISE telescope from 3.4 $\mu$m to 22 $\mu$m \citep{Wright2010}, a H$\alpha$ map of RCW 36 from the Southern H$\alpha$ Sky Survey Atlas (SHASSA) \citep{Gaustad2001}, and a large scale dust polarization map at 500 $\mu$m with a spatial resolution of 2.5$^{\prime}$ that was observed with the BLASTPol balloon-borne receiver \citep{Fissel2016,Fissel2019,Gandilo2016}.

\section{Results}\label{sec:results}

\subsection{The integrated intensity map}
\begin{figure}
    \centering
    \includegraphics[width=\hsize]{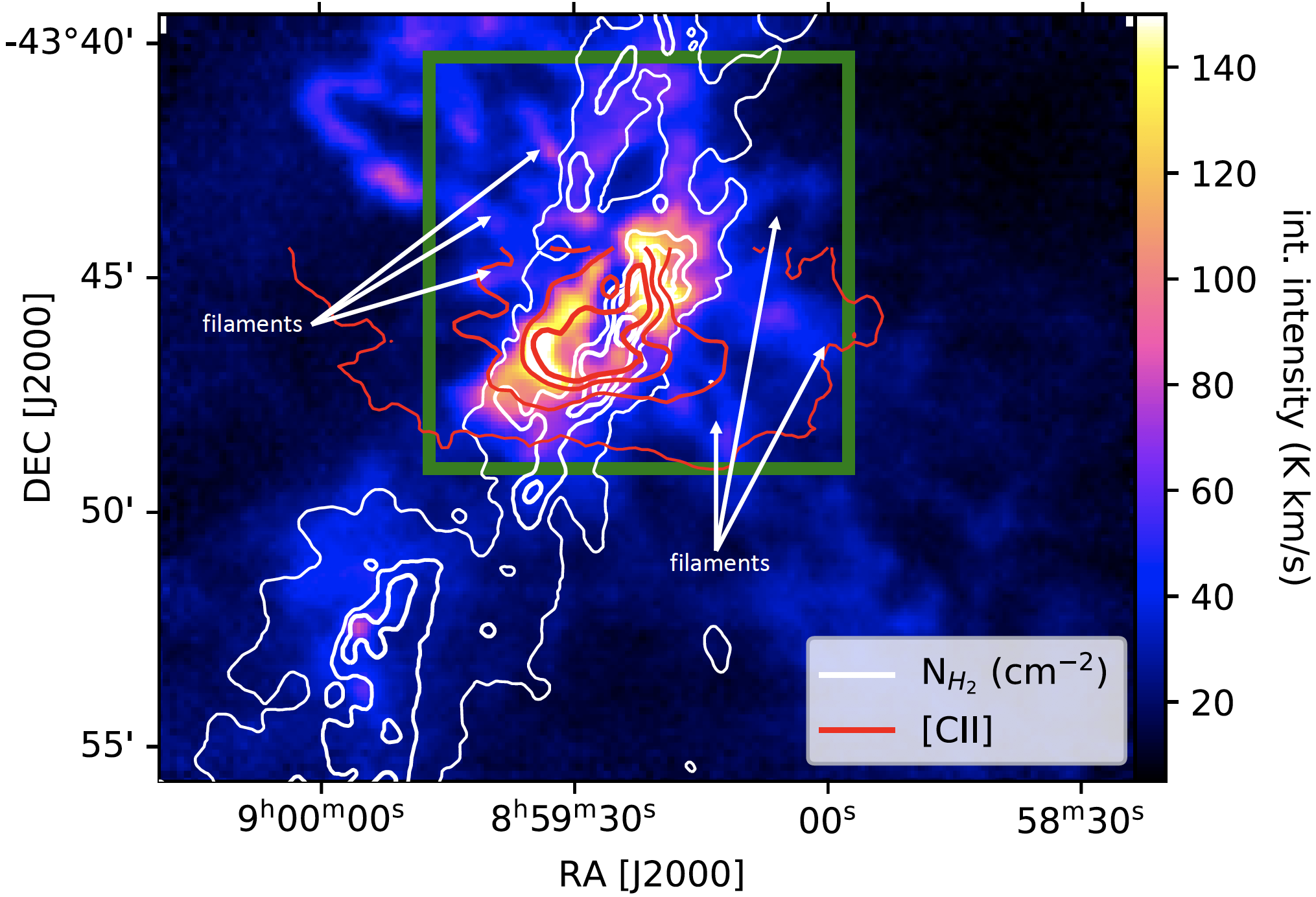}
    \includegraphics[width=\hsize]{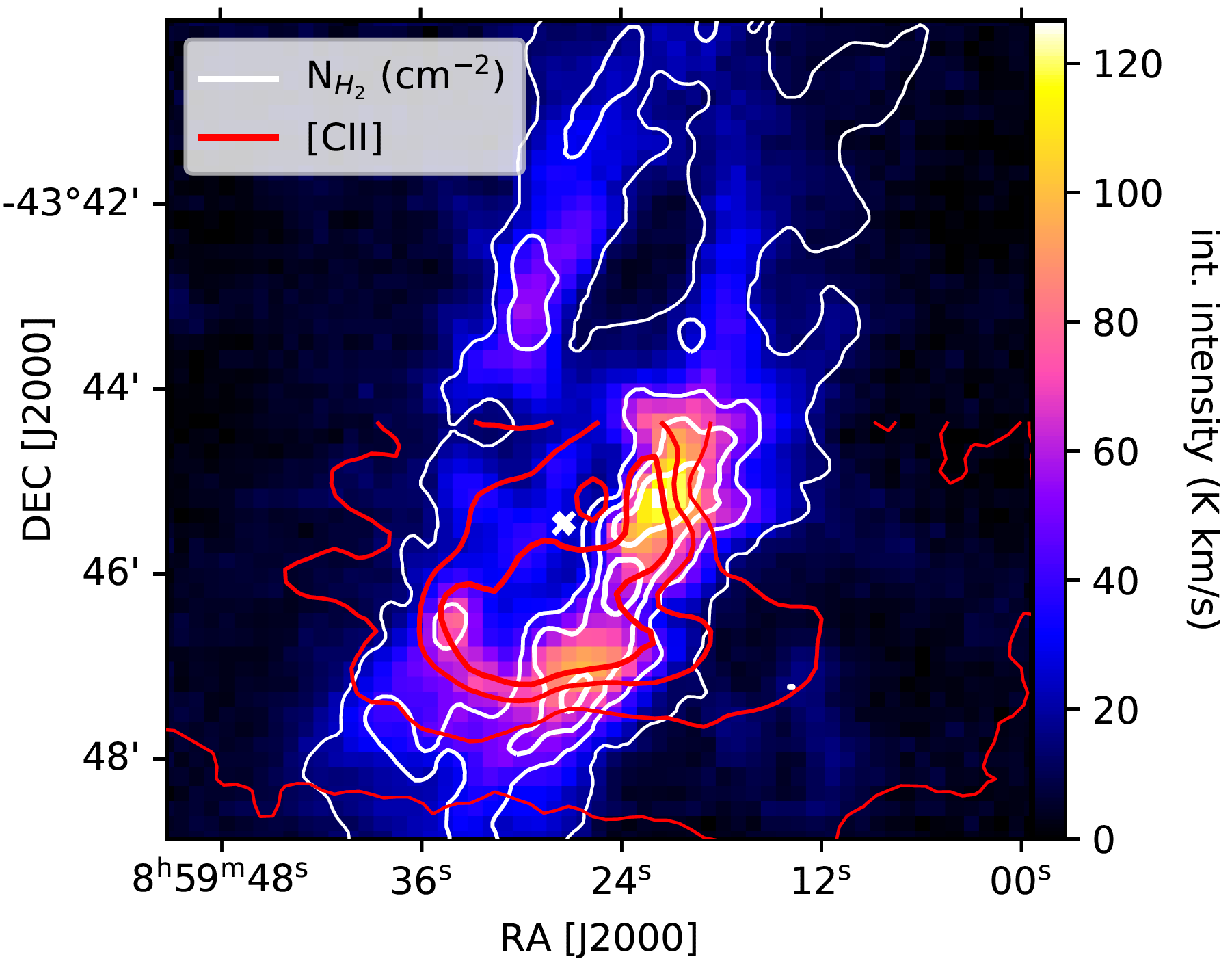}
    \caption{{\bf Top}: The integrated $^{12}$CO(3-2) emission map from -20 to 20 km s$^{-1}$ with \textit{Herschel} column density contours in white and \CII\ integrated intensity contours starting at 50 K km~s$^{-1}$ with increments of 100 K km~s$^{-1}$ in red to delineate their structure. Outside the molecular ring, the $^{12}$CO(3-2) emission shows filamentary structures that are oriented perpendicular to the ring (indicated by the white arrows). {\bf Bottom}: The same for $^{13}$CO(3-2), in the green box indicated in the figure above, which follows the column density contours of the high-column density ring closely. The \CII\ emission shows a structure perpendicular to the high-column density ring in the south-west at the location where there is a drop in $^{13}$CO(3-2) intensity. The white crosses indicate the location of the two O star candidates.}
    \label{fig:integratedMapsCO}
\end{figure}

 Figures~\ref{fig:integratedIntensities} and \ref{fig:integratedOI} present the integrated intensity map for \CII\ and \OI, respectively. \CII\ is brightest towards the region that is surrounded by the high-column density ring and shows more extended emission towards the two cavities. 
 Outside of the cavity walls, the \CII\ intensity decreases rapidly. \OI\ on the other hand is only detected towards the high column density ring where a high-density PDR can be expected.\\
The line integrated intensity maps of $^{12}$CO(3-2) and $^{13}$CO(3-2), presented in Fig. \ref{fig:integratedMapsCO}, show a different structure. The emission follows the ring-like \textit{Herschel} dust column density maps rather closely and shows that the CO emission is brightest at the edge around the bright \CII\ emission, which would be expected for a PDR irradiated by the central cluster. This is not a resolution effect since both beams have 
similar spatial resolution. 
To the south-west of the ring there is a region (local peak in dust column density), where the ring is significantly less bright in $^{12}$CO(3-2) and $^{13}$CO(3-2) (by a factor 2-3), but the \CII\ contours show a protrusion toward the west. This may indicate that there is a puncture in the ring so that the \CII\ emitting gas can freely expand into a lower density region, driven by the stellar feedback from the east. On the other hand, the weaker CO emission at this location (C3 in Fig.~\ref{fig:integratedIntensities}) can be partly caused by self-absorption. This self-absorption can be related to the protrusion as it would give rise to warm and cold gas layers along the line of sight while there still appears to be a significant column density based on the Herschel data. 
On a larger scale outside the ring in Fig. \ref{fig:integratedMapsCO}, $^{12}$CO(3-2) shows filamentary emission that is perpendicular to the dense ring. Several of these filamentary structures are also detected in \CII\ and show a curvature that appears to originate in the central cluster. This suggests that pre-existing filaments are being swept up. \\
Recently, \CII\ integrated intensity observations at an angular resolution of 90$^{\prime\prime}$ were presented in \citet{Suzuki2021} using a 100 cm balloon-borne infrared telescope. Smoothing the FEEDBACK integrated intensity map to a resolution of 90$^{\prime\prime}$, presented in App. \ref{sec:90arcsec}, the results look extremely similar to the maps presented in \citet{Suzuki2021}. The intensity maps appear to agree within 10\%, which corresponds to typical calibration errors. Comparing the smoothed map in Fig. \ref{fig:compSuzuki} with the FEEDBACK map in Fig. \ref{fig:integratedIntensities}, it is evident that the higher spatial resolution of the SOFIA observations is necessary to resolve e.g. the ring and filamentary structures in \CII. Another major advantage of the FEEDBACK observations is the high spectral resolution of the observations that allows to study the dynamics of the region.

\subsection{Multiple velocity components, absorption and high-velocity wings}\label{sec:velComps}

\begin{figure}
    \centering
    \includegraphics[width=\hsize]{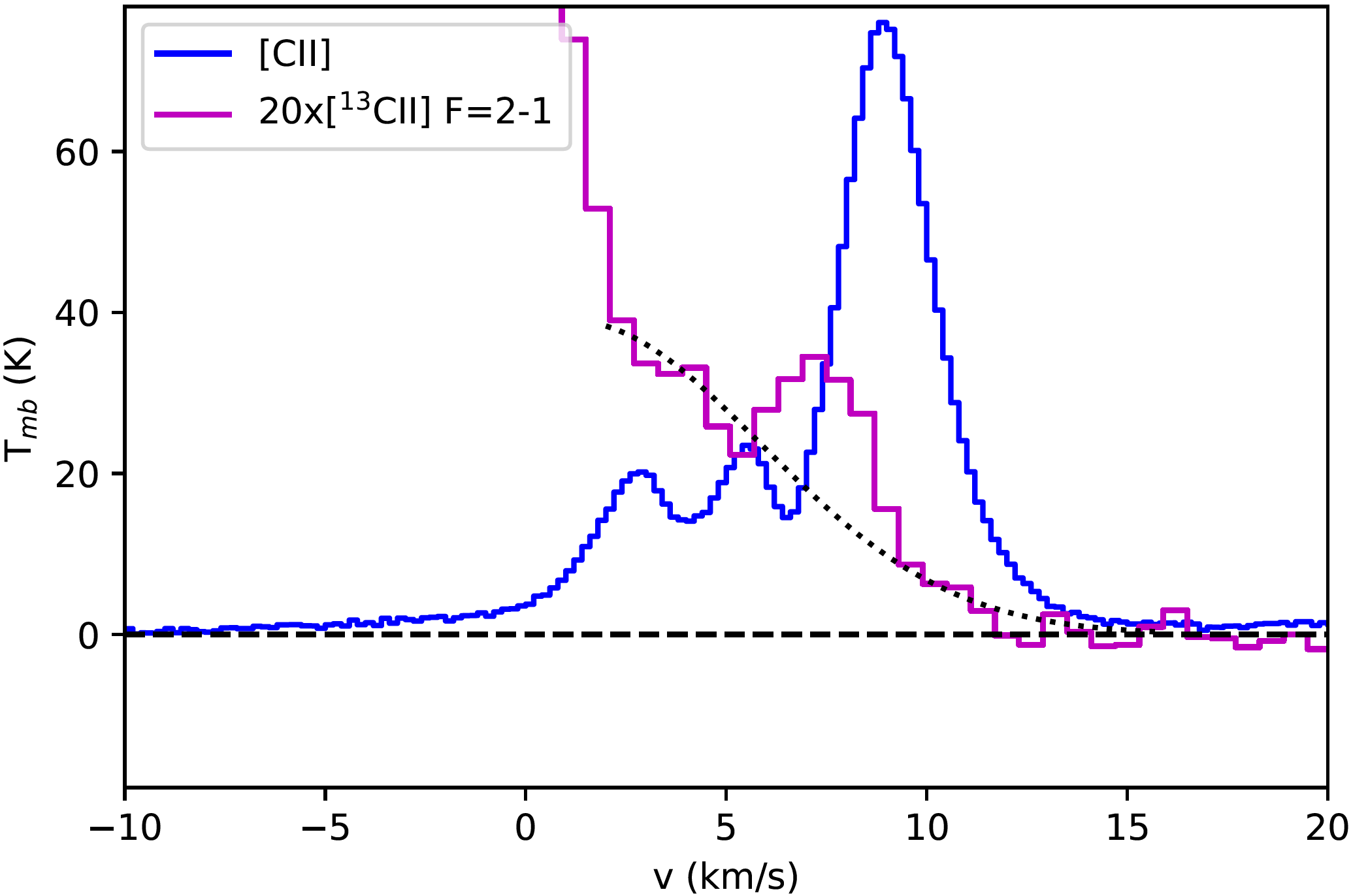}
    \caption{The \dCII\ F = 2-1 line emission (magenta), averaged over the dense molecular ring, compared to the \CII\ line profile (blue) which is averaged over the same area. Even though there is confusion by the redshifted high-velocity wing seen with \CII, indicated by the fitted curve to the wing (part of a gaussian), the \dCII\ emission profile clearly shows that the intensity dip at $\sim$ 7 km s$^{-1}$ in \CII\ and $^{12}$CO(3-2) is the result of self-absorption.}
    \label{fig:13CIIline}
\end{figure}

The \CII, CO, and \OI\ spectra shown in Fig. \ref{fig:integratedIntensities} are all very complex and cannot be described by a single Gaussian velocity component. 
The bulk emission of the molecular cloud/dense ring is best traced with the $^{13}$CO(3-2) line that is visible for all center positions (C1-5) around 7-8 km s$^{-1}$ but mostly absent for all western (W1-W3) and eastern (E1-E2) positions. The $^{12}$CO(3-2) line profile shows significant variation depending on the position. For the spectra at the central molecular ring, we observe two apparent line components that are caused by foreground or self-absorption because the $^{13}$CO line peaks in the $^{12}$CO dip at $\sim$7 km s$^{-1}$. Note that the optically thin dense gas tracers N$_{2}$H$^{+}$(2-1) and NH$_{3}$(1,1), presented in  \citet{Fissel2019},  also peak at velocities around 7-8 km s$^{-1}$. The absorption thus mimics two separate  
velocity components that were interpreted by \citet{Sano2018} as two velocity components tracing a cloud-cloud collision. 
We observe that the absorption in $^{12}$CO continues towards the east and west because the dip in the line profile is always around 7 km s$^{-1}$ and the $^{13}$CO emission clearly peaks in the $^{12}$CO dip for position W1 (but is too weak for W2 and E1 and E2). For all center positions, the self-absorption is dominant towards the blue-shifted part of the main velocity component which is generally associated with expansion \citep[e.g.][]{Castor1979}. \\
The \CII\ line has a bright, prominent component (main beam temperatures up to $\sim$100~K for the center positions) around 8-9 km s$^{-1}$. This bright component is also found in the eastern spectra, but is less strong for the three western positions (W1-W3). This component is also affected by absorption which we can prove with our \dCII\ line observations shown in Fig. \ref{fig:13CIIline} and Appendix \ref{sec:selfAbs}. The intensity of the \CII\ main component drops where \dCII\ has its emission peak, similar to what was observed for several other bright \CII\ sources with SOFIA \citep[e.g.][]{Graf2012,Guevara2020,Kirsanova2020,Mookerjea2021,Kabanovic2022}. There is also a separate second velocity component between 0 and 5 km s$^{-1}$ (hereafter the blue-shifted component) which is well visible in \CII\ and even brighter than the 7-8 km s$^{-1}$ component in the west of the ring (W1-W3). At some locations (C3, C4, W3) this component is also detected in $^{12}$CO(3-2), see Fig. \ref{fig:integratedIntensities}, but mostly at slightly higher velocities (3-5 km s$^{-1}$).\\
The \OI\ spectra, detected towards the dense ring, look slightly different than $^{12}$CO(3-2) and \CII. \OI\ is only detected at velocities higher than 7 km s$^{-1}$. At lower velocities, i.e. between 3 and 7 km s$^{-1}$, the spectra indicate absorption since the emission in this velocity interval goes under the baseline in C1-C5. This is demonstrated in more detail in App. \ref{sec:selfAbs} which shows that the integrated intensity of \OI\ from 3 to 7 km s$^{-1}$ becomes negative towards the dense ring, in particular towards the \textit{Herschel} column 
density peaks. This suggests significant \OI\ 63 $\mu$m foreground absorption at the same velocities where \CII\ and $^{12}$CO(3-2) are absorbed, similar to what was recently found towards several bright \OI\ 63 $\mu$m sources in \citet{Goldsmith2021}. Based on detailed studies of several clouds,  i.e.  M17 and MonR2 (Guevara et al., to be submitted) and RCW120 \citep{Kabanovic2022}, it has been proposed this could be due to a foreground cloud or cold atomic oxygen and C$^{+}$ in the low-density halo of the cloud.
Lastly, in several spectra of Fig. \ref{fig:integratedIntensities}, high-velocity blue- and red-shifted wings are observed in the intervals from -8 
to 0 km s$^{-1}$ and 14 to 23 km s$^{-1}$.  Even though these high-velocity wings are not always very clear on the presented figures because of the very bright emission at 5-12 km s$^{-1}$, associated with the central ring, they are well above the noise level (which is 0.8 K in individual spectral bins). The wings are particularly evident at positions C2, C3 and W3 in Fig. \ref{fig:integratedIntensities}. For the red-shifted wing, it has to be noted that the \dCII\ F=2-1 transition is located between 17 and 23 km s$^{-1}$ which provides a contribution (1.5 K km s$^{-1}$ on average) to the wing at these velocities.
\subsection{Gas dynamics in \CII\ and CO}

\begin{figure}
    \includegraphics[width=0.88\hsize]{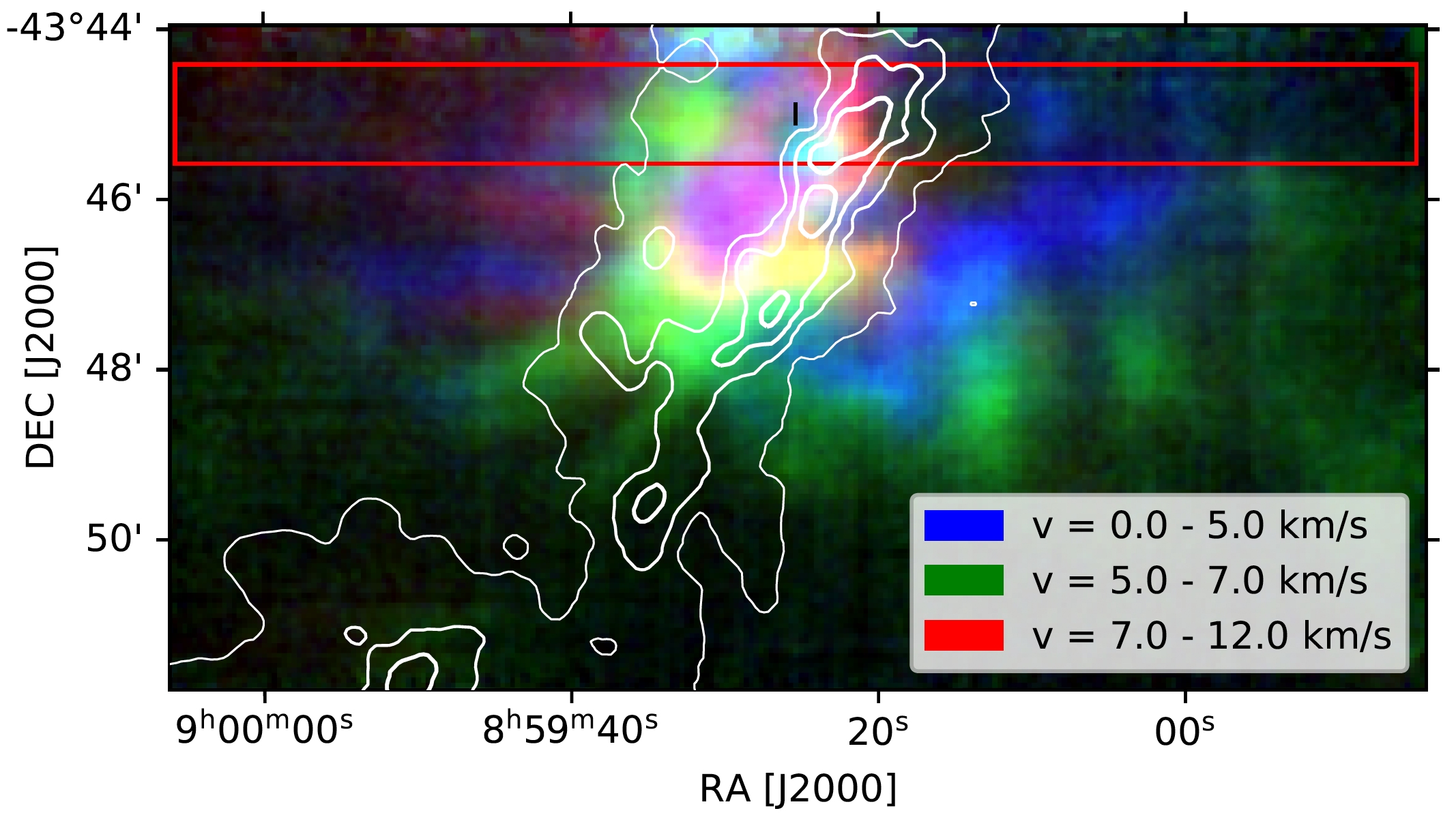}
    \includegraphics[width=0.88\hsize]{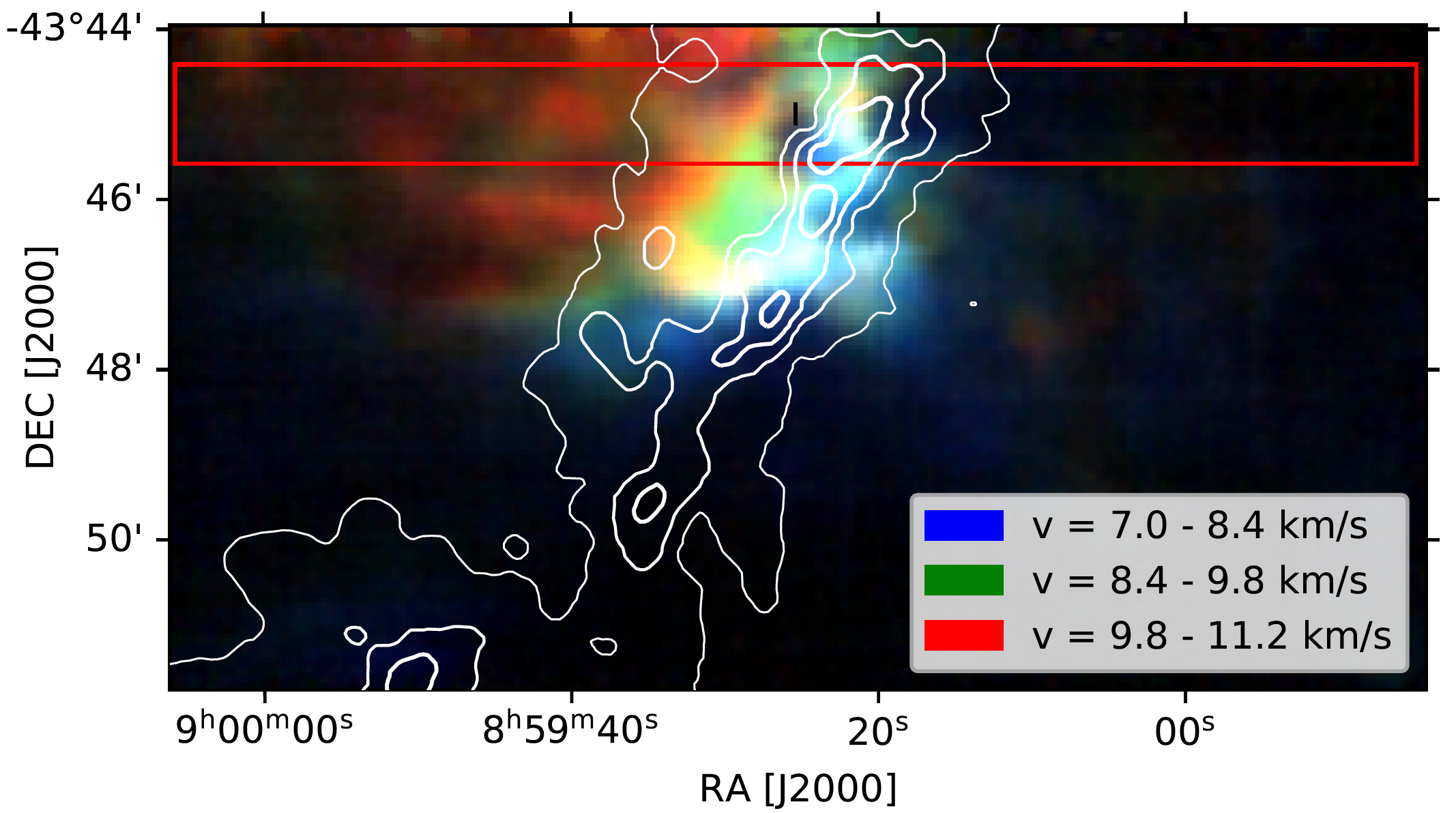}
    \includegraphics[width=\hsize]{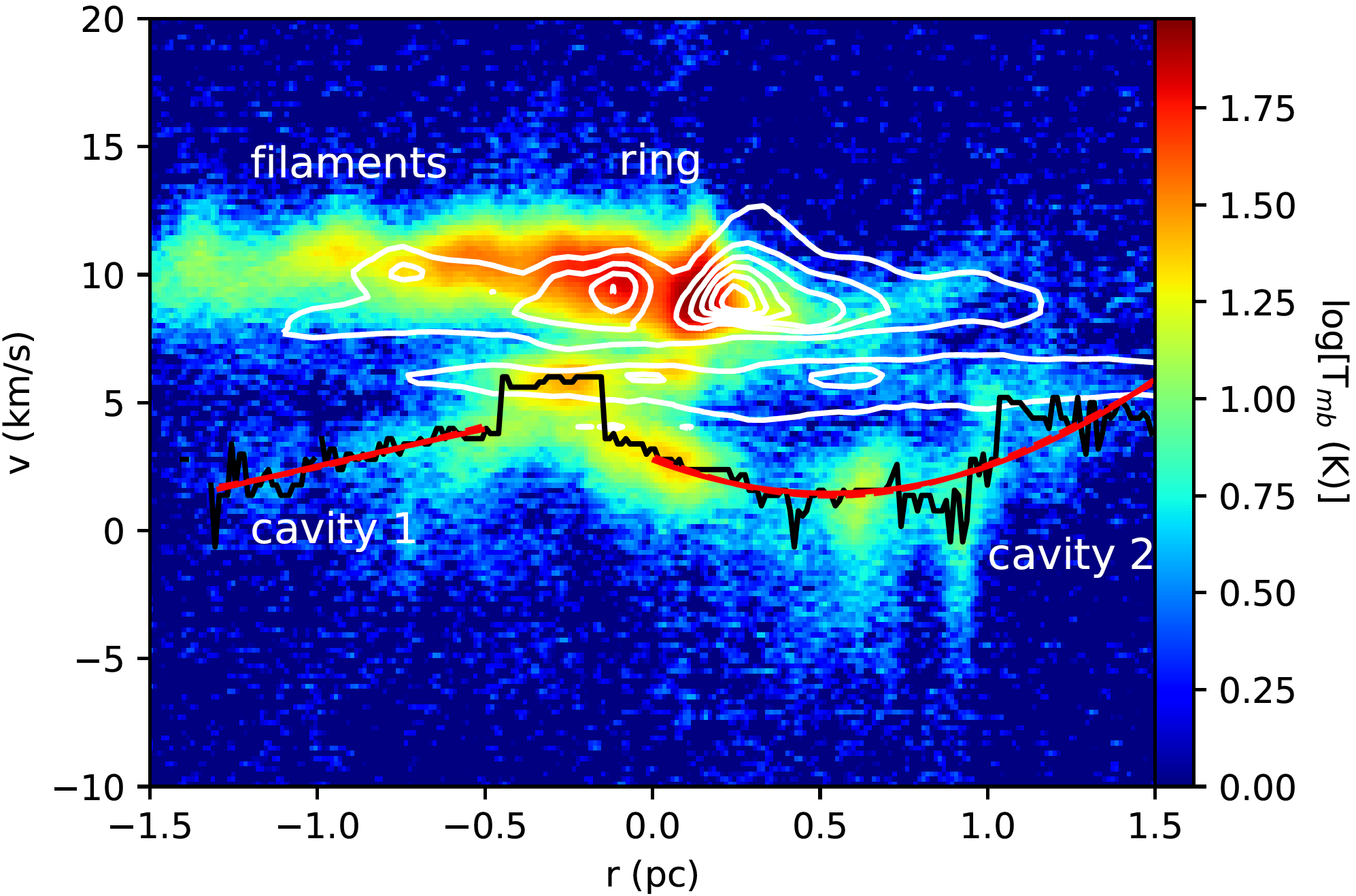}
    \caption{{\bf Top}:  
    \CII\ RGB map presenting the global kinematic structure of RCW 36 between 0 and 12 km s$^{-1}$. 
    The color coding is given in the panel. The white contours indicate the column density 
    N$_{H_{2}}$ at 1, 2, 4, and 8$\times$10$^{22}$ cm$^{-2}$.  
    The red bar outlines the area of the map used to produce the position-velocity (PV) diagram below. The small black vertical line inside the red bar indicates the center (0 pc) for the PV diagram. 
    {\bf Middle}: The same as above, but focusing on the velocity structure in the most red-shifted velocity interval between 7 and 11.2 km s$^{-1}$.
    {\bf Bottom}: The \CII\ PV diagram parallel to the bipolar cavities and the $^{12}$CO(3-2) emission in white contours (starting at 5 K km s$^{-1}$ with increments of 10 K km s$^{-1}$). The full and dashed red curves, which closely overlap and thus are difficult to separate on this figure, show the fitted functions (2$^{nd}$ order polynomial: full, sine function: dashed) to the two identified blue-shifted expanding \CII\ shells that are found in the two cavities. The black line follows the peak intensity along the expanding shells as a function of radius. 
    At velocities $<$ 0 km s$^{-1}$ and $>$ 12 km s$^{-1}$, the emission from the high-velocity wings is observed. This emission does not show an elliptical shape in the PV diagram, which would be predicted for expanding shells. 
    }
    \label{fig:pvCavities}
\end{figure}

\subsubsection{The expanding shells}\label{sec:expShells}
\begin{figure*}
    \centering
    \includegraphics[width=0.49\hsize]{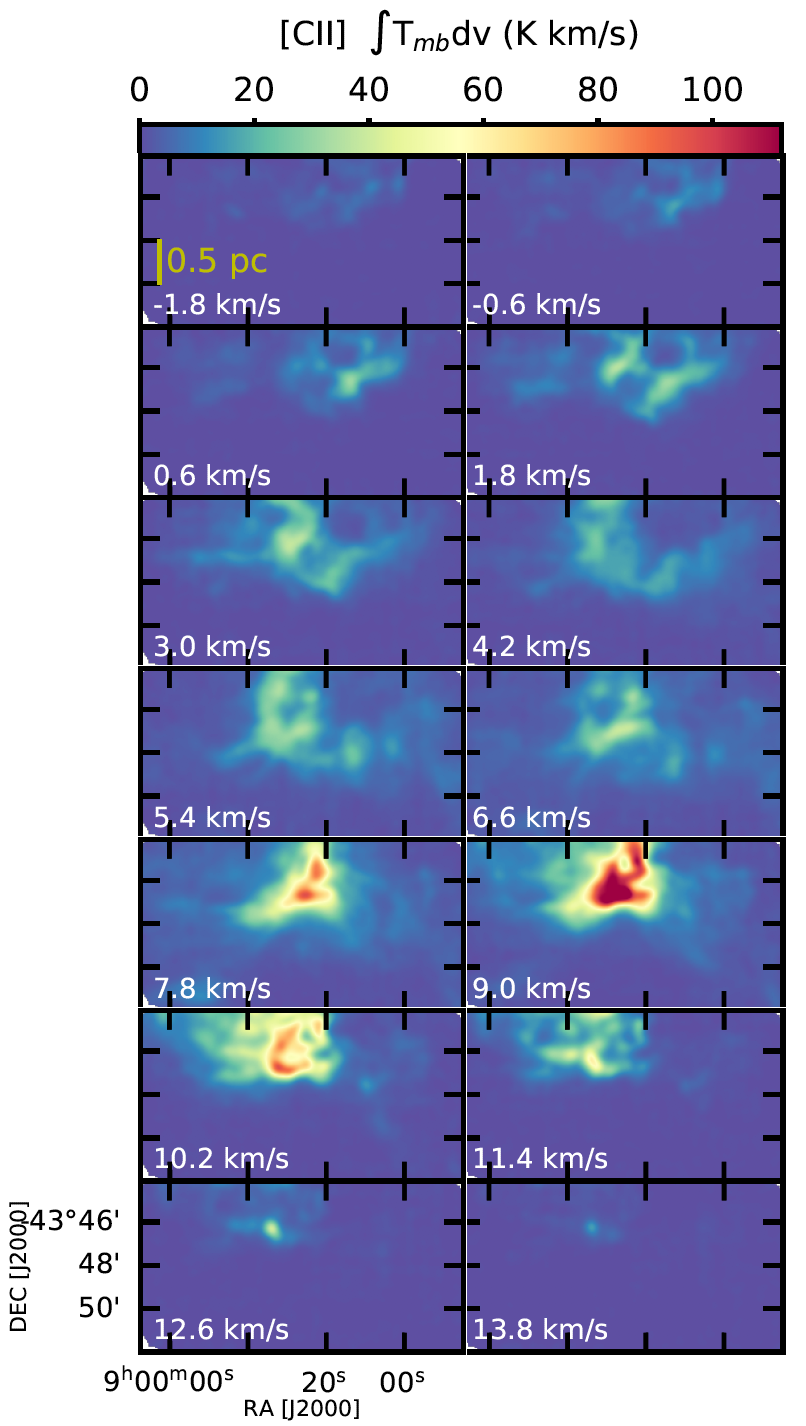}
    \includegraphics[width=0.49\hsize]{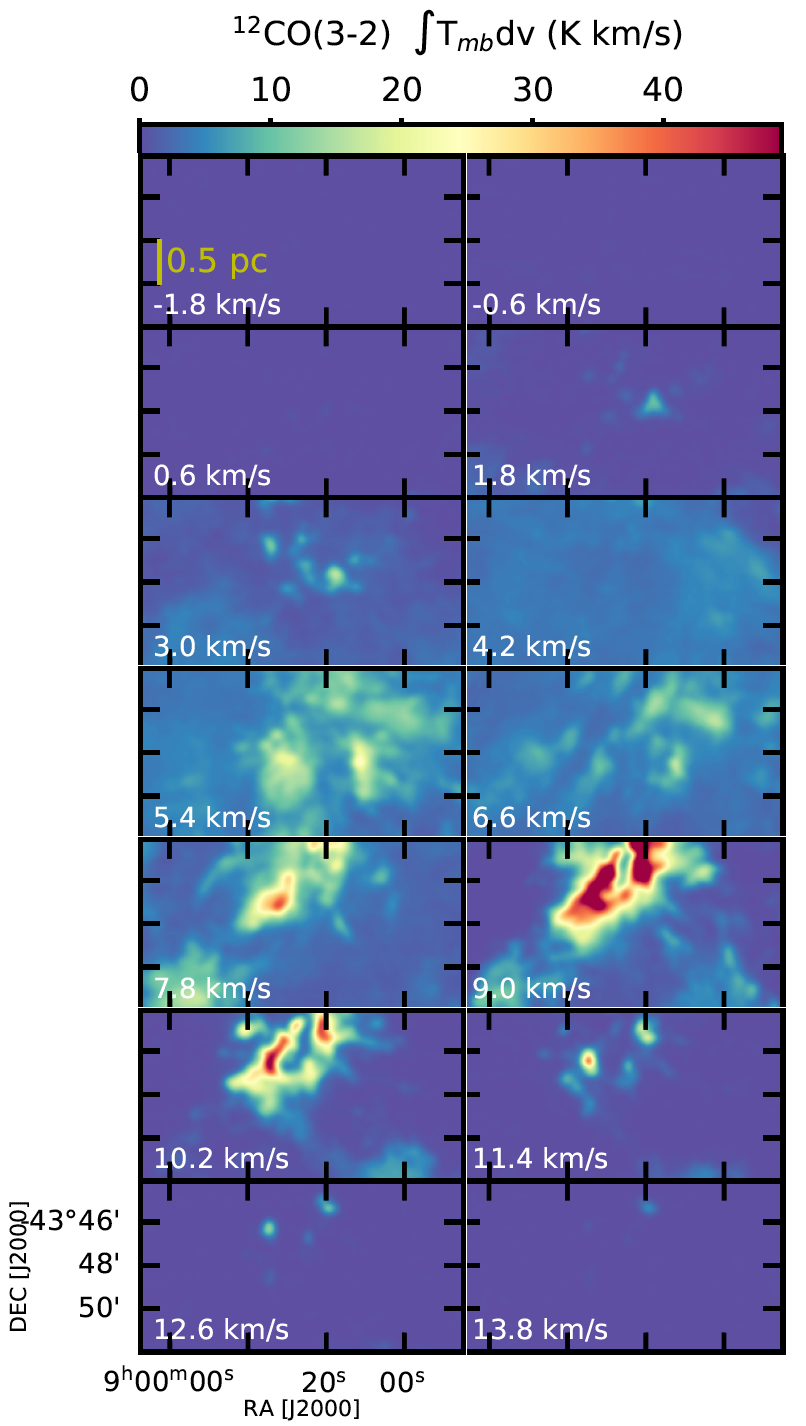}
    \caption{{\bf Left}: Channel maps of \CII\ from -1.8 km s$^{-1}$ to 13.8 km s$^{-1}$ with velocity intervals of 1.2 km s$^{-1}$ present the velocity structure of RCW 36. Below 6 km s$^{-1}$, the blueshifted expanding shells are observed in the colormap for \CII. {\bf Right}: The same for $^{12}$CO(3-2) covering the same area as the \CII\ map. It appears that $^{12}$CO(3-2) only marginally traces the expanding shell. At higher velocities, $^{12}$CO(3-2) mostly traces the kinematics of the Vela C molecular cloud.}
    \label{fig:chanMapsFull}
\end{figure*}

The \CII\ position-velocity (PV) diagram presented in Fig. \ref{fig:pvCavities} is a cut through RCW 36, parallel to the central axis of the bipolar cavities. This provides a view on the large scale kinematics in RCW 36. It displays a blue-shifted expanding shell signature in the western region of the observed map, i.e. r $>$ 0 pc in Fig. \ref{fig:pvCavities} (indicated with 'cavity 2'), where the western cavity is located. This expanding structure is similar to the ones observed towards bubble-like structures in Orion, RCW 120 and RCW 49 \citep[][]{Pabst2020,Luisi2021,Tiwari2021}. The expansion feature is observed between 1 and 6 km s$^{-1}$ which suggests that this expanding shell has an expansion velocity of $\sim$ 5 km s$^{-1}$. Later in this section we will constrain the expansion velocity quantitatively. The emission from the shell is also clearly identified in the individual spectra of Fig. \ref{fig:integratedIntensities} as a separate velocity component in the same velocity interval.
This feature is most prominent in the spectra of the western cavity (W1-3) and dominates over the component associated with the velocity of the molecular ring.  
The cavity to the east (r $<$ 0 pc) has a similar, though less bright, expanding shell that covers the same velocity range, see Fig. \ref{fig:pvCavities}. As the eastern cavity is less bright, it is not as clear in the PV diagram, but when inspecting the \CII\ channel maps in Fig. \ref{fig:chanMapsFull}, it becomes evident that the \CII\ emission between 1 and 6 km s$^{-1}$ shows the same morphology as the western cavity. This blue-shifted expanding shell morphology is also seen in the upper RGB image of Fig. \ref{fig:pvCavities}, where the center of both cavities shows blue-shifted emission confined within the green limb brightened cavity walls. Both the western and eastern cavity thus show a blue-shifted expanding shell in \CII\ with similar expansion velocity.  
To quantify the expansion velocity of these shells in the cavities, we determined the velocity of peak intensity for the shells (i.e. at v$_{LSR} <$ 6.5 km s$^{-1}$), see Fig. \ref{fig:pvCavities}. This curve was then fitted in each cavity by both a second order polynomial and a sine function which assumes that the reference velocity for expansion is v$_{LSR} =$ 6.5 km s$^{-1}$. To calculate the expansion velocity, we then determined the point of minimal v$_{LSR}$ obtained from the fit. The difference between this minimal velocity and v$_{LSR}$ = 6.5 km s$^{-1}$, is the expansion velocity. The results of these fits are summarized in Tab. \ref{tab:shellVelocities}. The expansion velocity in the cavities is then estimated by taking the average of all obtained expansion velocities, which gives: 5.2 $\pm$ 0.5 $\pm$ 0.5 km s$^{-1}$. The first estimated error corresponds to the spread of all obtained expansion velocities. The second estimated error is a systematic error caused by the uncertainty on the rest v$_{LSR}$ for the expanding shell which should be between v$_{LSR}$ = 6 and 7 km s$^{-1}$.  
The two expanding shells appear to spatially come together in projection at the central molecular ring, which is also observed in the channel maps in Fig. \ref{fig:chanMapsFull}. The $^{12}$CO(3-2) channel maps in Fig. \ref{fig:chanMapsFullCO} show indications that the outer layer of the western expanding shell is also detected in $^{12}$CO(3-2) between 1.4 and 3 km s$^{-1}$. This is not the case for all expanding \CII\ shells \citep[e.g.][]{Luisi2021}. 
In PDRs, CO generally exists in regions with A$_{\text{V}} \ge$ 2-3 \citep[e.g.][]{Tielens1985,Roellig2007} which implies that the expanding shell has a thickness of several A$_{\text{V}}$. This is consistent with the \textit{Herschel} column density map of Vela C which shows an A$_{\text{V}}$ of 3-6 towards the cavities.

\begin{table*}[]
    \centering
    \begin{tabular}{ccccc}
    \hline
    \hline
        function & v$_{LSR,min,west}$ (km s$^{-1}$) & v$_{exp,west}$ (km s$^{-1}$) & v$_{LSR,min,east}$ (km s$^{-1}$) & v$_{exp,east}$ (km s$^{-1}$)\\
        \hline
        sine & 1.4 $\pm$ 0.1 & 5.1 $\pm$ 0.1 $\pm$ 0.5 & 0.8 $\pm$ 0.9 & 5.7 $\pm$ 0.9 $\pm$ 0.5\\ 
        poly & 1.5 $\pm$ 0.1 & 5.0  $\pm$ 0.1 $\pm$ 0.5 & 1.6 $\pm$ 0.1 & 4.9 $\pm$ 0.1 $\pm$ 0.5 \\
        \hline
    \end{tabular}
    \caption{Minimal velocity (v$_{LSR,min}$) and corresponding expansion velocity (v$_{exp}$) obtained for the two fitted functions to the expanding shells in the west and the east, respectively. The fitted functions are a sine function (sine) and a second order polynomial (poly). The indicated errors for the minimal v$_{LSR}$ correspond to the error on the fit. The two indicated errors for the expansion velocities correspond to the fitting error and uncertainty on the assumed rest v$_{LSR}$ for the shells.}
    \label{tab:shellVelocities}
\end{table*}

\subsubsection{The kinematics of the molecular ring}\label{sec:ringExp}

\begin{figure}
    \centering
    \includegraphics[width=\hsize]{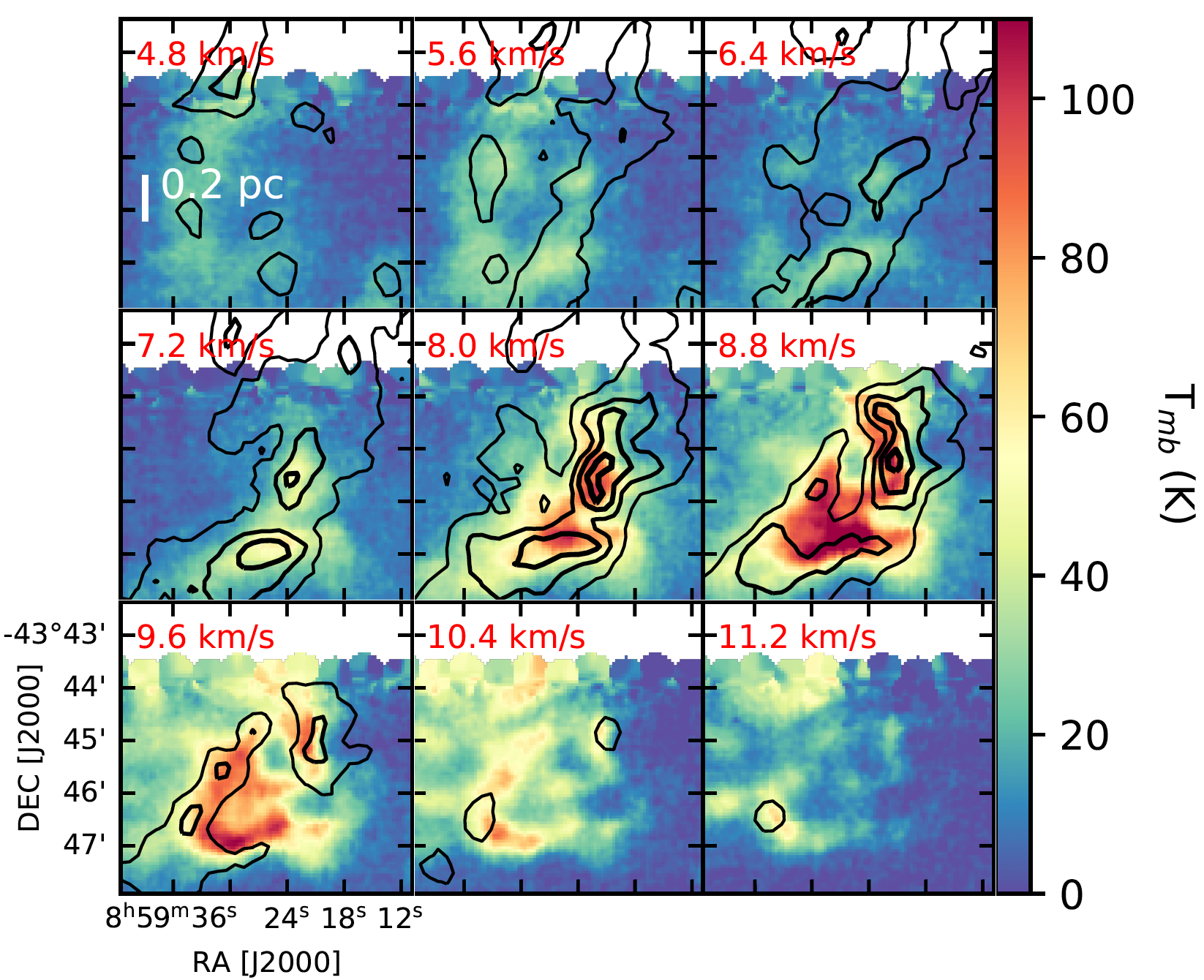}
    \caption{
    A zoom of the \CII\ channel maps on the ring around the young OB cluster, to focus on the velocity structure in the ring. The overlaid black contours indicate the $^{13}$CO(3-2) channel maps at the same velocity starting at 5 K km s$^{-1}$ with increments of 10 K km s$^{-1}$. It shows that the kinematics in \CII\ and $^{13}$CO(3-2) are very similar towards the ring, unlike the cavities. The eastern part of the ring is already detected starting at 5 km s$^{-1}$ while the western part of the ring is only detected starting at 7-8 km s$^{-1}$.} 
    \label{fig:chanMapRing}
\end{figure}
Our observations also resolve the central molecular ring which is best visible in the dust column density map and in the $^{13}$CO(3-2) emission (Fig.~\ref{fig:integratedMapsCO}). However, the ring is also associated with \CII\ emission as shown in the  
RGB images of Fig. \ref{fig:pvCavities}. The top panel indicates that the eastern part of the ring is more blue-shifted than the western part. Since it was proposed, by \citet{Minier2013} and from Fig. \ref{fig:observedMap}, that the eastern part of the ring is in front of the western part, this velocity structure fits with the proposed expansion of the molecular ring. A zoom into the ring kinematics by the channel maps in Fig. \ref{fig:chanMapRing} confirms this expansion and demonstrates that the expansion is also seen in $^{13}$CO(3-2). Unfortunately, it is very difficult to accurately estimate an expansion velocity for the ring. There are several reasons for this: self-absorption effects in all available lines limits an accurate determination of the typical velocity of the gas, stellar feedback breaking through the ring affects the velocity field, and the presence of additional velocity structure in the line of sight (see also Bij et al. in prep.) limits the possibility to use the moment 1 map. Therefore, we estimate the expansion velocity from inspecting the channel maps in Fig. \ref{fig:chanMapRing}. The eastern area of the ring is found at 5-6 km s$^{-1}$ whereas the western part is at $\sim$8.0 - 8.8 km s$^{-1}$. This suggests a typical expansion velocity of 1.0-1.9 km s$^{-1}$.
More generally, Fig. \ref{fig:chanMapRing} shows that \CII\ and CO have the same kinematic structure towards the ring which implies that they trace the same expansion dynamics. The full ring is thus swept up by the expansion from the cluster with the CO emission at the outer side of the \CII\ emission as would be expected in a PDR structure with internal FUV irradiation. Lastly, in the south-west of the ring, where Fig. \ref{fig:integratedMapsCO} shows that the integrated \CII\ emission crosses the molecular ring, Fig. \ref{fig:chanMapRing} now shows that $^{13}$CO and \CII\ have the same kinematic morphology between 8.0 and 9.6 km s$^{-1}$. 

\subsubsection{Filamentary structures in \CII}\label{sec:filsExp}
\begin{figure*}
    \centering
    \includegraphics[width=0.9\hsize]{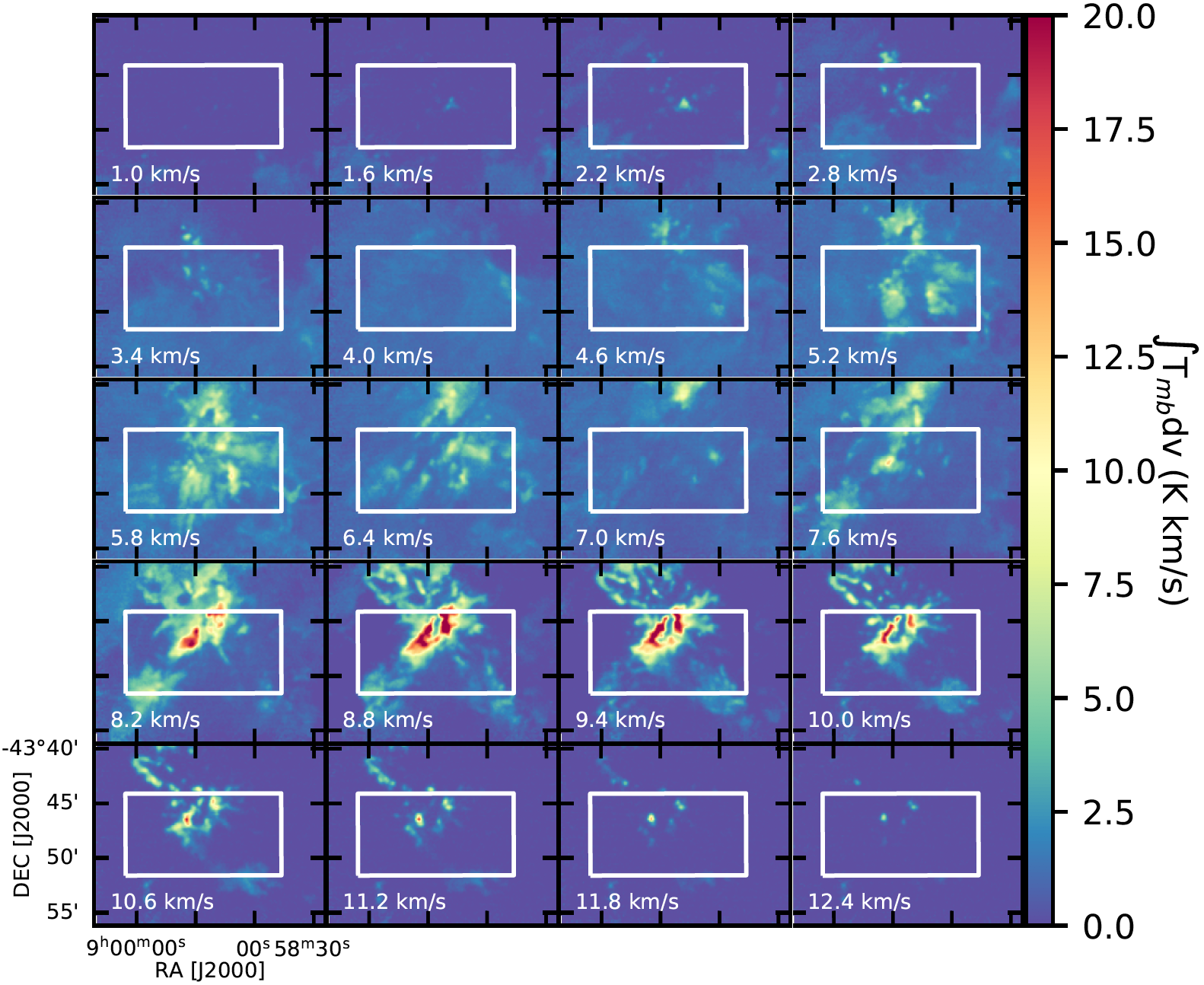}
    \caption{Channel maps of $^{12}$CO(3-2) from 1.0 km s$^{-1}$ to 12.4 km s$^{-1}$ over the full APEX map of RCW 36. The white box indicates the currently observed \CII\ map. Below 4 km s$^{-1}$, emission in an arc is observed that indicates that the shell is also (weakly) detected in $^{12}$CO(3-2). Note the clumpy nature of this $^{12}$CO emission originating from the shell. At velocities above 8 km s$^{-1}$, filamentary gas perpendicular to the central ring is observed.}
    \label{fig:chanMapsFullCO}
\end{figure*}

The PV diagram in Fig. \ref{fig:pvCavities} also shows significant \CII\ emission around 10 km s$^{-1}$, with corresponding $^{12}$CO(3-2) emission towards the eastern cavity.  This emission is indicated with the name $'$filaments$'$ in the PV diagrams. The typical elliptical signature expected in PV diagrams for expanding shells is not as evident for this \CII\ emission which could be due to the small velocity difference with the ring ($<$~2-3 km s$^{-1}$). In the channel maps in Figs. \ref{fig:chanMapsFull} and \ref{fig:chanMapsFullCO}, this 10 km s$^{-1}$ emission displays a filamentary structure which is not typical for expanding shells. However, Fig. \ref{fig:chanMapsFullCO} in particular also shows that these filaments have a curved morphology that appear to have their origin towards the central cluster. This \CII\ and CO emission might thus originate from low-density filaments, originally converging towards the ridge where RCW 36 formed, which are now being swept away at velocities $<$~3 km s$^{-1}$ by the feedback from the stellar cluster. Such filamentary gas, which might originally be flowing towards a central filament or ridge, was observed for a wide variety of low- to high-mass star formation regions such as e.g. Musca, B211/3 and DR21 \citep{Goldsmith2008,Schneider2010,Palmeirim2013,Cox2016,Bonne2020b}. The observed small expansion velocity for these filaments around RCW 36 is quite similar to the expansion velocity in the ring which suggest that overdensities, which are clearly detected in CO,  experience significantly slower expansion driven by stellar feedback. Lastly, no corresponding bright filamentary \CII\ emission at 10 km s$^{-1}$ is observed towards the western cavity, which might be the result of the molecular cloud morphology around RCW 36 before the onset of stellar feedback. 

\subsection{High-velocity wing emission}
\begin{figure}
    \centering
    \includegraphics[width=\hsize]{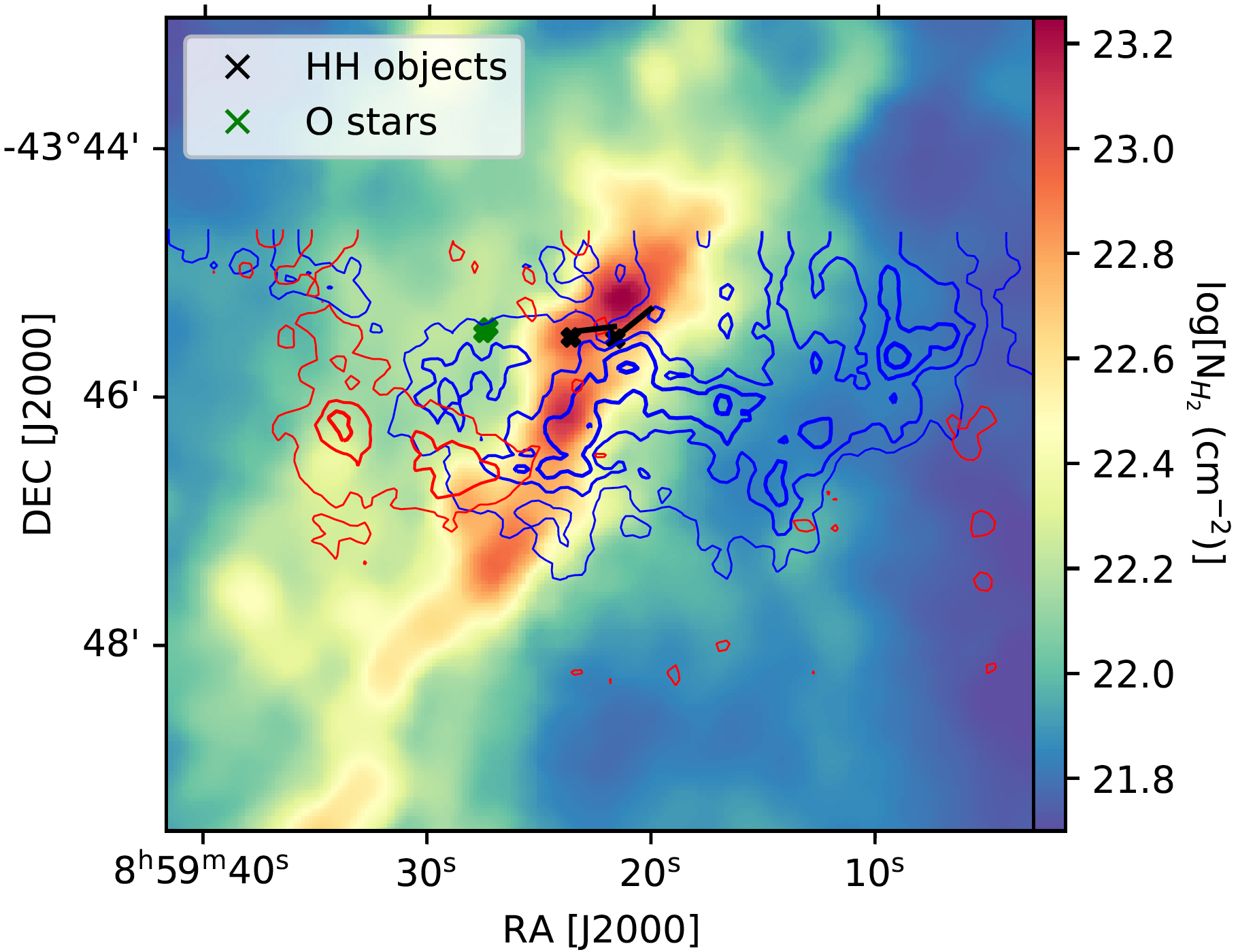}
    \caption{\textit{Herschel} column density map overlaid with the contours from the integrated blue-shifted high-velocity wing (in blue) and the red-shifted high-velocity wing (in red) of \CII. The location of the two O stars and HH objects, with the orientation of their observed jet (black lines), is also indicated. This shows that the observed high-velocity wings are not associated with the HH jets.}
    \label{fig:outflowColdens}
\end{figure}

The spectra in Sec. \ref{sec:velComps} showed the presence of blue- (v$_{LSR} <$ 0 km s$^{-1}$) and red-shifted (v$_{LSR} >$ 14 km s$^{-1}$) high-velocity \CII\ wings. These high-velocity wings have a brightness up to 5-10 K in individual spectral bins, see Fig. \ref{fig:integratedIntensities}, which is well above the detection limit for \CII\ with a noise rms of 0.8 - 1.0 K. In Fig. \ref{fig:outflowColdens} the integrated intensity map of the high-velocity wings shows that the blue- and red-shifted wings display a bipolar distribution, similar to what is observed for protostellar outflows in CO \citep[e.g.][]{Bachiller1996,Arce2007}. However, for RCW 36, there are no known driving protostars or young stellar objects \citep{Ellerbroek2013a} at the location where the blue- and red-shifted outflow wings come together in the plane of the sky.  
Furthermore, the high-velocity wings are only detected in \CII. Even at the high column density ring the wings are at no point detected in $^{12}$CO (see spectra C1-C3 in Fig. \ref{fig:integratedIntensities}), and higher angular resolution observations (Bij et al., in prep.) indicate there are no protostellar sources in the ring that could drive such powerful mass ejection. The observed \CII\ wings are thus most likely not of protostellar nature.\\
In the currently covered area of RCW 36, there is a significant disparity between the blue- and red-shifted integrated wings. Even when integrating over the full 14-22 km s$^{-1}$ interval, which contains contamination from \dCII, the red-shifted wing remains 2 times less bright and covers a significantly smaller area than the blue-shifted wing. These high-velocity wings also do not appear to be directly associated with expanding shells as it does not show an expanding shell velocity structure in the PV diagram in Fig. \ref{fig:pvCavities} (i.e. an elliptical velocity profile). 
The brightest part of the blue-shifted wing can be found relatively close to the two reported HH objects in \citet{Ellerbroek2013b}. However, Fig. \ref{fig:outflowColdens} also shows the position angle of the HH jets which indicates that the wings seen in \CII\ are not associated with these HH jets. Lastly, Fig. \ref{fig:outflowColdens} shows that the brightest wings are located around the dense molecular ring and towards the cavity walls. This suggests that the wings might be the result of gas that is entrained from the ring and cavity walls by the impact of feedback processes.

\section{Analysis}\label{sec:analysis}

\subsection{The expanding shell energetics from \textit{Herschel}}
\begin{figure}
    \centering
    \includegraphics[width=\hsize]{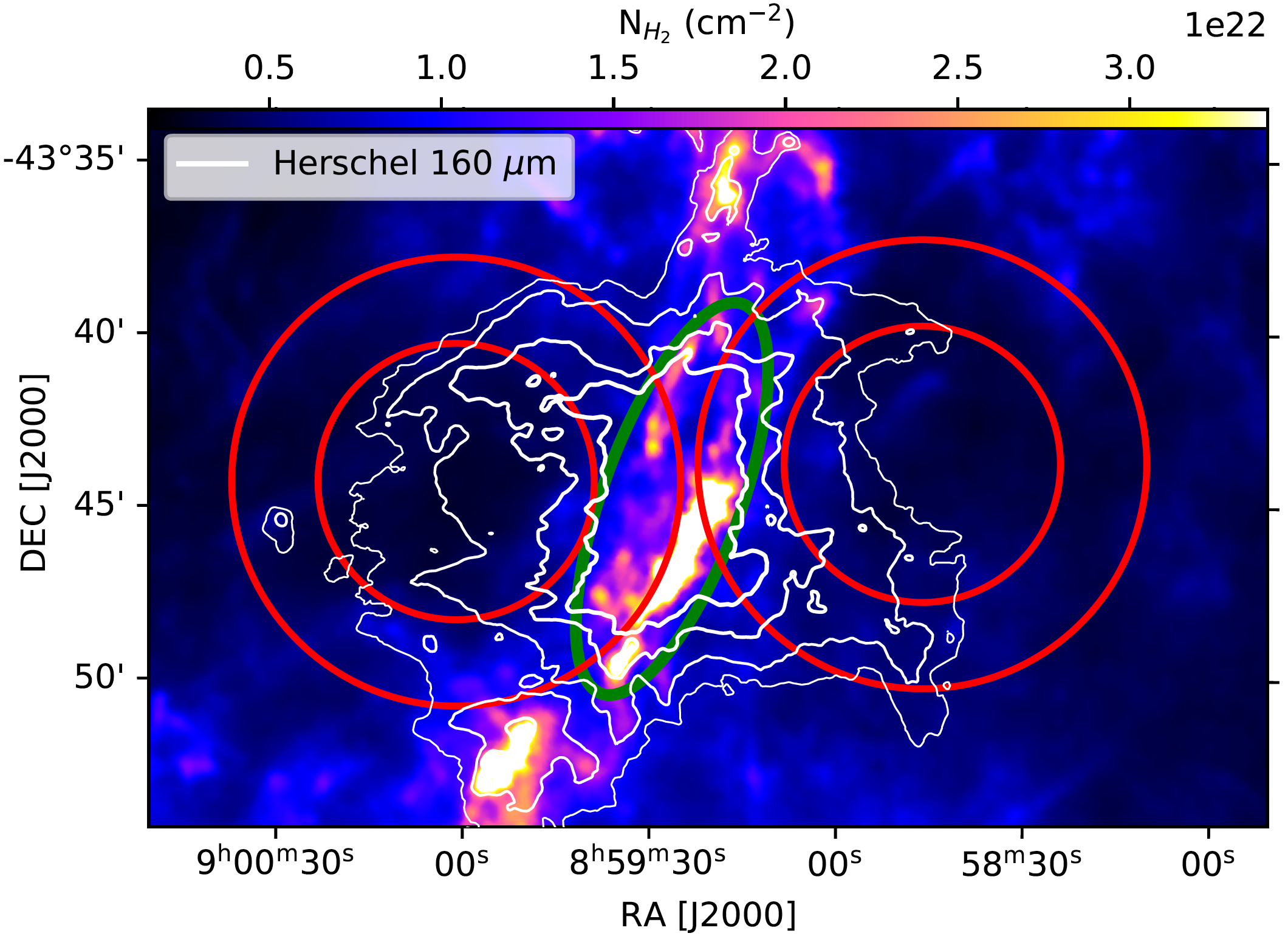}
    \caption{\textit{Herschel} column density map with overlaid \textit{Herschel} contours at 160 $\mu$m (in white) which is bright towards the ring and cavity walls. In order to calculate the masses for the expansion energetics, the ring is defined by the region encompassed by the green ellipse and the expanding shells in the cavities by the two red annuli. The regions where the annuli overlap with the green ellipse are excluded for the mass calculation of the expanding shells. 
    }
    \label{fig:defineShells}
\end{figure}

Both cavities have blue-shifted expanding \CII\ shells. In contrast to the shells in previously studied bubble \HII\ regions \citep[e.g.][]{Pabst2020,Luisi2021,Tiwari2021}, the analysis for RCW 36 is a bit more involved as it has two expanding cavities and an expanding molecular ring around the cluster. There are two approaches to estimate the expansion energetics. One is based on the \textit{Herschel} data and the second one on the \CII\ data. \\ 
First we will work with the \textit{Herschel} data of Vela C. 
The column density map produced from SED fitting to the 160 to 500 $\mu$m data might not be ideal to calculate the shell mass as most of 
these wavelengths particularly trace the cold dust ($<$ 20-30 K). Furthermore, the shells in the bipolar cavities are prominent at 160 $\mu$m and shorter wavelengths, which could be expected for the strongly irradiated expanding shells if they have a relatively low A$_{V}$ ($\lesssim$5). In this case, the shell is mostly dominated by the presence of PDRs which would lead to higher temperatures. A method that would be more sensitive to relatively warm shells was presented in \citet{Tiwari2021}, using a column density map produced based on the \textit{Herschel} 70 $\mu$m and 160 $\mu$m map. This makes use of the fact that the optical depth at 160 $\mu$m ($\tau_{160}$) can be converted to a hydrogen nucleus column density N$_{H}$ using the dust extinction cross section at this wavelength which is given by C$_{ext,160}$/H = 1.9$\times$10$^{-25}$ cm$^{2}$/H \citep{Draine2003}. The optical depth is related to the observed intensity by I$_{\nu}$ = B$_{\nu}$(T$_{d}$)$\tau_{\nu}$, assuming optically thin emission and limited fore/background emission. B$_{\nu}$(T$_{d}$) is the Planck function at a specific dust temperature T$_{d}$. In order to calculate the dust temperature map, the method presented in \citet{Peretto2016} and \citet{Bonne2022}, based on the intensity ratio of 70 $\mu$m (after smoothing to the resolution of 
the 160 $\mu$m data) and 160 $\mu$m, is used. From this dust temperature map, it is then possible to calculate the optical depth and the column density \citep{Draine2003}.\\
Based on a visual inspection of the \textit{Herschel} and WISE data at 160 $\mu$m and shorter wavelengths, the limb-brightened part of the shell in the cavities corresponds to a width of $\sim$0.3 pc, which is $\sim$30 \% of the 1.0 pc radius for both cavities, see Fig. \ref{fig:defineShells}. This figure also defines the area of the expanding ring. The estimated mass of the ring and shells are presented in Tab. \ref{tab:shellEnergetics}. To obtain the full mass of the expanding shells in the cavities, the calculated mass in the limb-brightened shells is multiplied by a geometric correction factor of 2.7. This correction factor was calculated using a code that determines the fraction of the shell located in the limb-brightened part of the shell in a 1D cut through the spherical half-shell and then assumes an axial symmetry. Using an expansion velocity of 1.0-1.9 km s$^{-1}$ for the ring and 5.2 km s$^{-1}$ for the two expanding shells in the cavities, we then arrive at a total kinetic energy of 2.4-2.6$\times$10$^{47}$ erg, see Tab. \ref{tab:shellEnergetics}.\\
We note that the expanding central ring contains dust that is also detected at wavelengths $>$ 160 $\mu$m. When using the mass for the ring from the column density map produced by the SED fitting between 160 $\mu$m and 500 $\mu$m, the mass agrees roughly within a factor 2 with the method based on the 160 $\mu$m and 70 $\mu$m ratio (i.e. the mass of the ring obtained from SED fitting is about a factor 2 lower, see Tab. \ref{tab:shellEnergetics}). In fact, the results for the bipolar cavities also agree within a factor 2. Identifying the best column density map to calculate the contribution from the central ring is a non-trivial issue, but these results indicate that an uncertainty of at least a factor 2 should be taken into account.

\begin{table}[]
    \centering
     \caption{The mass, and the resulting kinetic energy of the expanding ring and shells, calculated with the different methods. This was calculated using E$_{K}$ = $\frac{1}{2}$Mv$_{exp}^{2}$. An expansion velocity range of 1 - 1.9 km s$^{-1}$ was used for the ring, and expansion velocity of 5.2$\pm$0.5$\pm$0.5 km s$^{-1}$ was assumed for the shells in the cavities. The error on the energy values for the cavity was estimated based on the uncertainty on the expansion velocity. Note that this does not take into account errors on the mass. In the table, `full cavity' indicates that the eastern and western cavity were not separated. Based on the calculations using different C$_{ul}$/A$_{ul}$ values, we obtain an uncertainty of 30\% on the calculated mass from \CII. 
     }
     \label{tab:shellEnergetics}
     \begin{tabular}{ccc}
     \hline
     \hline
        Region & Mass & Kinetic energy \\
        \hline
        \hline
        \multicolumn{3}{c}{70 $\mu$m and 160 $\mu$m column density map}\\
        \hline
        ring &  9.1$\times 10^{2}$ M$_{\odot}$ & 0.9-3.3$\times 10^{46}$ erg\\ 
        east cavity &  4.5$\times 10^{2}$ M$_{\odot}$ & 1.2$\pm$0.2$\pm$0.2$\times 10^{47}$ erg \\
        west cavity &  3.9$\times 10^{2}$ M$_{\odot}$ & 1.1$\pm$0.2$\pm$0.2$\times 10^{47}$ erg \\
        total & & 2.4-2.6$\times 10^{47}$ erg \\
        \hline
        \hline
        \multicolumn{3}{c}{SED fitted column density map}\\
        \hline
        ring &  4.1$\times 10^{2}$ M$_{\odot}$ & 0.4-1.5$\times 10^{46}$ erg\\ 
        east cavity &  2.5$\times 10^{2}$ M$_{\odot}$ & 6.8$\pm$1.3$\pm$1.3$\times 10^{46}$ erg \\
        west cavity &  2.3$\times 10^{2}$ M$_{\odot}$ & 6.2$\pm$1.2$\pm$1.2$\times 10^{46}$ erg \\
        total & & 1.2-1.4$\times 10^{47}$ erg \\
        \hline
        \hline
        \multicolumn{3}{c}{\CII\ shell}\\
        \hline
        full cavity (100 K) & 1.5$\pm$0.5$\times 10^{2}$ M$_{\odot}$ & 4.1$\pm$1.4$\pm$0.8$\times 10^{46}$ erg\\
        full cavity (250 K)  & 1.1$\pm$0.3$\times 10^{2}$ M$_{\odot}$ & 3.0$\pm$1.1$\pm$0.6$\times 10^{46}$ erg\\
        full cavity (500 K) & 1.0$\pm$0.3$\times 10^{2}$ M$_{\odot}$ & 2.7$\pm$1.0$\pm$0.5$\times 10^{46}$ erg\\
        \hline
    \end{tabular}
 
\end{table}

\subsection{The expanding shell energetics from \CII}
\begin{figure}
    \centering
    \includegraphics[width=\hsize]{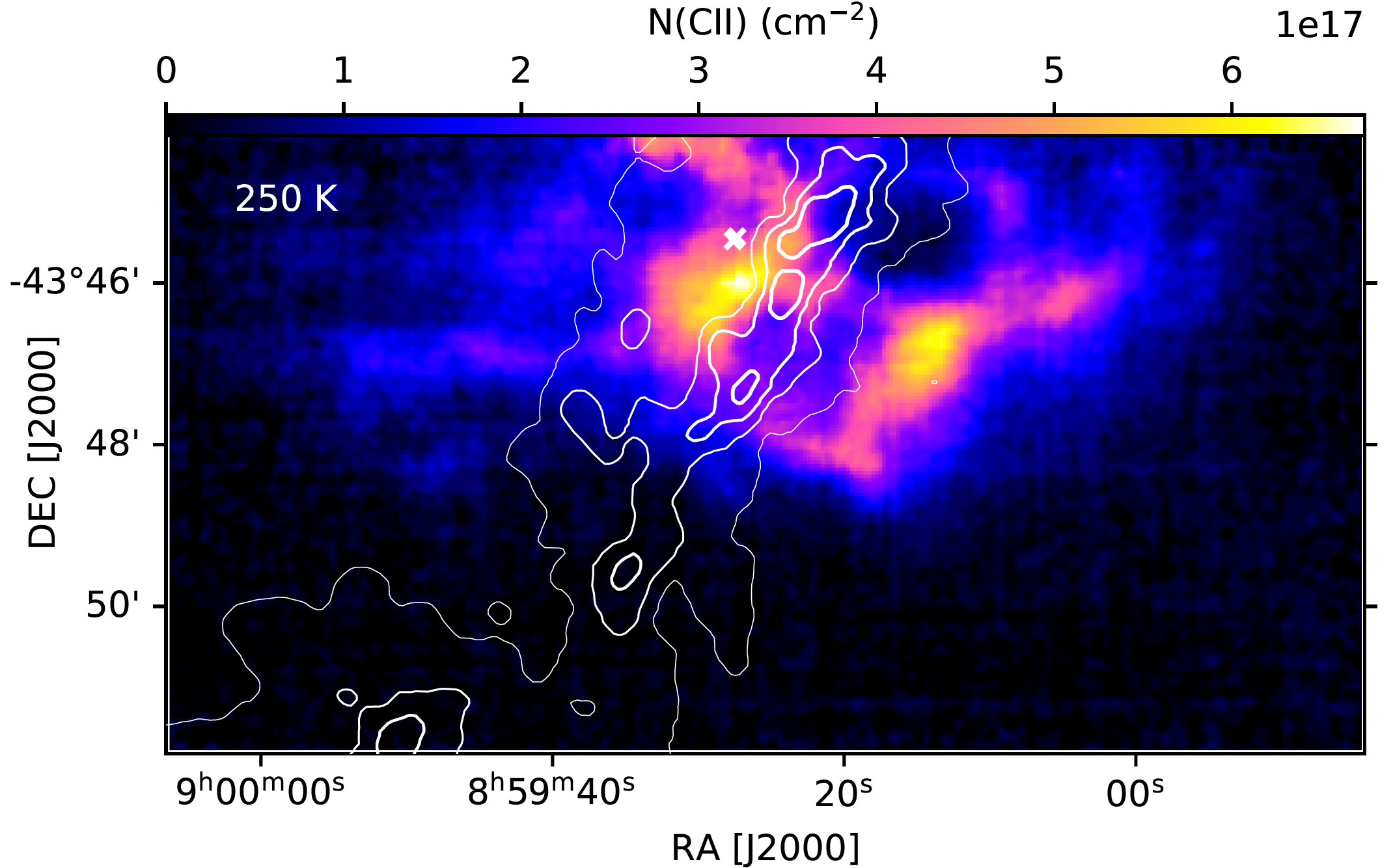}
    \caption{C$^{+}$ column density map of the blue-shifted expanding shells in the cavities, i.e. the \CII\ emission between 0 and 5 km s$^{-1}$, assuming T$_{kin}$ = 250 K. When working with a different temperature, the \CII\ column density structure is exactly the same only with different absolute values. The white contours indicate the \textit{Herschel} column density and the white crosses indicate the location of the two O stars.}
    \label{fig:CIIcolDensShell}
\end{figure}

\CII\ is a direct tracer of the expanding shells at velocities between 0 and 5 km s$^{-1}$ in the cavities and can thus be used to calculate its mass and kinetic energy. Note, however, that currently only 50 \% of the planned region is observed in \CII. Therefore, we apply a correction factor to the calculations. The correction factor is calculated by determining the fraction of the 70 $\mu$m emission from RCW 36 covered by the current \CII\ map.  Using different minimal 70 $\mu$m intensities (i.e. 500-1500 MJy/sr) to define RCW 36, we find that typically 47 - 55 \% of RCW 36 was covered. Therefore we apply a correction factor of 2.0$\pm$0.2. In order to calculate the C$^{+}$ column densities in the expanding shell we need the excitation temperature. Generally, this is calculated using the \dCII\ line \citep[e.g.][]{Guevara2020}. However, this is not possible for RCW 36 at the moment because there is no convincing \dCII\ detection corresponding to the emission between 0 and 5 km s$^{-1}$. 
Therefore, we calculate the kinetic temperature of the expanding shell from estimating the FUV-field strength at the location of the expanding shells and converting this into a C$^{+}$ kinetic temperature using the PDR Toolbox \citep{Kaufman2006,Pound2008}. This method is further specified in Appendix \ref{sec:fuvEstimate} and results in an estimated surface temperature range of 100-500 K which is a good proxy for the gas temperature that is traced by \CII\ emission and in agreement with work towards other \HII\ regions \citep[e.g.][]{Schneider2018,Pabst2020}. Assuming optically thin \CII\ emission for the shell, which is currently justified by the non-detection of \dCII\ emission at velocities lower than 6 km s$^{-1}$, we use Eq.~26 from \citet{Goldsmith2012} to calculate the C$^{+}$ column density. We assume that \CII\ emission is thermally excited and work with a ratio C$_{ul}$/A$_{ul}$ = 2 (C$_{ul}$ is the collision rate and A$_{ul}$ the Einstein-coefficient for radiative excitation) because H and H$_{2}$ number densities are expected to be around 3-6$\times$10$^{3}$ cm$^{-3}$ in the shells. 
Varying this ratio between 1 and 5, the resulting column density variations remain within 30\% at each temperature, which is not significantly larger than other uncertainties such as the temperature (leading to variations up to 50\%) or the optically thin assumption. Decreasing the ratio of Einstein coefficients down to C$_{ul}$/A$_{ul}$ = 0.3 would increase the mass by a factor 2. The column density map for the \CII\ emission between 0 and 5 km s$^{-1}$ is presented in Fig. \ref{fig:CIIcolDensShell}. 
Integrating the column density map, using the value C/H = 1.6$\times$10$^{-4}$ from \citet{Sofia2004}, the total mass of the expanding shells in the cavities are: 1.5$\times$10$^{2}$ M$_{\odot}$, 1.1$\times$10$^{2}$ M$_{\odot}$ and 1.0$\times$10$^{2}$ M$_{\odot}$, respectively at a mean kinetic temperature (T$_{\text{kin}}$) of 100 K, 250 K and 500 K. Note that this conversion factor might give a lower limit for the total column density since the expanding shell is also weakly detected in the $^{12}$CO(3-2) line and the densities towards the shell of RCW 36 are far higher than the typical densities reported in \citet{Sofia2004} where C/H = 1.6$\times$10$^{-4}$ is mostly found to be an upper limit for the higher densities in their sample. These uncertainties might be part of the explanation why these masses, and resulting expansion energetics, are significantly lower than the values obtained with the \textit{Herschel} methods. They are about a factor 4 smaller than the values obtained using the SED fitted column density map which gives the lowest values for the \textit{Herschel} based calculations. Another reason for this difference might be the spherical assumption when correcting the \textit{Herschel} limb-brightened shells to a total mass. Recently, \citet{Kabanovic2022} demonstrated with detailed radiative transfer and self-absorption analysis that the expanding shell in RCW 120 is expanding in a flattened, and not a spherical, molecular cloud. A similar point might be valid for the cavities in RCW 36, which would reduce the mass estimated based on the \textit{Herschel} data.

\subsection{The mass and momentum flux of the high-velocity wings}\label{sect:outflowAnalysis}

\begin{figure}
    \centering
    \includegraphics[width=\hsize]{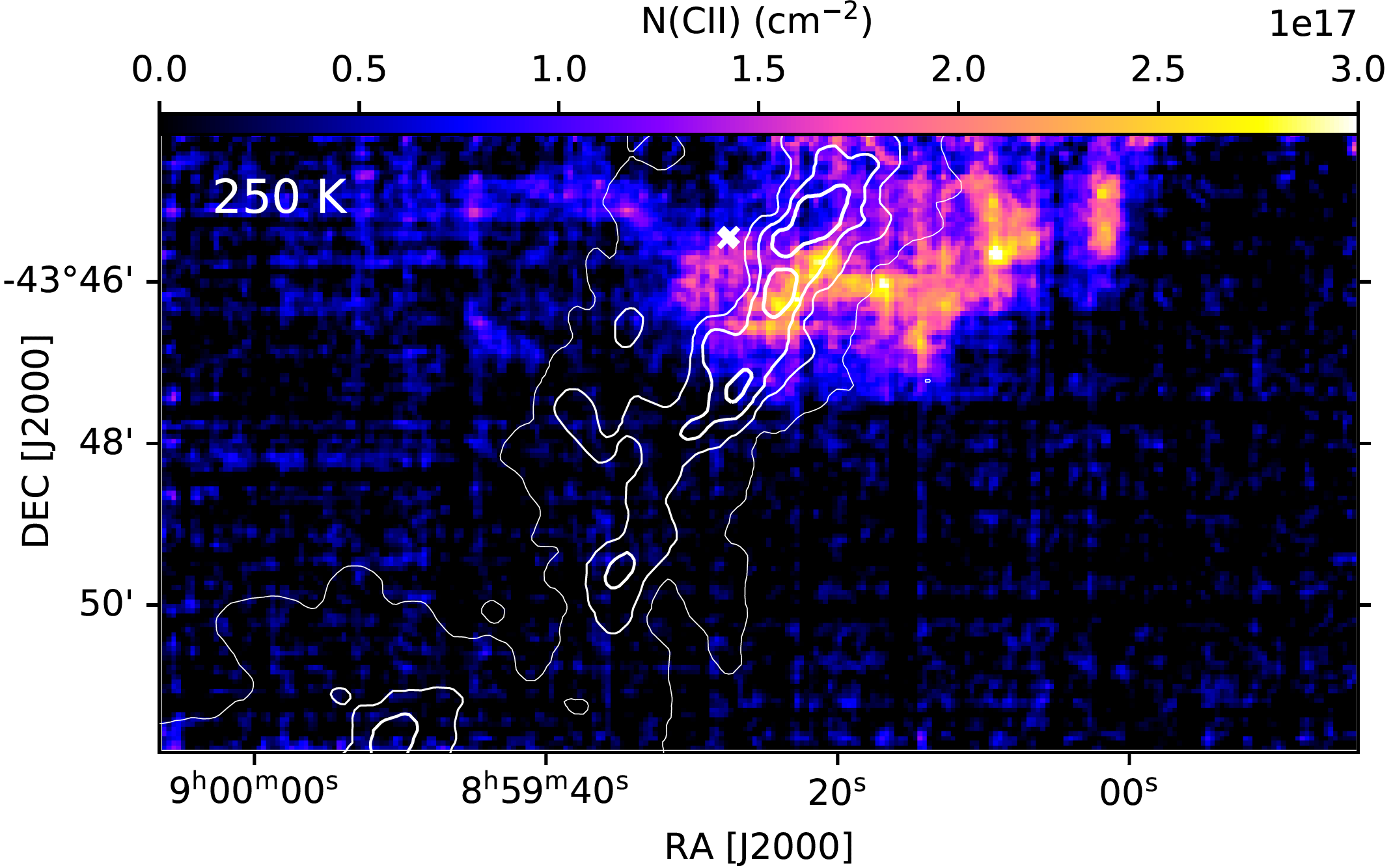}
    \includegraphics[width=\hsize]{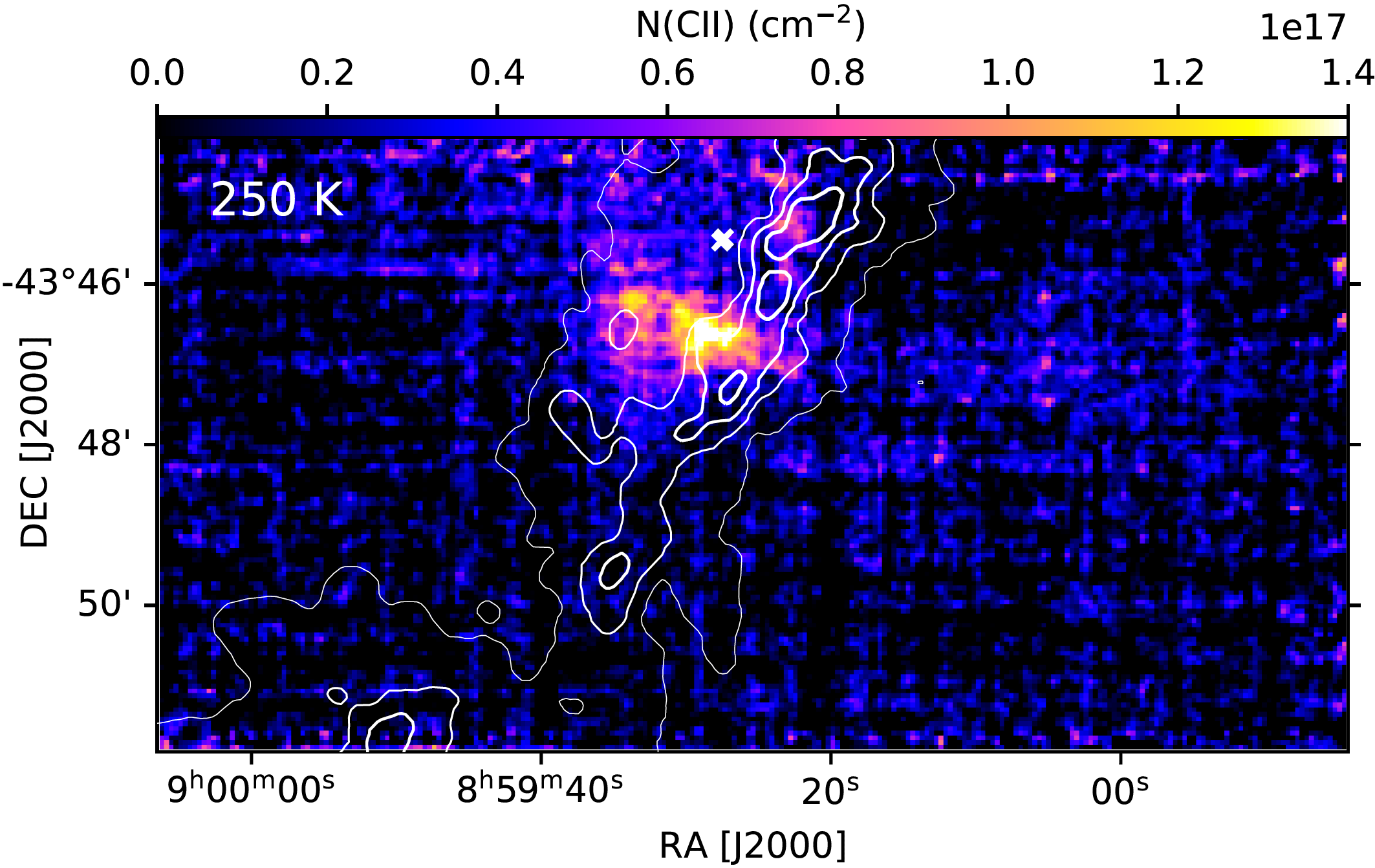}
    \caption{{\bf Top}: C$^{+}$ column density map of the blue-shifted high-velocity wing between -8 km s$^{-1}$ and 0 km s$^{-1}$, assuming T$_{kin}$ = 250 K. The white crosses indicate the location of the two O stars. {\bf Bottom}: The same for the red-shifted high-velocity wing between 14 km s$^{-1}$ and 22 km s$^{-1}$.}
    \label{fig:blueOutflowColdens}
\end{figure}

At velocities around -8 to 0 km s$^{-1}$ and 14-22 km s$^{-1}$ we observe \CII\ emission features that are not associated with expanding shell(s). These blue- and red-shifted high-velocity wings rather resemble what is observed for protostellar outflows in CO. 
However, there are no indications of classical outflow wings in $^{12}$CO(3-2) at any location in the map. In addition, \CII\ emission is not a typical tracer for protostellar outflows. Nevertheless, we perform a first order approximation to calculate the mass and momentum flux of these bipolar \CII\ high-velocity wings using the method presented in \citet{Cabrit1992,Beuther2002} for bipolar molecular outflows\footnote{See the equations in section 3.2 in \citet{Beuther2002}.}. We first calculate the C$^{+}$ column density maps associated with the high-velocity wings with the same method used in the previous section for the shells. Again, the resulting C$^{+}$ mass associated with the wings depends on T$_{\text{kin}}$ and C$_{ul}$/A$_{ul}$. T$_{\text{kin}}$, between 100 and 500 K, is again quite well constrained from the PDR models. C$_{ul}$/A$_{ul}$ is less constrained as the density and collision partners of C$^{+}$ in these high-velocity wings is uncertain. Therefore, taking C$_{ul}$/A$_{ul} \sim$ 2 will give a reasonable view on the plausible mass range (within a factor 2). 
The resulting C$^{+}$ column density maps of the velocity intervals from -8 to 0 km s$^{-1}$ (blue-shifted wing) and 14 to 22 km s$^{-1}$ (red-shifted wing) are presented in Fig. \ref{fig:blueOutflowColdens}. In the red-shifted wing a correction of 1.5 K km s$^{-1}$ is made to take into account the contribution from the \dCII\ line at velocities above 17 km s$^{-1}$. Measured with respect to the central velocity of 7 km s$^{-1}$, these velocity ranges give an observed velocity of 15 km s$^{-1}$ for both the blue- and red-shifted wings, respectively. Since the driving mechanism, and thus the inclination, of the mass ejection is uncertain, we work with an average angle of 57$^{o}$ \citep{Bontemps1996}, which also seems reasonable for mass flows in the cavities that are close to the plane of the sky. From the integrated intensity map of the high-velocity wings, we get a radial extent of 1.5 pc. Taking again into account that currently only half of the cavity is observed, the obtained values are corrected by a factor 2.0$\pm$0.2. The resulting mass, mass ejection rate and momentum flux of the bipolar high-velocity wings, assuming a kinetic temperature of 100-500 K for \CII, are presented in Table~\ref{tab:outflowParams}.


\begin{table*}[]
    \centering
     \caption{The calculated mass, momentum flux (F$_{m}$) and mass ejection rate ($\dot{\text{M}}_{out}$) of the bipolar wings observed with \CII\ in the RCW 36 cavities for kinetic temperatures of 100 K, 250 K and 500 K. Based on arguments presented in the text, the errors, which are difficult to quantify for all sources of uncertainty, on these values are easily a factor 2.}
    \label{tab:outflowParams}
    \begin{tabular}{cccc}
    \hline
    \hline
         T$_{kin}$ & Mass & F$_{m}$ & $\dot{\text{M}}_{out}$ \\
         \hline
         100 K & 4$\times$10$^{2}$ M$_{\odot}$ & 2$\times$10$^{-2}$ M$_{\odot}$ km s$^{-1}$ yr$^{-1}$ & 7$\times$10$^{-4}$ M$_{\odot}$ yr$^{-1}$\\
         250 K & 3$\times$10$^{2}$ M$_{\odot}$ & 1$\times$10$^{-2}$ M$_{\odot}$ km s$^{-1}$ yr$^{-1}$ & 5$\times$10$^{-4}$ M$_{\odot}$ yr$^{-1}$\\
         500 K & 3$\times$10$^{2}$ M$_{\odot}$ & 1$\times$10$^{-2}$ M$_{\odot}$ km s$^{-1}$ yr$^{-1}$ & 4$\times$10$^{-4}$ M$_{\odot}$ yr$^{-1}$\\
         \hline
    \end{tabular}
 
\end{table*}

\section{Chandra}\label{sec:chandra}
Chandra observations trace the presence of hot plasma formed in the winds of massive stars. A map of diffuse X-ray emission for RCW 36 was presented in \citet{Townsley2014}. The methods to extract this diffuse X-ray emission are described in detail in  \citet{Broos2010,Broos2012,Broos2018}. Analysing the Chandra data will thus allow a more comprehensive view of the feedback processes that are driving the observed dynamics around RCW 36.

\subsection{The diffuse X-ray emission map}\label{sec:diffXrayEmission}
\begin{figure}
    \centering
    \includegraphics[width=\hsize]{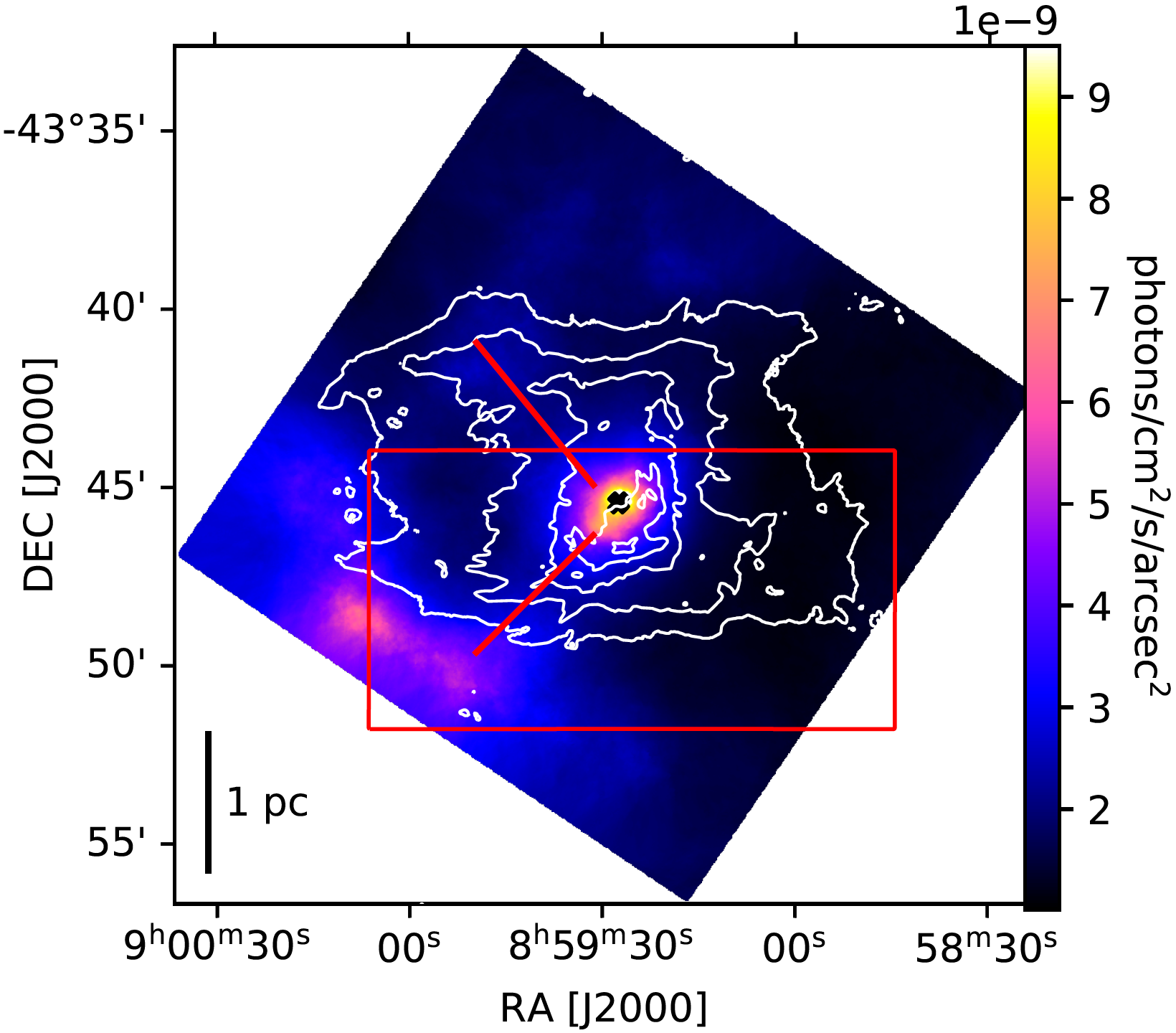}
    \includegraphics[width=0.9\hsize]{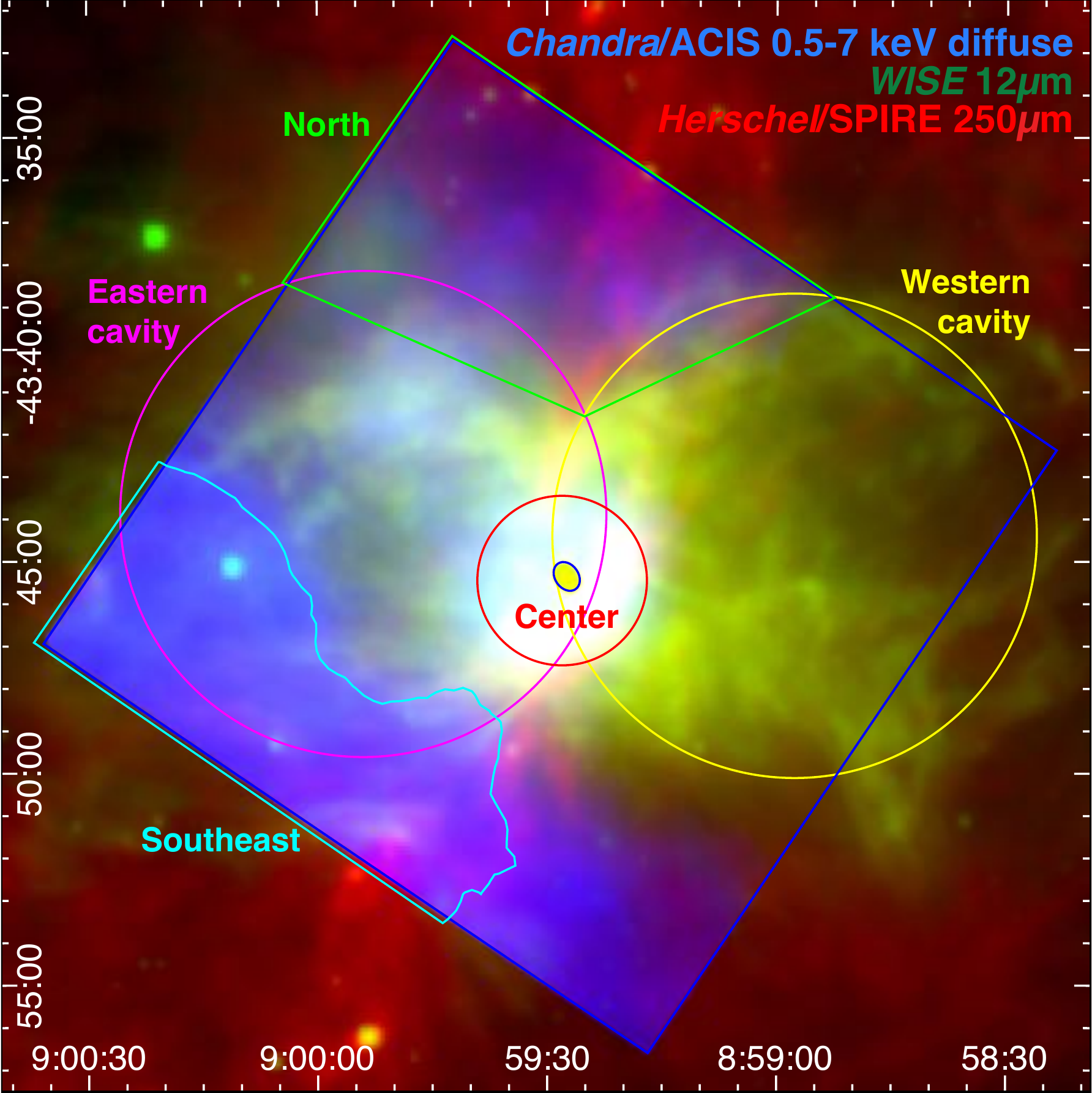}
    \caption{{\bf Top}: Diffuse X-ray emission observed with the ACIS instrument on Chandra. The white contours show the 70 $\mu$m emission of RCW 36 from Fig. \ref{fig:observedMap} starting at 10$^{-1.5}$ Jy pixel$^{-1}$ with increments of 0.5 in the exponent. The brightest Chandra emission is centered on the O stars of the cluster, which are indicated by the black cross. In the cavities, there is diffuse emission detected which is significantly weaker than extended emission outside the cavities (outlined by the 70 $\mu$m emission). 
    The diffuse emission outside the cavities likely is the result of hot plasma leaking out of the \HII\ region, which fits with the observed physical connection to the emission inside the \HII\ region and the fact that the hot plasma fitting blindly converged to very similar properties for the different regions. Two locations where this physical connection is seen, are indicated with the red lines. The red box indicates the outline of the currently observed \CII\ map.
    {\bf Bottom}: A smoothed image of Chandra/ACIS diffuse X-ray emission in the context of WISE and Herschel images.  The five regions used to extract diffuse X-ray spectra are overlaid.  The small blue ellipse at field center marks a region excluded from the $'$Center$'$ diffuse X-ray spectrum because it is dominated by emission from unresolved young stars at the center of the cluster.  See Appendix D for details of the X-ray spectral fitting.}
    \label{fig:DiffuseXrayEmission}
\end{figure}

The diffuse X-ray emission is presented in Fig. \ref{fig:DiffuseXrayEmission}. The brightest peak is centered on the two O star candidates of the young cluster. Fainter emission is seen inside the bipolar cavities, which is explained by extinction from the cavity walls when analyzing the X-ray spectra (see App. \ref{sec:xpecFit}). The morphology of the diffuse X-ray emission in Fig. \ref{fig:DiffuseXrayEmission} hints that the hot plasma formed in the OB cluster is leaking out of the \HII\ region into the ISM beyond the cavities of RCW 36.\\\\
To constrain the excitation conditions of the diffuse X-ray emission, X-ray spectra of several regions in RCW 36 are fitted using plasma models in XSPEC \citep{Arnaud1996}. Five diffuse extraction regions, shown in Fig. 14, were defined to obtain spectra with sufficient S/N for fitting.  X-ray events from hundreds of X-ray point sources were excised from these regions, to minimize their influence on the diffuse X-ray spectra.  Details on our methods for fitting diffuse X-ray emission in and around massive star-forming regions are given in \citet{Townsley2019}.\\
The extracted diffuse X-ray spectra and fitting results are presented in App. \ref{sec:xpecFit} and Tab. \ref{tbl:diffuse_spectroscopy_style2}. The central region is filled by a soft, single-temperature thermal plasma plus residual emission from unresolved pre-main sequence stars in the young cluster. The eastern and western cavities are filled with a thermal plasma of similar temperature. The intrinsic surface brightness of this emission, which corrects for the intervening absorption, is higher than that outside the cavities, confirming that the morphology of the apparent surface brightness map is influenced by intervening absorption.  Hot plasma in the ISM surrounding RCW 36 shows two thermal plasma components, one similar in temperature to the RCW 36 emission and one substantially hotter.  While the hotter component is likely unrelated background emission (often seen along Galactic Plane sightlines), the cooler component is strong evidence that the plasma generated by massive star winds in RCW 36 is leaking out of its \HII\ region into the surrounding molecular cloud.

\subsection{Properties of the hot plasma}
From the fitted spectra of the hot plasma, the physical properties (plasma density, pressure,...) can be calculated with an assumed emissivity curve and a filling factor which was taken to be unity \citep{Townsley2003}. The assumed emissivity curve from \citet{Landi1999} assumes collisional ionization equilibrium (CIE) for the plasma. This assumption is justified since non-equilibrium effects only become notable below plasma temperatures of 10$^{6}$ K \citep{Vasiliev2011,Vasiliev2013} whereas the hot plasma temperature in RCW 36 (1.7-2$\times$10$^{6}$ K) is above this value, see also Tab. \ref{tab:PlasmaFit}. The calculation of the physical properties was not done for the regions outside the bipolar \HII\ region (north and southeast) as there is no evident assumption for the 3D geometry at these locations. This calculation of the physical properties also was not done for the central region directly around the OB cluster because of the non-equilibrium ionization (NEI) of the plasma and relatively low temperature. As a result, the bandpass correction factor necessary for the limited ACIS bandpass is larger than 40 which would make the results for the center highly uncertain. For the cavities, the bandpass correction factors are 4.4 and 3.4 for the east and west, respectively, which makes their estimated properties less uncertain. The resulting physical properties of the plasma are presented in Tab. \ref{tab:PlasmaProperties}. Both cavities have an estimated temperature around 2$\times$10$^{6}$ K and electron densities around 0.2 cm$^{-3}$, which results in an estimated pressure of 0.6-0.9$\times$10$^{6}$ K cm$^{-3}$.\\
The resulting estimated energy of the plasma in both cavities match the expansion energy of the expanding shells, presented in Tab. \ref{tab:shellEnergetics}, well (i.e. it is slightly higher than the estimated expansion energy based on \CII\ and slightly lower than the estimates based on \textit{Herschel}). This suggests that under adiabatic conditions the expanding shells in the cavities might be driven by the hot plasma \citep[e.g.][]{Weaver1977,Lancaster2021a}. This will be examined in more detail later on as other processes such as photo-ionized gas pressure and direct radiation pressure could also play a role. Note that the unity filling factor is an assumption. The appropriate scaling factor ($\eta^{-1/2}$, with $\eta$ the filling factor) to correct for a lower filling factor is also presented in Column 2 of Tab. \ref{tab:PlasmaProperties}.  This indicates that the resulting plasma properties are not extremely sensitive to the filling factor.

\begin{deluxetable}{@{\hspace{2em}}lcccc@{}}
\centering  \tabletypesize{\footnotesize} \tablewidth{0pt}

\tablecaption{Physical Properties of the Diffuse X-ray Plasma Components for RCW~36, D = 900~pc. \label{tbl:physics}}

\tablehead{
\colhead{Parameter} & \colhead{~Scale factor~~} & 
\multicolumn{2}{c}{RCW 36 Diffuse Region}\\[-7pt]
\colhead{}{\hrulefill} & \colhead{}{\hrulefill} &
\multicolumn{2}{c}{\hrulefill}\\[-7pt]
\colhead{} & \colhead{} &
\colhead{East Bubble}  & \colhead{West Bubble}
}

\startdata
\multicolumn{4}{l}{Observed X-ray properties} \\
\hline
$kT_x$ (keV)                & \nodata     &  0.15                &  0.17                 \\
$L_{X,bol}$ (erg~s$^{-1}$)  & \nodata     &$4.4 \times 10^{33}$  & $1.9 \times 10^{33}$  \\ 
$V_x$ (cm$^3$)              & $\eta$      &$4.1 \times 10^{56}$  & $4.1 \times 10^{56}$  \\
                            &             &                      &                       \\
\hline
\multicolumn{3}{l}{Derived X-ray plasma properties} \\
\hline
$T_x$ (K)                   & \nodata      & $1.7 \times 10^6$    & $2.0 \times 10^6$     \\
$n_{e,x}$ (cm$^{-3}$)       &$\eta^{-1/2}$ &  0.22                &  0.15                 \\
$P_x/k$ (K~cm$^{-3}$)  	    &$\eta^{-1/2}$ & $0.9 \times 10^6$    & $0.6 \times 10^6$     \\
$E_x$ (erg)                 &$\eta^{1/2}$  & $7.2 \times 10^{46}$ & $5.4 \times 10^{46}$  \\
$\tau_{cool}$ (Myr)         & $\eta^{1/2}$ & 0.5                  & 0.9                   \\
$M_x$ (M$_\odot$)           & $\eta^{1/2}$ & 0.05                 & 0.03                  \\
\enddata

\tablecomments{Equations detailing how the derived properties were obtained from the observed properties are given in \citet{Townsley2003}.  
The quantity $\eta$ is a ``filling factor,'' $0 < \eta < 1$, accounting for partial filling of the volume with the X-ray-emitting plasmas.  The parameters in the table should be multiplied by the appropriate scale factor (Column 2) to account for this filling factor.  Derived plasma properties are proportional to $\eta^{1/2}$ and are thus only weakly sensitive to this correction.}

\label{tab:PlasmaProperties}
\end{deluxetable}

\section{Discussion}\label{sec:discussion}
\subsection{3D structure of RCW 36}
\begin{figure}
    \centering
    \includegraphics[width=\hsize]{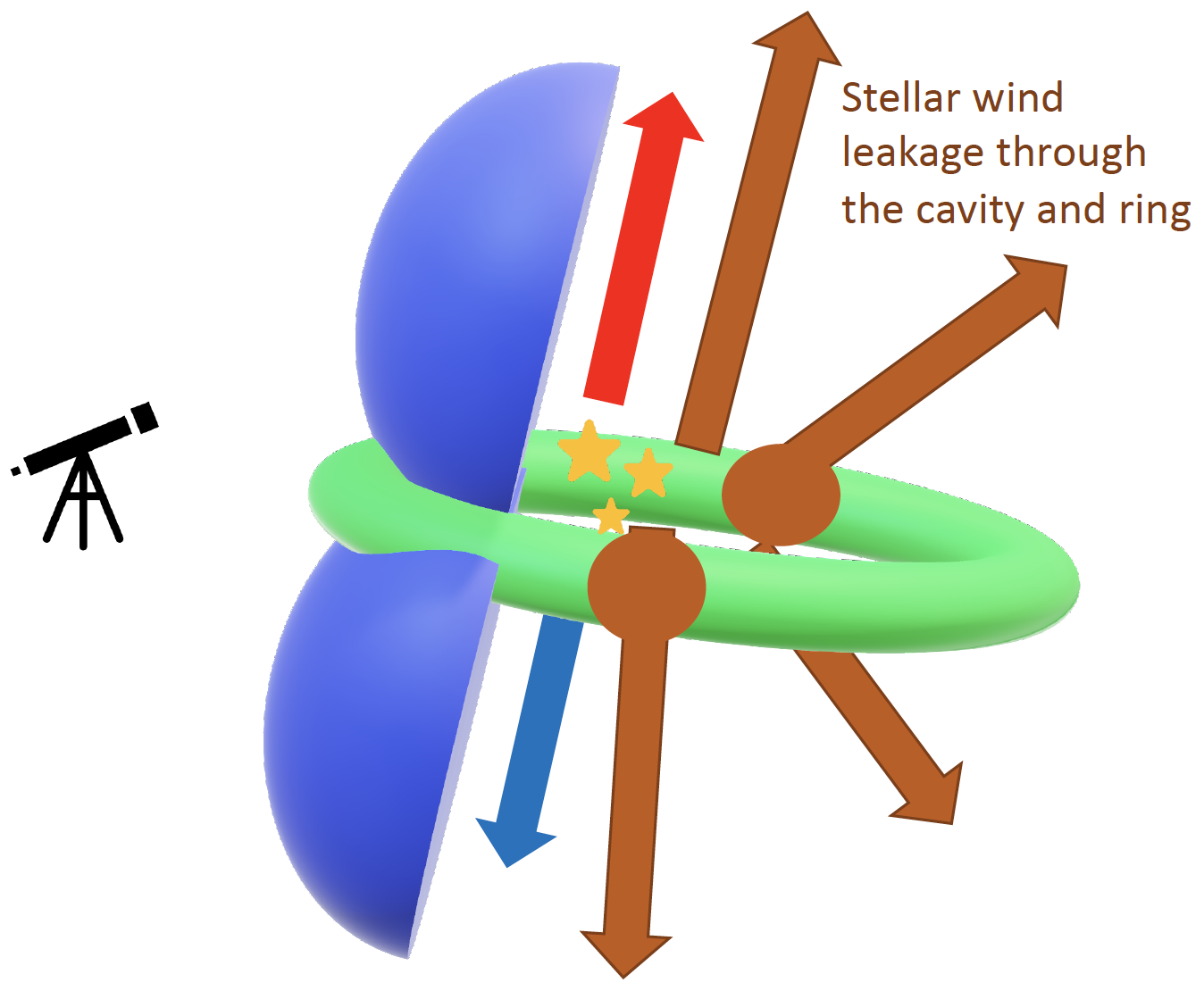}
    \caption{Sketch of RCW 36 seen from the side and indicating our point of view (telescope). With \CII, we observe blue-shifted expanding shells (blue) in the bipolar cavities. The ring (green) is expanding at a lower velocity into the Vela C molecular cloud. The blue and red arrows indicate the high-velocity mass ejection in the cavities associated with the wings (blue and red) in the spectra. The brown arrows indicate the hot plasma leakage which is presumably most effective towards the back of RCW 36 where no red-shifted expanding shells are observed with the \CII\ data, but we also observed it can channel through punctures in the ring as well (brown patches on the ring).}
    \label{fig:3dMorph}
\end{figure}

RCW 36 consists of a bipolar cavity and a dense central ring surrounding the waist of the \HII\ region created by the central OB cluster. In the cavities, the expanding \CII\ shell is only observed at blue-shifted velocities, which implies it is expanding towards us.  Fitting the foreground column to the Chandra data towards the cavities, in particular the eastern one, suggests that most column density, observed with {\sl Herschel}, is associated with and on the near side of RCW 36 from our point of view. The observed \CII\ shells are thus expanding into the predominant nearby column density reservoir of RCW 36. This can offer an explanation why no corresponding red-shifted expanding shell is detected for the cavities even though a few expanding filamentary structures are observed towards slightly more redshifted velocities. As most of the column density appears to be located in front, the gas behind the ridge might be mostly diffuse gas. As a result, the hot plasma created by the OB stars would be able to leak from RCW 36 in that direction, as sketched in Fig. \ref{fig:3dMorph}. This is in essence the same scenario observed for RCW 49 in \citet{Tiwari2021}. Interestingly, in contrast to RCW 36 (2 O star candidates and $\sim$ 1 Myr), RCW 49 contains 37 OB identified stars and is slightly older with an estimated age $\sim$ 2 Myr.\\
To obtain a first idea of the expansion timescale for the dense molecular ring, we use the estimated expansion velocity between 1.0 and 1.9 km s$^{-1}$ and an estimated radius of 0.9 pc. 
As it is unclear what drives the expansion of the ring, we calculated the Spitzer expansion timescale for expansion driven by ionizing radiation \citep{Spitzer1978}, the Weaver solution for expansion driven by stellar winds \citep{Weaver1977} and the ballistic expansion timescale which can be expected if shell expansion is driven by stellar winds in an isothermal density field with a power law index of -2 \citep{Geen2022}. The Spitzer solution gives an expansion timescale of 1.2-1.7 Myr, the Weaver solution gives a timescale of 0.3-0.6 Myr and the ballistic expansion give a timescale of 0.5-0.9 Myr. The calculations of the Spitzer and Weaver solutions are further specified in App. \ref{sec:SpitzerWeaverCalc}. Broadly speaking, the Spitzer and ballistic solution for the dense ring appear to be in better agreement with the cluster age than the Weaver solution \citep{Ellerbroek2013a}. However, the upper value of the Spitzer solution, which might be the most representative one (see App. \ref{sec:SpitzerWeaverCalc}), has an equally large deviation from the estimated cluster age as the Weaver solution. The cavities on the other hand have a radius of $\sim$1.0 pc and an expansion velocity of 5.2$\pm$0.5$\pm$0.5 km s$^{-1}$ which results in an estimated Spitzer expansion timescale of 0.2-0.4 Myr (see App. \ref{sec:SpitzerWeaverCalc}), a Weaver expansion timescale of 0.1 Myr and a ballistic expansion timescale of $\sim$ 0.2 Myr. This is roughly a factor 3-5 shorter than for the ring and suggests that the expansion in the cavities is a recent feature. Note that this assumes the cavity expands from the center of the cavity, which might not be fully correct. Considering the expansion starts from the center of RCW 36, we get a radius between 1.5-2.0 pc. This results in an estimated ballistic expansion timescale of 0.3-0.4 Myr which still is shorter than the expansion of the dense ring. RCW 36 thus shows inhomogeneous expansion: relatively slow expansion in the dense gas followed by increasing expansion velocity once the expansion reaches the lower density gas. This fits with relatively low ($<$ 3 km s$^{-1}$) expansion velocity of the denser filamentary structures perpendicular to the ring.\\\\
\citet{Samal2018} noted that only 16 of the 1377 \HII\ regions considered in their sample are visibly identified as bipolar \HII\ regions. A correction factor has to be taken into account based on inclination selection effects but, as mentioned in \citet{Samal2018}, this still implies that bipolar \HII\ regions are rather rare. From the faster expansion in the bipolar cavities of RCW 36 compared to the ring, it can be argued that the morphology of RCW 36 is changing rapidly (on $\sim$ 0.2 Myr timescales) and thus that a \HII\ region might only be observed as bipolar for a part of its evolution. 
As a short exercise, assuming a correction factor 5 for the inclination \footnote{A factor 5 correction for the inclination assumes that we can only identify a bipolar \HII\ region if the inclination angle with the line of sight is larger than 70-75$^{o}$.  This might be a bit of an overestimate, but most identified bipolar nebulae in \citet{Samal2018} have clearly separated cavities which suggests their identification is constrained to high inclination angles.}, and a potential correction factor 5 for the short timescale\footnote{Based on the expansion timescale results, a bipolar \HII\ region would be identified for $\sim$0.2-0.5 Myr. Most mass around a \HII\ region might be removed in a few Myr (e.g.  1-3 Myr), see for example later sections in the discussion. This suggests a correction factor of the order 5, assuming the presence of some mass is required to identify the structure of a \HII\ region in the study of Spitzer images by \citet{Samal2018}.} would imply that up to 30 \% of the \HII\ regions might go through a bipolar phase. However, this remains speculative at the moment since this is based on a single case. More generally, the observation of expanding half-shells, instead of fully spherical shells, in \CII\ emission towards \HII\ regions \citep[e.g.][]{Anderson2019,Pabst2020,Luisi2021,Tiwari2021}, and the evidence of important hot plasma leakage, 
 can fit in the view of the ISM where (massive) stars form in a complex organization of filaments and sheets. The creation of bipolar \HII\ regions can then occur when an O star forms close to the mid-plane of a sheet-like structure such that the radiation and wind breaks out on both sides at about the same time. The hot plasma can then start driving expanding shells in the lower density gas outside the sheet. This is however speculative since it is based on a single case while the observations also indicate that the expansion strongly depends on the density structure. Furthermore, simulations by \citet{Wareing2017} suggest that the bipolar morphology remains present for a significant part of their simulation. This is due to the apparent low expansion velocity ($<$ 1 km s$^{-1}$) for the cavities in their simulation which is not in agreement with the observed velocities in the cavities for RCW 36. As the expansion dynamics might depend from region to region, a detailed analysis of age and dynamics of a sample of bipolar \HII\ regions will be required to fully address this question.

\subsection{The pressure and energy budget in RCW 36}\label{sec:pressEnergyBudget}

\begin{figure}
    \centering
    \includegraphics[width=\hsize]{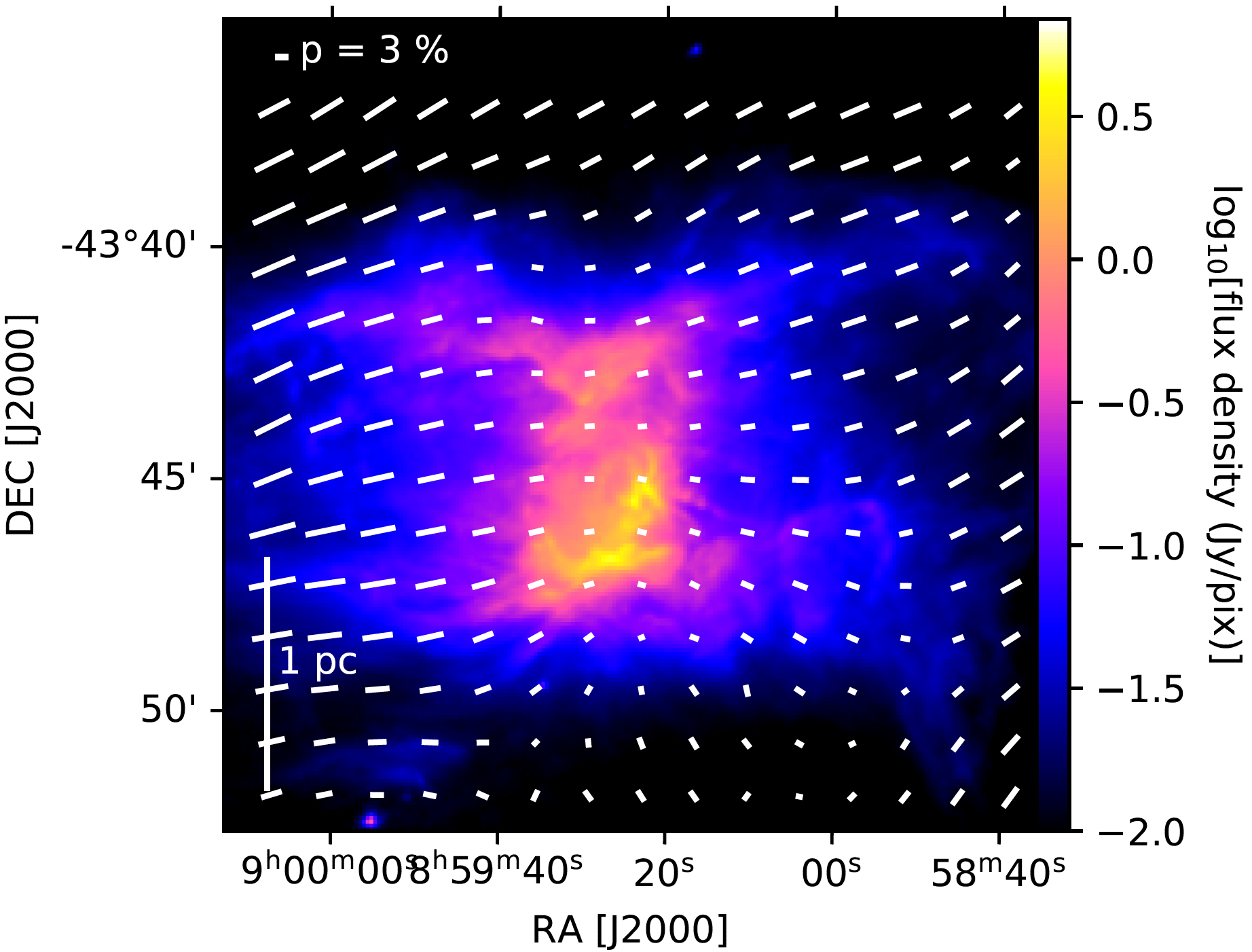}
    \caption{\textit{Herschel} 70 $\mu$m map of RCW 36 overlaid with the magnetic field orientation obtained from BLASTPol at 500 $\mu$m \citep{Fissel2016,Fissel2019}. The magnetic field orientation tends to follow the edges of the shells and associated kink near the central ring. It is also observed that the polarization fraction quickly drops towards the central ring which might be associated with the complex morphology in this region that remains unresolved with the BLASTPol data.}
    \label{fig:blastPolIm}
\end{figure}

Analysis of the Chandra data shows that a hot plasma, created by the stellar winds, pervades RCW 36 and that it has the required energy to drive the observed expanding shells in both cavities. To discuss the role of different physical process, the associated pressure terms were estimated and are presented in Tab. \ref{tab:pressureTable}. Specific assumptions made for these calculations are specified in App. \ref{sec:linewidthShell}.
The thermal pressure from the hot plasma in both cavities is $\sim$ 0.6-1.0$\times$10$^{6}$ K cm$^{-3}$, which is slightly larger than the estimated thermal pressure of the photo-ionized gas, i.e. 4.6-5.0$\times$10$^{5}$ K cm$^{-3}$, derived from the SHASSA H$\alpha$ observations \citep{Gaustad2001}. However, as for the hot plasma, if the photo-ionized gas is confined in a sub-region of the cavities, this will increase the estimated pressure. As a result, these terms probably dominate over the direct radiation pressure in the cavities, see Tab. \ref{tab:pressureTable}. Since cavities are filled by both the hot plasma and photo-ionized gas, the photo-ionized and collisionally ionized gas can thus be in pressure equilibrium across the contact discontinuity.\\
The physical processes that drive expanding motion of \HII\ regions is still a matter of debate. Simulations generally predict that ionizing radiation is the main driver of expansion compared to stellar winds \citep[e.g.][]{Haid2018,Geen2021,Ali2022}. Yet, simulating the high dynamic range of temperatures and densities associated with the presence of stellar winds in the ISM is challenging \citep[e.g.][]{Dale2015}. Observations by \citet{Lopez2014} also indicate that the ionized gas pressures appear to dominate most \HII\ regions, however other observational studies \citep[e.g.][]{Pabst2019,Luisi2021,Tiwari2021} have proposed that stellar winds drive the high-velocity expansion detected with spectrally resolved \CII\ observations. Above we already noted that thermal pressure of the ionized gas and hot plasma seem to be of the same order for RCW 36. Here, we will discuss the associated energy terms. Based on the SHASSA H$\alpha$ intensity, the ionizing photon flux in each cavity is $\sim$8.5$\times$10$^{46}$ s$^{-1}$, see App. \ref{sec:linewidthShell}. This results in a  total injected energy of $\sim$ 6.2$\times$10$^{49}$ erg over the estimated lifetime of the cluster (i.e. 1.1 Myr) when assuming an energy of 13.6 eV for the ionizing photons. Taking into account that the coupling efficiency of ionizing radiation energy to radial expansion typically is $<$ 10$^{-4}$ \citep{Haid2018}, this is about an order of magnitude too small to explain the expanding motion of the shells in the cavities. We do note that at very early evolutionary stages in \citet{Haid2018}, the coupling efficiency for ionizing photons is significantly higher. However, considering a coupling efficiency of 10$^{-3}$ over the first 0.3 Myr remains insufficient to explain the observed expansion. Therefore, unless the coupling efficiency for ionizing photons in \citet{Haid2018} is about an order of magnitude lower than in RCW 36, we propose that the hot plasma created by stellar winds drives the expansion of the blueshifted shells in the cavities.\\
On the other side of the shell, ram pressure, thermal pressure and magnetic pressure from the ambient cloud confine the shell.  The values in Tab. \ref{tab:pressureTable} indicate that in this low-density region, the ram pressure confines the shell. Inside the compressed shell, turbulent pressure seems to be the dominant term to prevent the shell from being further compressed between the hot plasma on one side and ram pressure on the other side. 
Even though the estimated magnetic pressure in the shell is significantly smaller than the turbulent pressure, it might play a role in limiting the initial compression rate of the propagating shock that is driven by the expansion. Inspecting the BLASTPol magnetic field map towards RCW 36 in Fig. \ref{fig:blastPolIm}, it is observed that the magnetic field morphology follows the bipolar cavities and forms a kink near the ring where the polarization fraction drops. This behaviour where the magnetic field orientation is aligned with the shell morphology around a \HII\ region, as it is dragged along by the expansion, was recently also observed in the Keyhole Nebula \citep{Seo2021} and NGC 6334 \citep{Arzoumanian2021}. The magnetic field morphology in the ring and whether it significantly affects the expansion remains unclear with current data. This requires higher angular resolution observations (Bij et al. in prep.). In the ring, it is also unclear which process (ionizing radiation or the hot plasma) drives the expansion. The number of ionizing photons and coupling efficiency is sufficient to drive the ring expansion, but the hot plasma in the ring remains unconstrained. As a result, we cannot reach a firm conclusion.

\subsection{Feedback driven turbulence}\label{sec:turbRing}
\begin{figure}
    \centering
    \includegraphics[width=\hsize]{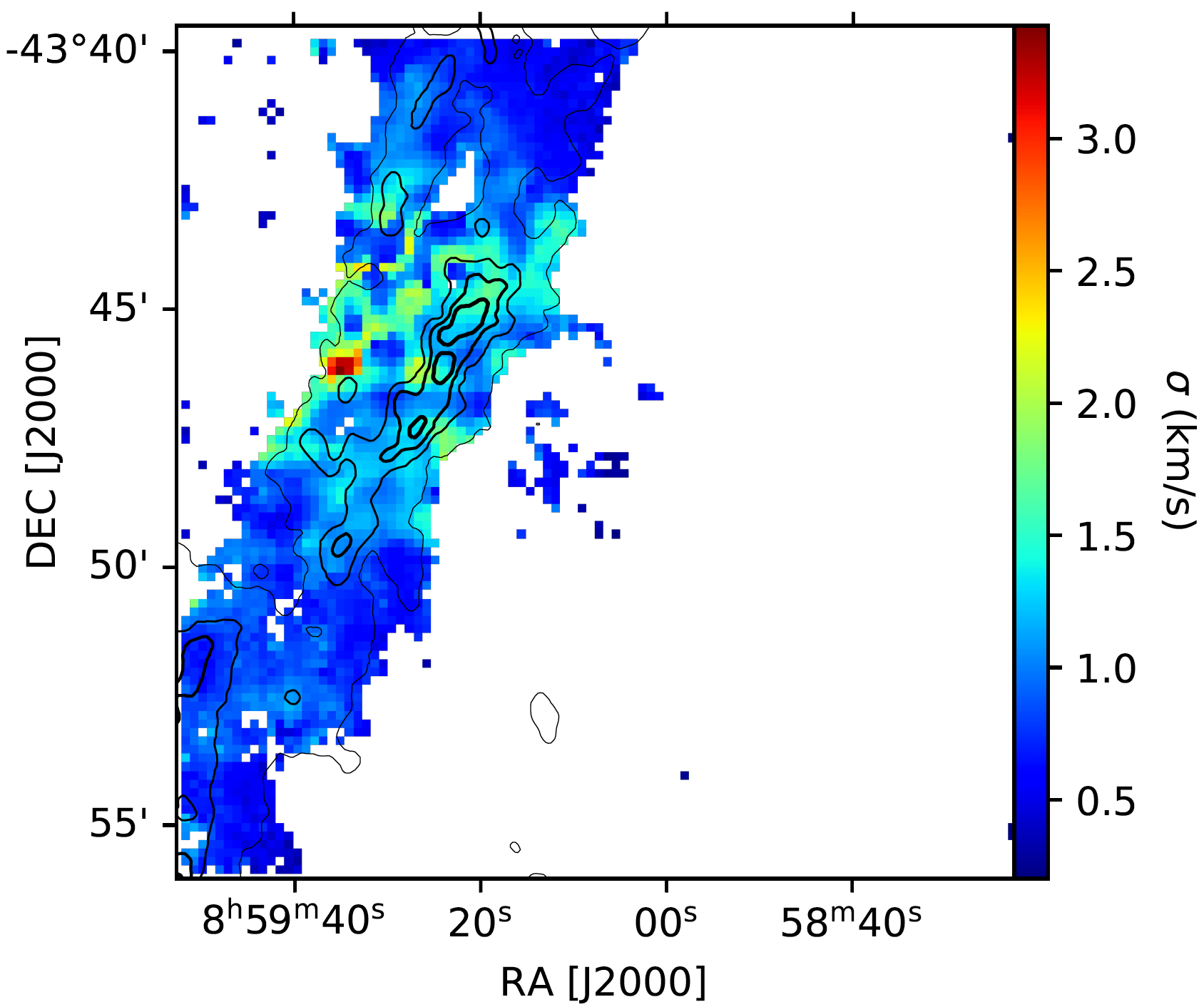}
    \caption{The velocity dispersion map of $^{13}$CO(3-2) for RCW 36. The contours indicate the \textit{Herschel} column density map. Despite the relatively limited extent of the map, it becomes clear that there is a significant velocity dispersion increase at the ring surrounding the OB cluster.}
    \label{fig:13coSigmaMap}
\end{figure}

In Fig. \ref{fig:13coSigmaMap}, a noteworthy increase in $^{13}$CO(3-2) velocity dispersion is observed in the ring of RCW 36 compared to the surrounding regions of the Vela C molecular ridge which does not seem to be explained by opacity broadening, see App. \ref{sec:linewidthShell}. This indicates that the mechanical energy input from stellar feedback can significantly increase the turbulent energy in the compressed gas. This energy injection might thus play a role in setting the thickness of the shell, as it provides a rough pressure equilibrium with both sides of the shell. Specifically, the turbulent pressure in the ring, derived from the spectral linewidth, is similar to or higher than the ram pressure provided by the ambient gas, see Tab. \ref{tab:pressureTable}. Otherwise, the shell would be further compressed until an increased temperature (or magnetic field strength or turbulence) due to this compression would halt it. From the information provided by the $^{13}$CO(2-1) linewidth, corrected for optically thick line broadening (App.~\ref{sec:linewidthShell}), it is possible to estimate the turbulence injection by stellar feedback in the molecular ring. The calculated total turbulent energy injection in the ring and cavity shells is given in Tab. \ref{tab:turbulenceTable}. From Fig. \ref{fig:13coSigmaMap}, it is reasonable to assume a linewidth for this gas of 0.7 km s$^{-1}$ before the onset of stellar feedback, which would give a total turbulent energy of 2.5$\times$10$^{2}$ M$_{\odot}$ (km s$^{-1}$)$^{2}$ for the gas now located in the ring. Comparing this with the value for the ring in Tab. \ref{tab:turbulenceTable}, this implies that the turbulent energy of the gas, after opacity broadening correction, has roughly increased by more than 60 \% compared to the onset of stellar feedback. Turbulence typically dissipates over a few crossing timescales (t$_{cross}$ = d/$\sigma_{los}$) \citep[][]{Ostriker2001}. For the ring, with a width of $\sim$0.1 pc, this gives a crossing timescale of 1.1$\times$10$^{-1}$ Myr. Using this timescale results in a turbulence dissipation rate of 2.4$\times$10$^{33}$ erg s$^{-1}$ that has to be injected to maintain the turbulent linewidth. This is only a small fraction of the probable energy injection rate by the O star candidates, which will be discussed in more detail in the next section. 

\begin{table*}[]
\caption{The estimated pressure (range) for the different components associated to the expanding shells in the cavities and the dense molecular ring around the central cluster. The calculation of the values are presented in more detail in App. \ref{sec:linewidthShell}.}
\footnotesize
    \begin{center}
    \setlength{\tabcolsep}{12pt}
    \begin{tabular}{cccccc}
        \multicolumn{5}{c}{Bipolar cavities}\\
        \hline
        \hline
         & P$_{th}$/k (K cm$^{-3}$) & P$_{rad}$ (K cm$^{-3}$)  & P$_{turb}$/k (K cm$^{-3}$) & P$_{B}$/k (K cm$^{-3}$) & P$_{ram}$/k (K cm$^{-3}$) (1)\\
         \hline
        Hot plasma &  0.6-0.9$\times$10$^{6}$ & & & & \\
        \HII\ region & 4.6-5.0$\times$10$^{5}$ & 4.5$\times$10$^{5}$ & & & \\
        Ambient cloud & 0.5-2.5$\times$10$^{4}$ & & $<$1.3-6.7$\times$10$^{5}$ & 0.3-1.4$\times$10$^{5}$ & 0.7-3.5$\times$10$^{6}$\\
        Shell (internal) & 2.5-7.5$\times$10$^{5}$ & & 0.5-1.7$\times$10$^{6}$  & $<$ 0.8-3.8$\times$10$^{5}$ \\
        \hline
    \end{tabular}
    \newline
    \vspace*{0.5 cm}
    \newline
    \begin{tabular}{cccccc}
    \multicolumn{5}{c}{Center + Dense ring}\\
    \hline
    \hline
     & P$_{th}$/k (K cm$^{-3}$) & P$_{rad}$ (K cm$^{-3}$)  & P$_{turb}$/k (K cm$^{-3}$) & P$_{B}$/k (K cm$^{-3}$) & P$_{ram}$/k (K cm$^{-3}$) (1)\\
    \hline
    Hot plasma & \\
    \HII\ region & 2$\times$10$^{6}$ (2) & 1.2$\times$10$^{6}$ \\
    Ambient cloud & 0.2-2.0$\times$10$^{5}$ & & $<$1.3-13$\times$10$^{6}$ & 0.4-7.8$\times$10$^{6}$ & 0.3-2.8$\times$10$^{6}$ \\
    Ring (internal) & 2.4-3.6$\times$10$^{5}$ & & 3.0-3.7$\times$10$^{6}$ & & \\
    \hline
    \end{tabular}
    \end{center}
    NOTE\hspace{0.3 cm}
    (1) Ram pressure provided by the ambient cloud on the expanding shell/ring. 
    (2) \citet{Minier2013}
    \label{tab:pressureTable}
\end{table*}

\begin{table*}[]
\caption{The turbulent energy (E$_{turb}$), dissipation timescale (t$_{diss}$), turbulent dissipation rate ($\dot{E}_{turb}$) and the total injected turbulent energy (E$_{turb,tot}$) for the dense ring and the full expanding shell associated with the bipolar cavities.}
    \begin{center}
    \setlength{\tabcolsep}{12pt}
    \begin{tabular}{ccccc}
        \hline
        \hline
         & E$_{turb}$ (erg) & t$_{diss}$ (Myr) & $\dot{E}_{turb}$ (erg s$^{-1}$) & E$_{turb,tot}$ (erg)\\
         \hline
        Ring & 8.2$\times$10$^{45}$ & 1.1$\times$10$^{-1}$ & 2.4$\times$10$^{33}$ & 8.2$\times$10$^{46}$ (over 1.1 Myr)\\
        Shells (from \CII) & 2.2$\times$10$^{46}$ & 2.0$\times$10$^{-1}$ & 3.2$\times$10$^{33}$ & 2.2$\times$10$^{46}$ (over 0.2 Myr)\\
        \hline
    \end{tabular}
    \end{center}
     NOTE\hspace{0.3 cm} These calculations for the shell make use of the typical velocity dispersion of 1.5 km s$^{-1}$ in the expanding shell and the shell mass deduced from \CII.
    \label{tab:turbulenceTable}
\end{table*}

\subsection{Leakage of the hot plasma}\label{sec:plasmaLeakage}
\begin{figure*}
    \includegraphics[width=0.95\hsize]{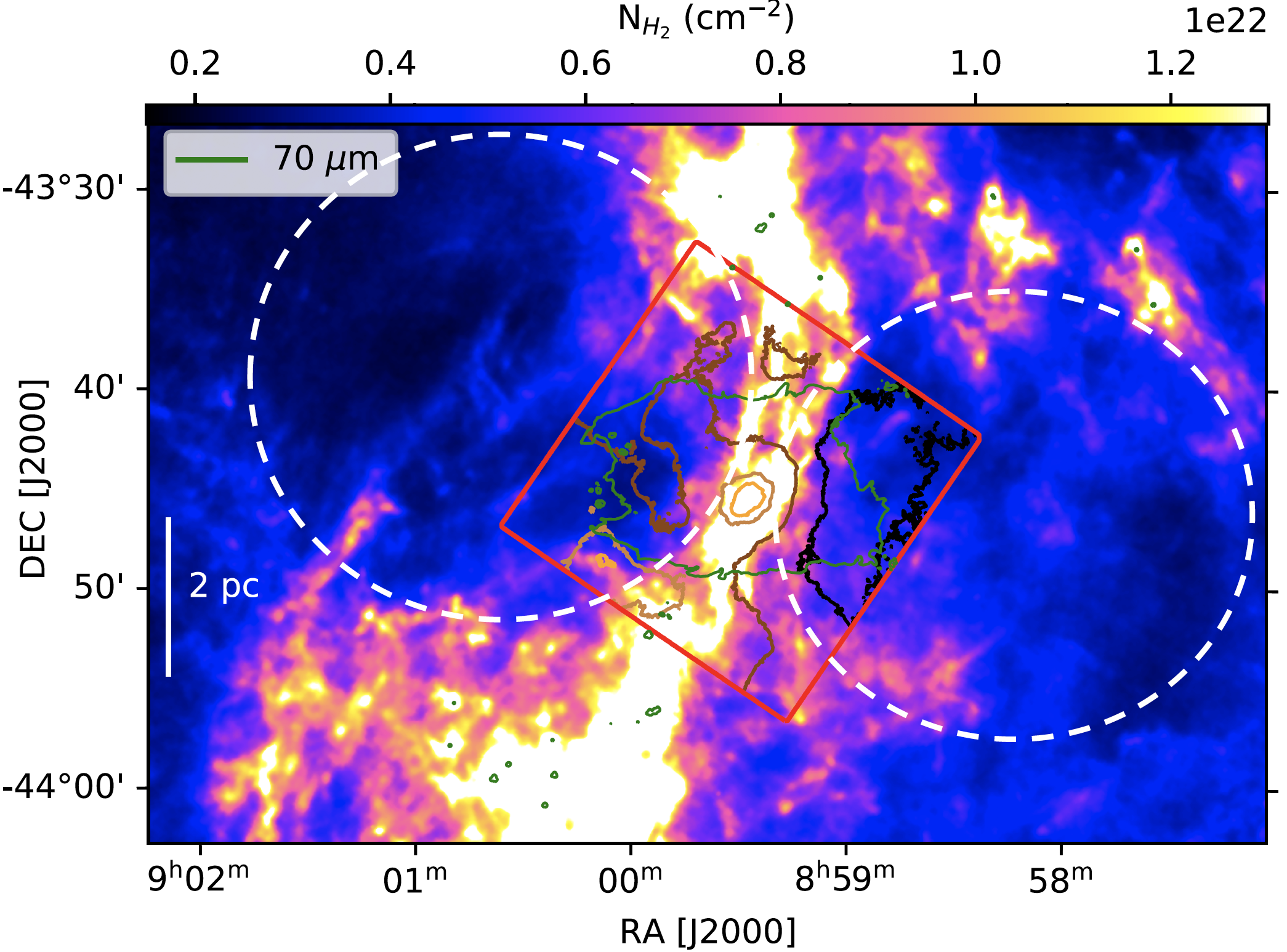}
    \caption{\textit{Herschel} column density map of the region in Vela C around RCW 36 (outlined by the green 70 $\mu$m contour). The bipolar cavity only fills up a part of the large low column density region around RCW 36, indicated by the dashed white circles. The black to orange contours indicate the increasing diffuse Chandra X-ray emission. This shows some hints of a correlation between the diffuse X-ray emission and the column density structure in the larger low-density regions surrounding RCW 36
    , e.g. in the north and south-east of the map. The red box indicates the region covered by the Chandra data.}
    \label{fig:largeCavities}
\end{figure*}

\begin{figure*}
    \centering
    \includegraphics[width=0.85\hsize]{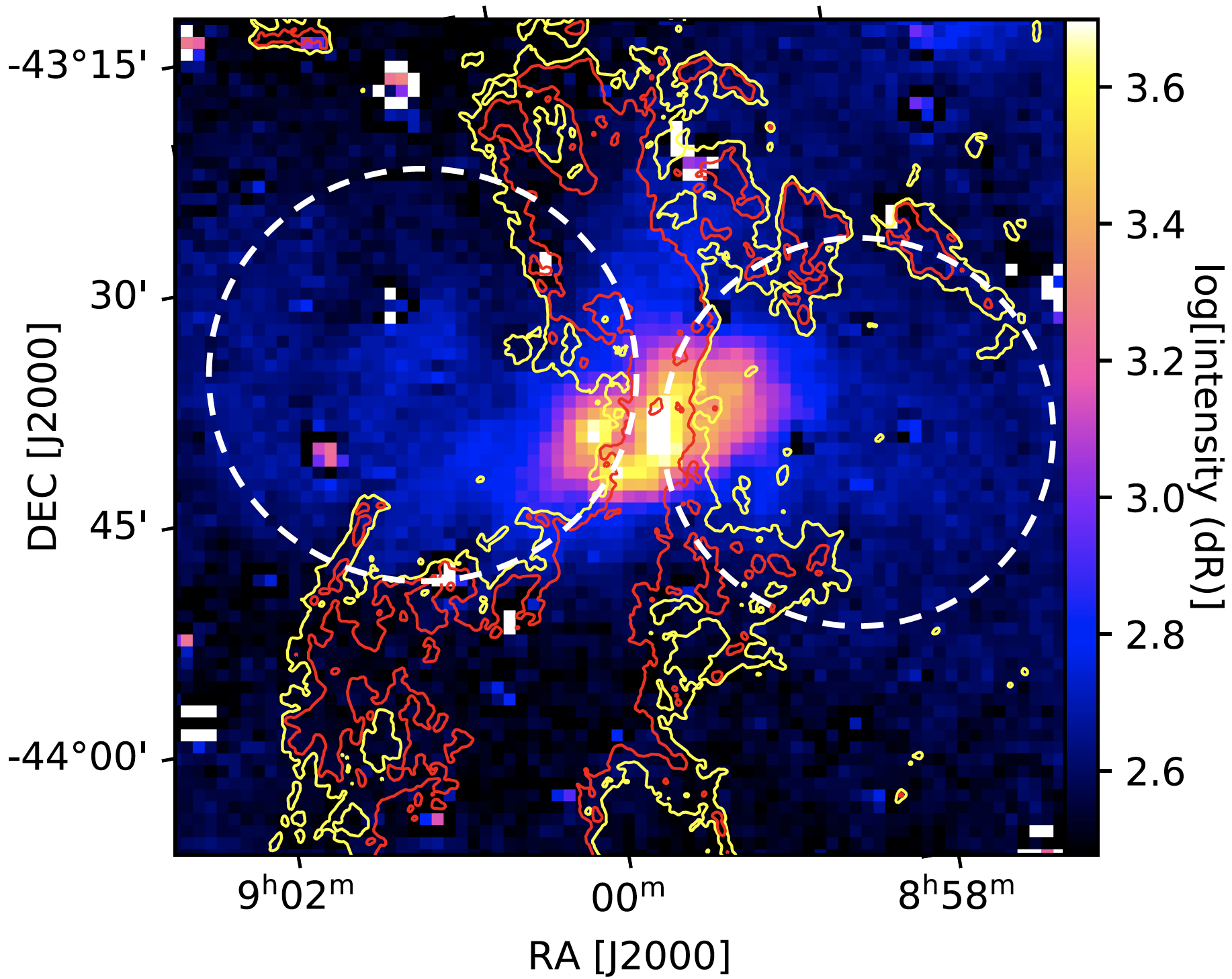}
    \caption{SHASSA H$\alpha$ map of the larger region around RCW 36. The peak H$\alpha$ emission corresponds to the center of RCW 36. The yellow and red contours indicated the \textit{Herschel} column density contours at N$_{H_{2}}$ = 4$\times$10$^{21}$ cm$^{-2}$ and 6$\times$10$^{21}$ cm$^{-2}$. The white dashed circles indicate the same large-scale low density cavities as in Fig. \ref{fig:largeCavities}. These cavities are clearly emitting H$\alpha$ emission, associated with the leakage of ionizing photons, compared to unperturbed regions in the Vela C molecular cloud.}
    \label{fig:largeCavitiesHalpha}
\end{figure*}

In Sect. \ref{sec:diffXrayEmission} we proposed that there is leakage of the hot plasma through the diffuse medium around RCW 36 which leads to the hot plasma found outside RCW 36 in Fig. \ref{fig:DiffuseXrayEmission}. To quantify the leakage, we have to estimate the total energy injected in the ISM by the stellar winds of the OB cluster. This is done using several stellar atmosphere models as there is a large uncertainty on the actual stellar wind mass-loss rates which from observations appear to be lower than predicted by the standard models \citep[e.g.][]{Martins2004,Martins2005b,deAlmeida2019}. However, recently it was proposed that the observed values are underestimated because observations miss a significant part of the stellar winds as a large fraction of the stellar wind might be in a hot phase that is not traced by the observations \citep{Lagae2021}. To cover the range of possibilities, we will use the relations that were put forward in Eq. 12 of \citet{Vink2000}, Eq. 3 of \citet{Lucy2010}, Eq. 11 of \citet{Krticka2017} and Eq. 20 of \citet{Bjorklund2021}, all with a terminal velocity of 2500 km s$^{-1}$ \citep[e.g.][]{Lamers1995,Krticka2017,Lagae2021}. The formulas differ in the choice of critical parameters and their dependencies (luminosity, metallicity, effective temperature, terminal velocity etc.), but it is beyond the scope of this paper to discuss their differences in more detail. In any case, employing a range of 
relations allows to cover the standard predicted mass ejection rate from stellar models \citep{Vink2000} to predictions that are more adapted to fit with the current observations \citep{Bjorklund2021}. For the 09V and 09.5V star candidates in the cluster we use the stellar parameters from \citet{Martins2005a} and \citet{Pecaut2013} to take into account the effects of different stellar models. The results are summarized in Tab. \ref{tab:SWmodelsTable}, which indicates a significant change depending on the stellar model used and an even larger effect related to the the stellar wind mass loss rate prescription used. Only taking into account the two O star candidates of RCW 36, this results in a total injected energy in the range of 4.7$\times$10$^{47}$-8.3$\times$10$^{48}$ erg over 1.1 Myr, see Tab. \ref{tab:SWmodelsTable}. This order of magnitude difference depending on the model might be overcome by the missing hot phase proposed in \citet{Lagae2021}, however future observations will have to address if there is indeed such a large fraction of the stellar winds in the hot phase. The energy in the hot plasma of RCW 36, from Tab. \ref{tab:PlasmaProperties}, is of the order of 1-2$\times$10$^{47}$ erg, implying that $>$50\%, and potentially up to 97\%, of the total injected stellar wind energy is no longer found in the hot plasma.\\\\ 
Stellar wind energy loss methods were listed in \citet{Townsley2003}, \citet{HarperClark2009} and \citet{Rosen2014}. Here we will explore the options. From Tab. \ref{tab:PlasmaProperties}, it is found that the hot plasma luminosity is a factor 3 to more than an order of magnitude lower than required to explain the missing energy in the hot plasma. The injected turbulent energy in the molecular ring, presented in Tab. \ref{tab:turbulenceTable}, is a factor 5 to two orders of magnitude smaller than total injected energy as well. Thermal conduction at the interface of the hot plasma and the turbulent shell layer is another potential way to lose the plasma energy. Recently, work by \citet{Lancaster2021a,Lancaster2021b} indicated that cooling can be very efficient at the interface because of turbulent substructure and that this could even explain all the observed missing hot plasma energy. However, it was also noted that thermal conduction is strongly reduced when a (weak) magnetic field is parallel to the shell \citep{Soker1994,Rosen2014}. This is the case for RCW 36 based on Fig. \ref{fig:blastPolIm} and it should be noted that \citet{Lancaster2021a,Lancaster2021b} currently do not include the magnetic field. Simulations in \citet{Arthur2011} indicate that the magnetic field will be mostly perpendicular to the shell in the ionized gas and aligned with the shell in the neutral gas. \citet{Rosen2014} note that a parallel magnetic field with 1 $\mu$G would decrease the thermal conduction by several orders of magnitude. In RCW 36, a magnetic field strength of at least several $\mu$G is expected \citep{Crutcher2012}, 
which seems to exclude thermal conduction based on this work. However, more in depth theoretical studies of energy loss at the turbulent interface, including the magnetic field, would be invaluable to better constrain this. Lastly, wind energy transfer to dust grains through collisions was also excluded in \citet{Rosen2014} as a plausible explanation. This suggests that the majority of the stellar wind energy might have leaked through lower-density regions out of RCW 36. Therefore, we here try to estimate the plasma energy leakage from RCW 36 based on Eq 8 from \citet{HarperClark2009}. As we propose the region is fully open at one side of the region we use a shell cover fraction (C$_{f}$) of 0.5 for RCW 36 and radius of 1 pc. Plugging in these values with the hot plasma parameters from Tab. \ref{tab:PlasmaProperties} gives a current energy leakage per cavity of -2.0$\times$10$^{35}$ erg s$^{-1}$ which is almost two orders of magnitude larger than derived values from other processes in this work. Since leakage probably occurred over the evolution of the \HII\ region, we use an intermediate radius of 0.5 pc to estimate the leakage over the lifetime of RCW 36. Note that this is only an approximation as we have no detailed information on the time evolution of RCW 36. It should also be noted that the C$_{f}$ definition in \citet{HarperClark2009} assumes equal leakage in all directions, whereas in RCW 36 we propose that leakage has a preferential direction. This then gives a total leaked plasma energy of -1.7$\times$10$^{48}$ erg over 1.1 Myr. Compared to the values in Tab. \ref{tab:SWmodelsTable}, this suggests that leakage indeed has the potential to account for most missing hot plasma energy.\\ 
If there is indeed such strong hot plasma leakage, it could be wondered if the leaking plasma still has an effect on the molecular cloud. Studying the large scale molecular cloud structure around RCW 36 with sensitive dust continuum observations (i.e. \textit{Herschel} \& BLASTPol), it can be noted that RCW 36 is located at the center of a large low-column density region in Vela C \citep{Hill2011,Fissel2016}, see Fig. \ref{fig:largeCavities}. Furthermore, inspecting Fig. \ref{fig:largeCavities} we observe that the leaking hot plasma shows hints it correlates with the structure of this large low-column density region. It is thus possible that the potential large fraction of leaking hot plasma clears the low-density ambient cloud on a larger scale. When the injected feedback energy escapes into the diffuse ISM through such lower-density regions or $'$chimneys$'$ in the cloud it can thus further disrupt these regions and their mass provision to the ridge around RCW 36. From the large scale SHASSA H$\alpha$ map presented in Fig. \ref{fig:largeCavitiesHalpha}, we confirm that this low density region is indeed a channel of feedback energy leaking into the larger cloud. Estimating the number of photons in these larger cavities is relatively uncertain as their extent is not strongly constrained. Excluding the H$\alpha$ emission in the cavities defined by the 70 $\mu$m emission, we obtain ionizing photon fluxes as high as 6-7$\times$10$^{46}$ s$^{-1}$ in these larger cavities. This is fairly similar to the values obtained inside the 70 $\mu$m cavities and thus supports significant leakage from RCW 36 into the diffuse cloud.\\ 
As Vela C still is a relatively young molecular cloud where RCW 36 is the first \HII\ region, this suggests that the leakage is the result of the molecular cloud morphology before feedback sets in. This presence of lower-density gas (or voids) is expected if molecular clouds are indeed built out of filaments and sheets. 
\begin{table*}[]
\caption{{\bf Top}: The stellar wind mass ejection rates ($\dot{\rm M}_{\rm SW}$) for the individual O stars. {\bf Bottom}: The resulting total ejected energy over the lifetime of the cluster (1.1 Myr) using terminal wind velocities of 2500 km s$^{-1}$.}
    \centering
    \begin{tabular}{cccccc}
    \hline
    \hline
       \multicolumn{2}{c}{}  & \citet{Vink2000} & \citet{Lucy2010} & \citet{Krticka2017}  & \citet{Bjorklund2021}\\
        Model & star & log[$\dot{\rm M}_{\rm SW}$ (M$_{\odot}$ yr$^{-1}$)] & log[$\dot{\rm M}_{\rm SW}$ (M$_{\odot}$ yr$^{-1}$)] & log[$\dot{\rm M}_{\rm SW}$ (M$_{\odot}$ yr$^{-1}$)] & log[$\dot{\rm M}_{\rm SW}$ (M$_{\odot}$ yr$^{-1}$)]\\
        \hline
         \citet{Martins2005a} & O9V & -7.35 & -8.56 & -7.78 & -8.31 \\
         \citet{Martins2005a} & O9.5V & -7.57 & -9.00 & -7.94 & -8.53\\
         \hline
         \citet{Pecaut2013} & O9V & -7.12 & -8.56 & -7.61 & -8.10 \\
         \citet{Pecaut2013} & O9.5V & -7.35 & -8.58 & -7.78 & -8.31\vspace{0.4 cm}\\
         \hline
         \hline
         & & E$_{\rm SW}$ (erg) & E$_{\rm SW}$ (erg) & E$_{\rm SW}$ (erg) & E$_{\rm SW}$ (erg)\\
         \hline
         \citet{Martins2005a} & O9V & 2.7$\times$10$^{48}$ & 1.7$\times$10$^{47}$ & 1.0$\times$10$^{48}$ & 3.0$\times$10$^{47}$\\
         \citet{Martins2005a} & O9.5V & 1.6$\times$10$^{48}$ & 5.7$\times$10$^{46}$ & 6.7$\times$10$^{47}$ & 1.7$\times$10$^{47}$\\
         \citet{Martins2005a} & total & 4.3$\times$10$^{48}$ & 2.3$\times$10$^{47}$ & 1.7$\times$10$^{48}$ & 4.7$\times$10$^{47}$\\
         \hline
         \citet{Pecaut2013} & O9V & 5.4$\times$10$^{48}$ & 1.9$\times$10$^{47}$ & 1.7$\times$10$^{48}$ & 5.4$\times$10$^{47}$\\
         \citet{Pecaut2013} & O9.5V & 2.9$\times$10$^{48}$ & 1.7$\times$10$^{47}$ & 1.1$\times$10$^{48}$ & 3.1$\times$10$^{47}$\\
         \citet{Pecaut2013} & total & 8.3$\times$10$^{48}$ & 3.6$\times$10$^{47}$ & 2.8$\times$10$^{48}$ & 8.5$\times$10$^{47}$\\
         \hline
    \end{tabular}
    \label{tab:SWmodelsTable}
\end{table*}

\subsection{The implication of the high-velocity \CII\ wings}\label{sec:massEjection}

The bipolar wings at high velocities in \CII\ show no clear expanding shell morphology and are not driven by protostars in the ring, but rather by feedback from the OB cluster. This could explain why the wings become less clear at the high-column density region of the ring in Fig. \ref{fig:outflowColdens}. \CII\ high-velocity wings were also found in the bipolar regions S106 \citep{Simon2012} and NGC 7538 IRS1 \citep{Sandell2020}. Unlike S106 \citep{Schneider2018}, \OI\ is not clearly detected in the wings of RCW 36 which suggests that the density in these wings is comparatively low. In NGC 7538 IRS1, the wings are also detected in $^{12}$CO(3-2) which suggests that for NGC 7538 IRS1 the \CII\ wings might trace a protostellar outflow there.\\
To investigate whether ablation by stellar winds can drive the mass ejection observed with the \CII\ wings, we compare the energy (and momentum) rates for the stellar winds and the \CII\ wings.  
The energy injection rate of the most favorable stellar wind models ($\sim$3.9-7.5$\times$10$^{42}$ erg yr$^{-1}$, from \citet{Vink2000}) is of the same order as the outflow energy rate seen with the \CII\ wings ($\sim$ 3-5$\times$10$^{42}$ erg yr$^{-1}$). As the \CII\ outflow parameters should be accurate within a factor 2-3 according to our evaluation in Sec. \ref{sect:outflowAnalysis}, this suggests that stellar wind ablation might be able to continuously drive the observed \CII\ wings. If these high-velocity wings are continuously ejected from RCW 36, it might account for a significant fraction of the missing plasma energy. However, it has to be taken into account that we also observe actual leakage of the hot plasma and that there are other potential stellar wind energy sinks discussed in previous sections.  In addition this conclusion is only valid for the \citet{Vink2000} models. The estimated energy injection rate by all other models would not be able to explain the observed wings.When considering the total momentum injection rate of the stellar winds, this is about two orders of magnitude lower than the momentum flux observed for the \CII\ wings.\\
Another potential driver would be the radiation from the OB cluster. In the simulations by \citet{Kim2018}, which focus on radiation feedback in molecular clouds, they found mass ejection with velocities between $\sim$ 6 and 40 km s$^{-1}$ as a result of thermal and radiation pressure from ionizing radiation. Radiation might thus be able to drive the observed mass ejection from the center of RCW 36. The calculated radiation pressure (I/c) at r = 0.9 pc (corresponding to the ring) is 1.2$\times$ 10$^{6}$ K cm$^{-3}$ and thus not significantly higher than the pressure from the hot plasma in the cavities. If the acceleration originates in the ionized gas closer to the OB cluster the radiation pressure would be higher and could thus more easily drive high-velocity mass ejection into the low density cavities.  As the ionized gas does not cover the full region, because of the hot plasma, this also increases the thermal pressure of the ionized gas. However, there is very little mass inside the central ring \citep{Minier2013} which implies that the mass ejection should happen at the surface of the ring and/or bipolar cavities. The mass ejection seen with the wings might thus be driven by a combination of stellar wind and pressure from the ionizing radiation. This combination would leave more room for the observed hot plasma leakage and other energy sinks. In general it thus remains challenging to explain the observed wings with the constrained feedback mechanisms for RCW 36.\\\\
The mass ejection rate derived from the \CII\ wings in Tab. \ref{tab:outflowParams} is high. It suggests that 4.8-7.4$\times$10$^{2}$ M$_{\odot}$ can be removed from the region confined by the ring in RCW 36 over the current lifetime of the OB cluster (or 0.9-1.3$\times$10$^{2}$  M$_{\odot}$ over 0.2 Myr). This could have important implications for the star formation efficiency (SFE) in the cloud. 
If this mass ejection is going on for 1.1 Myr, which could fit with the missing hot plasma energy, it would have already removed about the same mass from the original star forming region as what is now found in the molecular ring and more than an order of magnitude more than the $\sim$ 20 M$_{\odot}$ of ionized gas confined by the ring \citep{Minier2013}. We found no indications of inflow towards RCW36 from the ambient gas as it is actively being swept away, so the main mass inflow would happen along the ridge of Vela C. Using $\pi$R$^{2}$v$_{\rm inf}\rho$ it is then possible to calculate the remaining mass inflow to the RCW 36 ring through the ridge, where R is the radius, v$_{inf}$ the inflow velocity and $\rho$ the density. Here, we assume that the 1.5 km s$^{-1}$ velocity gradient along RCW 36 \citep{Fissel2019} is associated with inflow though it is equally plausible this is mainly the result of expansion. This gives $\dot{\rm M}_{\rm inf}$ = 0.7$\times$10$^{-4}$ M$_{\odot}$ yr$^{-1}$ when we further assume R = 0.1 pc \citep{Minier2013} and n$_{H_{2}}$ = 2.4$\times$10$^{4}$ cm$^{-3}$ (converted to the density assuming a mean molecular mass $\mu$ = 2.33). The inflow rate is thus more than a factor 5 lower than the mass ejected from RCW 36 and therefore has a small effect on the disruption. The observed mass ejection rate could thus stop any new star formation around RCW 36 on the timescale of 1-2 Myr.\\
It is also interesting that the mass ejection rate is within a factor two of the typical proposed mass inflow rates ($\sim$ 10$^{-3}$ M$_{\odot}$ yr$^{-1}$) in globally collapsing hubs and ridges before massive stars have formed \citep[e.g.][]{Schneider2010,Peretto2013}. This implies that the observed mass ejection can remove the mass from the central region on roughly the same timescale that the collapsing hub/ridge provides mass inflow (i.e. a few free fall timescales of the hub), which is consistent with several simulations that emphasize the importance of stellar feedback to maintain a low SFE in molecular clouds \citep[e.g.][]{Dale2012,Walch2012}. The observations of RCW 36 thus suggest that stellar feedback is essential in preserving the low SFE in proposed dynamic high-mass star forming scenarios due to effective mass removal from the originally collapsing pc scale region.\\ 
There are still several uncertainties related to this rapid mass removal (e.g. when the high-velocity mass removal started exactly, the currently incomplete map, etc.), but future studies using the larger statistical sample that will be provided by the FEEDBACK Legacy program will give further insight into the timescale of this rapid mass removal and how it affects the SFE in high-mass star forming regions.


\section{Conclusions}
We presented the currently obtained \CII\ and \OI\ observations from the SOFIA FEEDBACK Legacy program towards RCW 36, a bipolar \HII\ region in the young Vela C molecular cloud. These observations are complemented with $^{12}$CO(3-2) and $^{13}$CO(3-2) observations from APEX, Chandra X-ray observations, \textit{Herschel} photometric data, and the BLASTpol dust polarization observation at 500 $\mu$m. The main results can be listed as follows:\\\\
$\bullet$ The \CII\ and $^{13}$CO(3-2) data show that the central molecular ring is expanding with an estimated velocity of 1.0-1.9 km s$^{-1}$ (see Sec.  \ref{sec:ringExp}).\\
$\bullet$ The \CII\ and $^{12}$CO(3-2) data show the presence of several filaments in the ambient cloud that are being swept away with velocities $<$ 3 km s$^{-1}$ (see Sec.  \ref{sec:filsExp}).\\
$\bullet$ The \CII\ line unveils blueshifted expanding shells in the bipolar cavities with a velocity of  $\sim$5.2$\pm$0.5$\pm$0.5 km s$^{-1}$  (see Sec.  \ref{sec:expShells}).\\
$\bullet$ The \CII, \OI, and $^{12}$CO(3-2) spectra have important absorption signatures towards the bright molecular ring (see Sec.  \ref{sec:velComps}).\\
$\bullet$ The \CII\ emission indicates that the warm gas from the PDR layer has broken through the dense swept-up ring in the south-west of this ring (see Sec.  \ref{sec:ringExp}).\\
$\bullet$ The BLASTPol data shows that the magnetic field, dragged along by the expansion, follows the curvature of the shells around the bipolar cavities. However, the magnetic field orientation might have set the direction of the bipolar cavities. Higher resolution observations are needed to address this (see Sec. \ref{sec:pressEnergyBudget}).\\
$\bullet$ The $^{13}$CO(3-2) emission has an increased linewidth in the swept-up ring compared to the unaffected ridge in Vela C which might alter the outcome of the second generation of star formation towards RCW 36 (see Sec. \ref{sec:turbRing}).\\
$\bullet$ Diffuse Chandra emission traces a hot plasma, created by stellar winds, that fills RCW 36 and appears to be able to drive the 5.2 km s$^{-1}$ blueshifted expanding shells in the cavities (see Sec. \ref{sec:chandra}).\\
$\bullet$ Fitting the Chandra data also demonstrates that the observed hot plasma outside the \HII\ region is likely hot plasma leaking from the \HII\ region.\\
$\bullet$ We find that between 50\% and 97\%, depending on the stellar wind model, of the injected stellar wind energy is missing in RCW 36. Hot plasma leakage can provide a consistent explanation for this, but we can't fully constrain the contribution from energy sinks at the hot plasma-shell interface (see Sec. \ref{sec:plasmaLeakage}).\\
$\bullet$ We showed that the hot plasma and ionizing photon leakage has cleared a larger region around RCW 36 in the Vela C molecular cloud. (see Sec. \ref{sec:plasmaLeakage}).\\
$\bullet$ The \CII\ spectra have bipolar high-velocity wings in the cavities without a visible expanding shell velocity structure, nor a counterpart in CO. These wings trace high-velocity (v $>$ 15 km s$^{-1}$) and effective ($\sim$ 5$\times$10$^{-4}$ M$_{\odot}$ yr$^{-1}$) mass ejection from RCW 36 (see Sec. \ref{sec:massEjection}).\\\\
The observed rapid shell expansion in the cavities compared to the ring indicates that the expansion accelerates when it reaches lower density gas. This suggests that this clear bipolar stage might only be a relatively short phase in the evolution of RCW 36 and thus even that a bipolar morphology at some point in expanding \HII\ regions might be more common than previously thought. The observed multistage expansion morphology, as well as the proposed important leakage, point to a scenario where stellar feedback evolves in a structure of filaments, sheets and lower-density ambient gas that make up the Vela C molecular cloud. Lastly, the expansion timescale of the ring and rapid mass ejection indicate that stellar feedback can rapidly suppress local massive star formation in a cloud (in 1-2 Myr). This could even allow to maintain a low star formation rate in gravitationally collapsing regions of a molecular cloud where massive stars form. Additionally, the proposed clearing of the ambient low-density cloud can prevent further provision of cold gas necessary to continue high-mass star formation.

\acknowledgments
We thank the anonymous referee for detailed and very insightful comments that significantly improved the quality of this paper. We thank F. Martins and J-C Bouret for providing comments on the modeling and observations of stellar winds related to this work. We thank F. Comeron for fruitful discussions on the interpretation of the star cluster observations towards RCW 36. We further thank M. Pound and M. Wolfire for discussions related to PDR structure and the PDR Toolbox. \\
This work is based on observations made with the NASA/DLR Stratospheric Observatory for Infrared Astronomy
(SOFIA). SOFIA is jointly operated by the Universities Space Research Association, Inc. (USRA),
under NASA contract NNA17BF53C, and the Deutsches SOFIA Institut (DSI) under DLR contract 50 OK 0901
to the University of Stuttgart.\\
We thank the USRA and NASA staff of the Armstrong Flight Research
Center in Palmdale and of the Ames Research Center in Mountain View,
and the Deutsches SOFIA Institut for their work on the observatory. We
acknowledge the support by the upGREAT team for operating the
instrument, for planning the detailed observing scenarios and in the
calibration of the data. The development and operation of upGREAT was
financed by resources from the MPI f\"ur Radioastronomie, Bonn, from
Universit\"at zu K\"oln, from the DLR Institut für Optische Sensorsysteme,
Berlin, and by the Deutsche Forschungsgemeinschaft (DFG) within the
grant for the Collaborative Research Center 956 as well as by the Federal Ministry
of Economics and Energy (BMWI) via the German Space Agency (DLR)
under Grants 50 OK 1102, 50 OK 1103 and 50 OK 1104.\\
The FEEDBACK project is supported by the BMWI via DLR, Projekt Number 50 OR 1916
(FEEDBACK). \\
L.B. was supported by a USRA postdoctoral fellowship, funded through the NASA SOFIA contract NNA17BF53C. N.S. acknowledges supported by the Agence National
de Recherche (ANR/France) and the Deutsche Forschungsgemeinschaft
(DFG/Germany) through the project ‘GENESIS’ (ANR-16-CE92-0035-
01/DFG1591/2-1).

\vspace{5mm}
\facilities{SOFIA (upGREAT), Chandra, APEX, Herschel}


\software{astropy \citep{Astropy2013},  
          cygrid \citep{Winkel2016},
          }
          

\bibliography{ms}{}
\bibliographystyle{aasjournal}

\newpage
\newpage
\appendix

\section{The integrated \CII\ map at 90$^{\prime\prime}$ resolution}\label{sec:90arcsec}
\begin{figure}
    \centering
    \includegraphics[width=0.5\hsize]{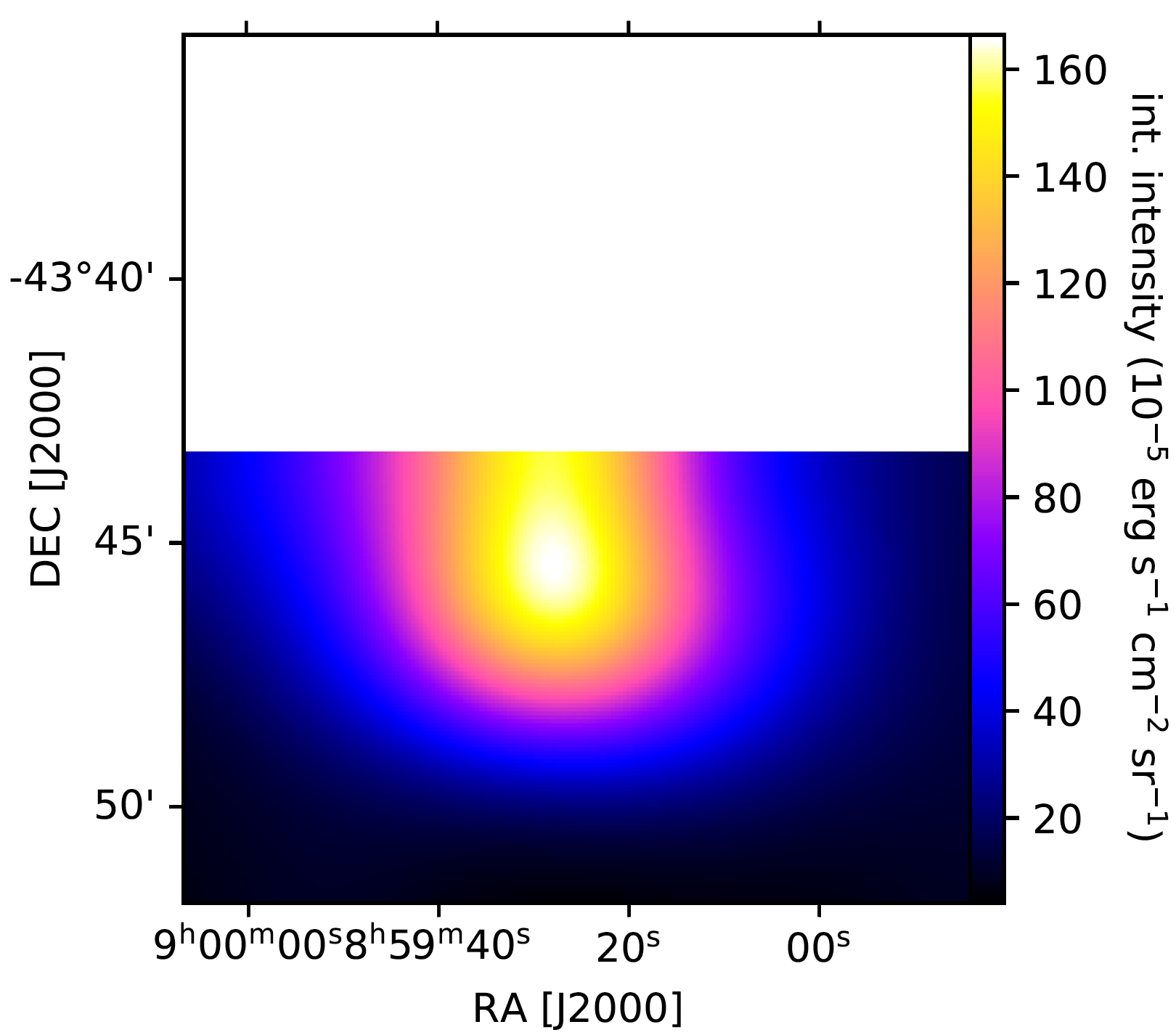}
    \caption{The FEEDBACK \CII\ integrated intensity map smoothed to a spatial resolution of 90$^{\prime\prime}$. It looks exactly like the 90$^{\prime\prime}$ map presented in Fig. 2 of \citet{Suzuki2021}.}
    \label{fig:compSuzuki}
\end{figure}

In order to compare the FEEDBACK observations with the recent observation of RCW 36 at 90$^{\prime\prime}$ resolution by \citet{Suzuki2021}, the integrated intensity map in Fig. \ref{fig:integratedIntensities} was smoothed to a resolution of 90$^{\prime\prime}$. Then the units were converted using I(\CII) (erg s$^{-1}$ sr$^{-1}$ cm$^{-2}$) = 7.0354$\times10^{-6}$ I(\CII) (K km s$^{-1}$). The resulting map is presented in Fig. \ref{fig:compSuzuki}. Comparing this map with Fig. 2 in \citet{Suzuki2021}, it is found that the structure and observed intensity of the region covered so far by the FEEDBACK observations is extremely similar. This is an important validation for the data presented in this paper.

\section{\CII\ self-absorption demonstrated with \dCII }\label{sec:selfAbs}
\begin{figure}
    \centering
    \includegraphics[width=0.49\hsize]{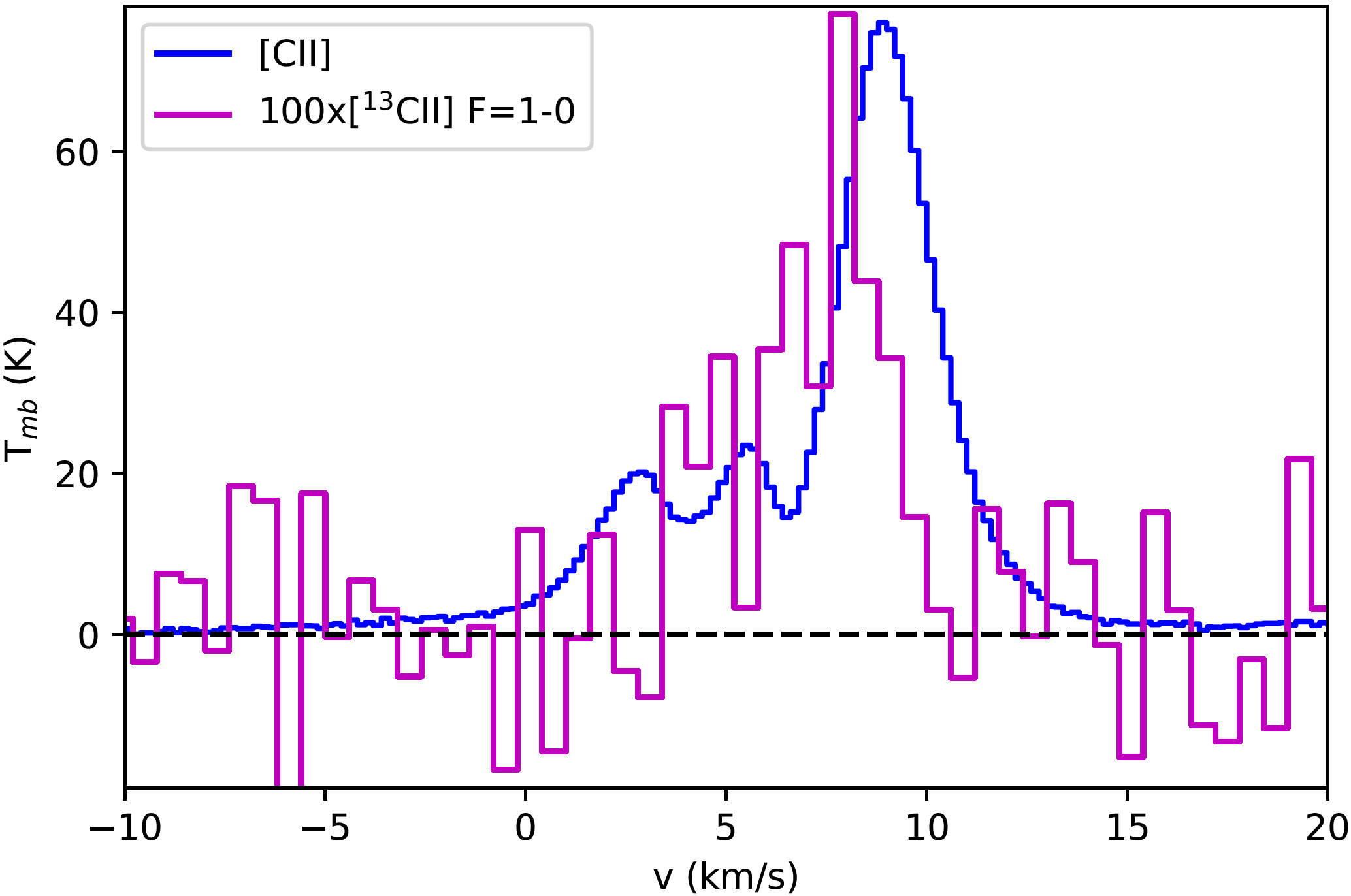}
    \includegraphics[width=0.49\hsize]{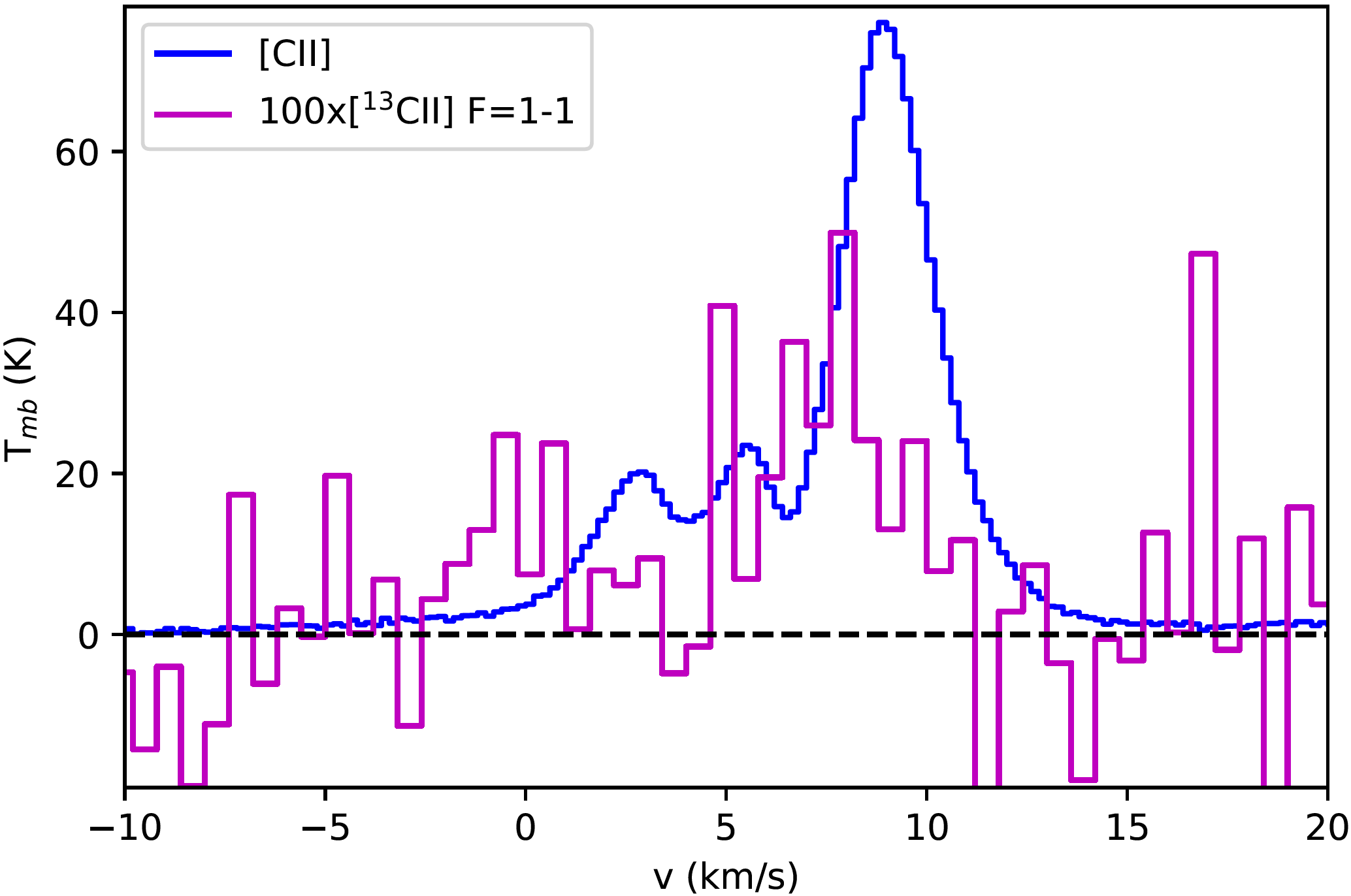}
    \caption{The \CII\ line compared with the \dCII\ F=1-0 and F=1-1 hyperfine transitions for the ring of RCW 36. Both \dCII\ transitions are detected and indicate that the \dCII\ emission is centered on the dip at 7 km s$^{-1}$ of the \CII\ line, implying self-absorption of the \CII\ line.}
    \label{fig:13CIIfig2}
\end{figure}

\begin{figure}
    \centering
    \includegraphics[width=0.6\hsize]{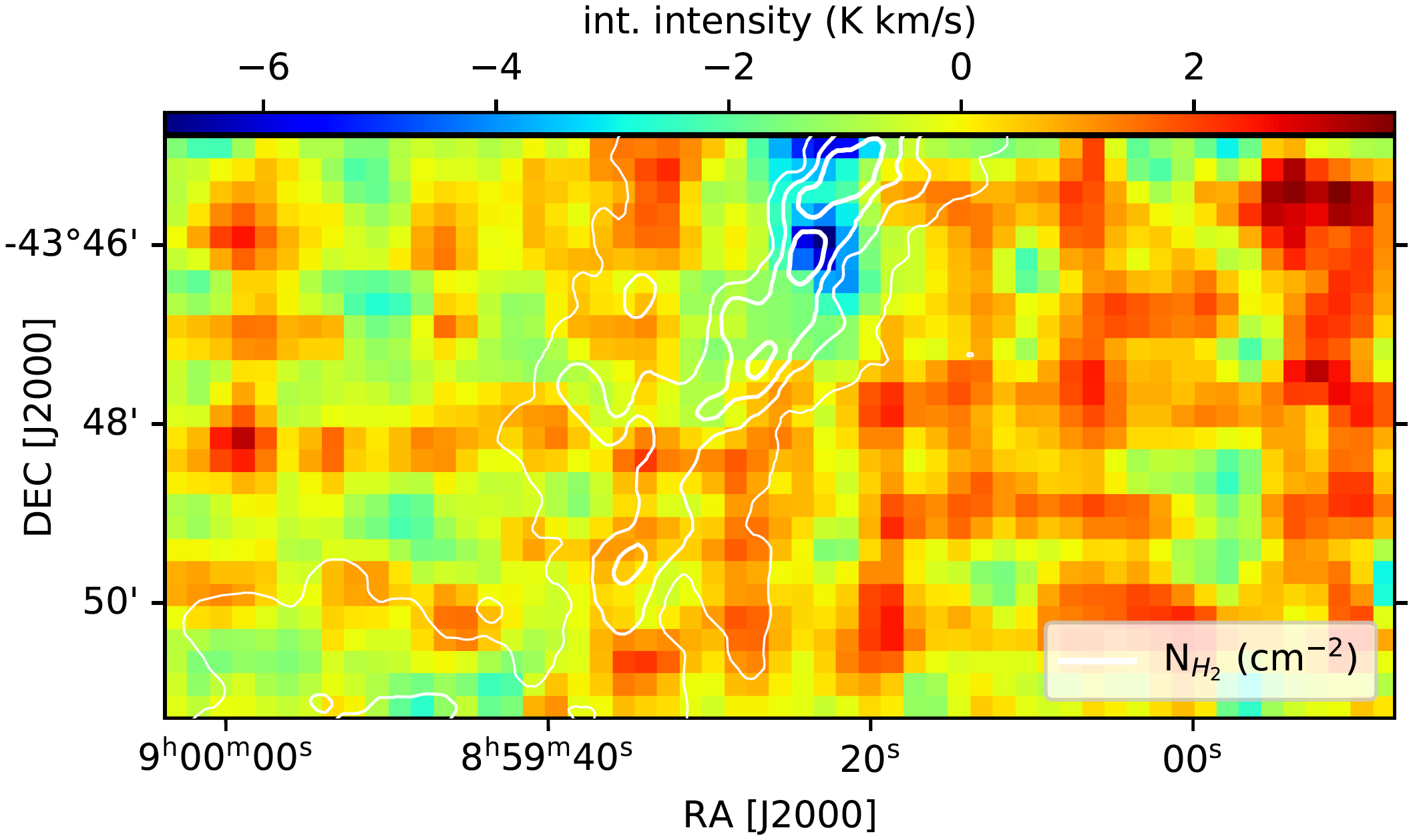}
    \caption{The \OI\ integrated intensity map from 3 to 7 km s$^{-1}$ with the white contours indicating the column density. Towards the dense ring, and in particular the highest column density region, there is absorption in this velocity interval.}
    \label{fig:OIselfabsorptionMap}
\end{figure}

To verify whether \CII\ self-absorption affects the observed spectrum, the three \dCII\ hyperfine structure lines were used as they are expected to be optically thin \citep{Ossenkopf2013}. As the \dCII\ hyperfine structure lines are at a relative velocity of -65.2 km s$^{-1}$ (F=1-0), 11.7 km s$^{-1}$ (F=2-1) and 62.4 km s$^{-1}$ (F=1-1) compared to the \CII\ line, these three hyperfine transitions are also covered by the upGREAT receiver. The brightest transition (F=2-1) is closest to the \CII\ and contaminated by the \CII\ outflow wing for RCW 36. The F=2-1 transition is presented in Fig. \ref{fig:13CIIline} and the two other hyperfine transitions are presented in Fig. \ref{fig:13CIIfig2}. These figures show, as far as possible for F=1-0 and F=1-1 because of the noise, that there is one main velocity component at 7 km s$^{-1}$ which leads to strong absorption that is seen with \CII, \OI, $^{12}$CO(3-2) and even a bit with $^{13}$CO(2-1). This result is also consistent with Fig. 2 in \citet{Fissel2019} which shows that the optically thin N$_{2}$H$^{+}$(1-0) dense gas tracer is centered on 7 km s$^{-1}$.\\
To visualize the \OI\ absorption, the spectra were integrated between 3 and 7 km s$^{-1}$ (the velocities where the \OI\ emission goes under the baseline in Fig. \ref{fig:integratedIntensities}). The resulting map is shown in Fig. \ref{fig:OIselfabsorptionMap} which shows that the strongest absorption is located at the densest clumps of the ring.

\section{Estimation of the FUV field and \CII\ temperature}\label{sec:fuvEstimate}
\begin{figure}
    \centering
    \includegraphics[width=0.6\hsize]{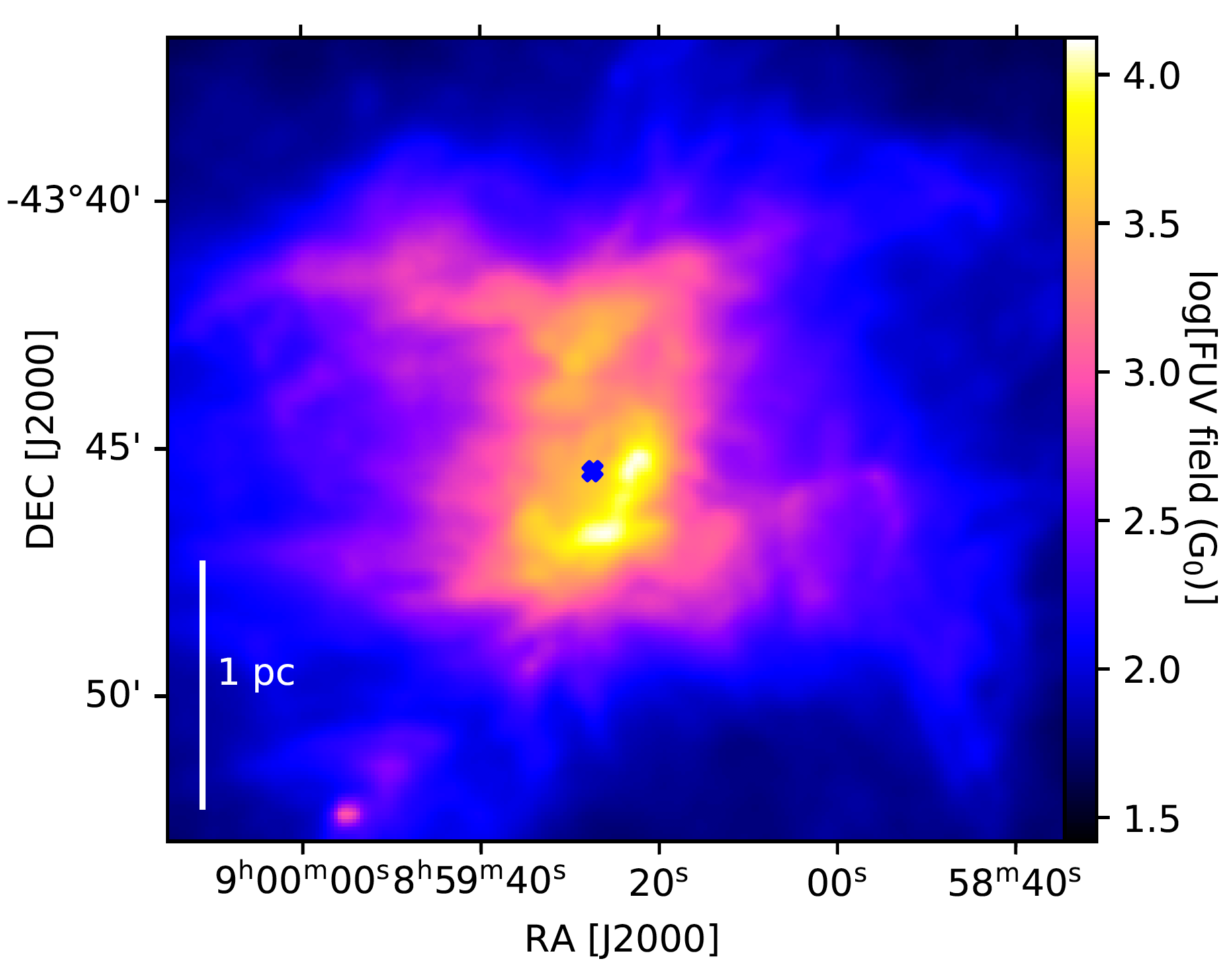}
    \caption{The FUV field map of RCW 36 on a log scale calculated from the \textit{Herschel} 70 $\mu$m and 160 $\mu$m maps. The blue crosses indicate the locations of the O stars.}
    \label{fig:fuvMap}
\end{figure}

\begin{figure}
    \centering
    \includegraphics[width=0.6\hsize]{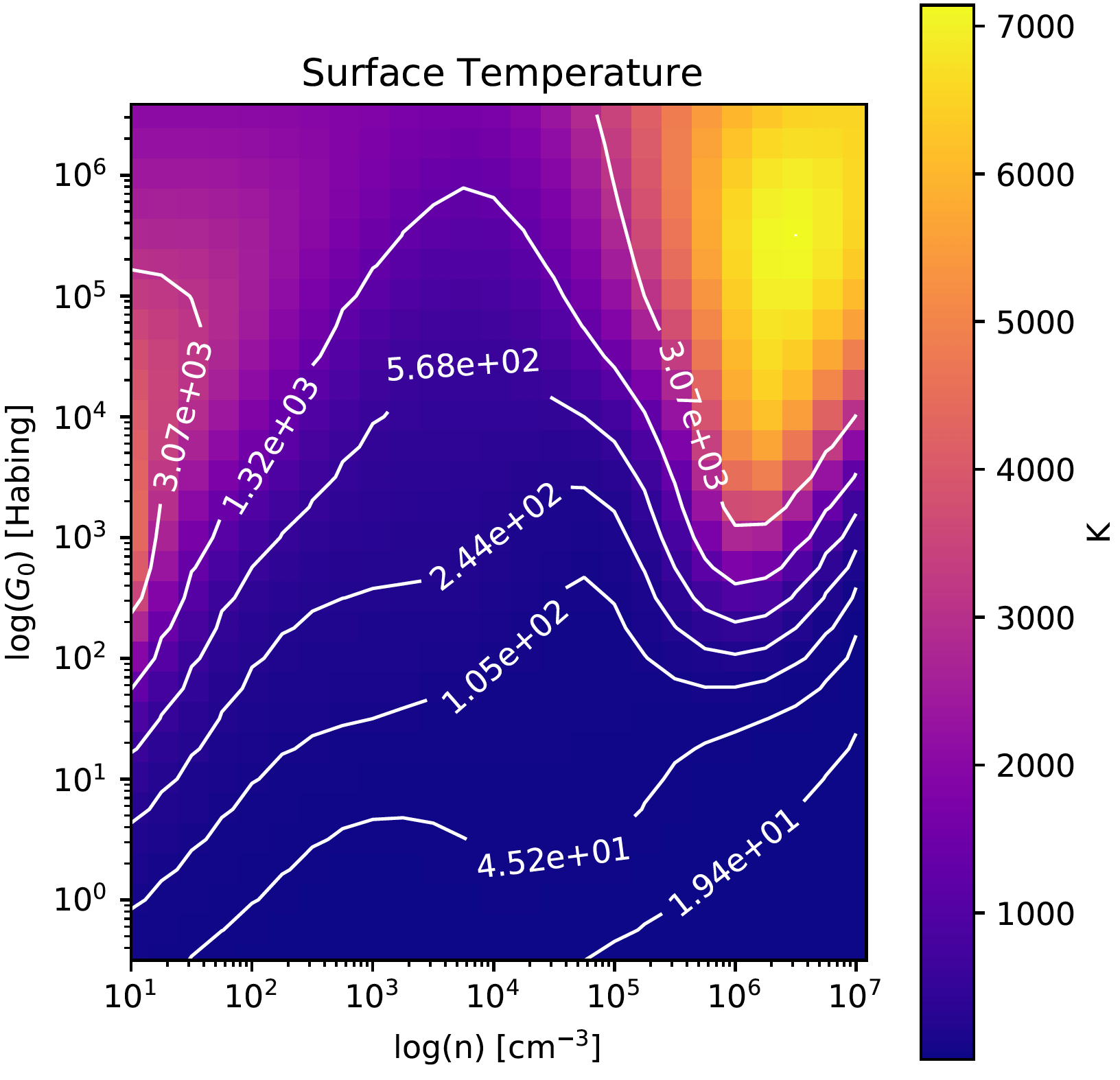}
    \caption{The surface temperature as a function of the density and FUV field in the PDR Toolbox \citep{Kaufman2006,Pound2008}. Working with G$_{0}$ = 500-2000 and possible densities in the shells and cavities (10$^{2}$ and 10$^{5}$ cm$^{-3}$) points to temperatures between 100 and 500 K for \CII.}
    \label{fig:temperaturePDRtoolbox}
\end{figure}

In order to estimate a temperature for the gas traced by \CII\ we will work with the PDR Toolbox \citep{Kaufman2006,Pound2008},  but first we must estimate the strength of the FUV field. Two methods are used to estimate the field strength independently. First we use the method presented in \citet{Kramer2008,Schneider2016} which makes use of the \textit{Herschel} intensity at 70 $\mu$m and 160 $\mu$m to estimate FUV field in Habing \citep{Habing1968} using
\begin{equation}
    \text{F}_{\text{FUV}}[\text{G}_{0}] = 4000\pi \text{I}_{\text{FIR}}/1.6
\end{equation}
where I$_{\text{FIR}}$ is the far-infrared intensity from adding the intensity at 70 $\mu$m and 160 $\mu$m. The resulting map of the estimated FUV field strength is presented in Fig. \ref{fig:fuvMap} which shows that the FUV field strength is generally between G$_{0}$ = 300-2000 for RCW 36 within 1-2 pc from the cluster center. This method assumes that the \CII\ emission originates from roughly the same region as this warm dust emission, which likely is a relatively good assumption.\\
The second method to calculate the FUV field at a specific location, presented in \citet{Bonne2020a}, directly works with the ionizing radiation intensity from the O stars in the central cluster. The FUV irradiation from these O stars is calculated from the \citet{Kurucz1979} models with solar abundances \citep{Grevesse1998}. In order to select the Kurucz models to calculate the FUV field, the stellar parameters for the spectral types of the O stars were taken from \citet{Pecaut2013}. This results in an FUV field of G$_{0}$ = 1.9$\times 10^{3}$ at a distance of 1 pc and G$_{0}$ = 4.7$\times 10^{2}$ at a distance of 2 pc. These calculations of the FUV field take no extinction on the path of the FUV photons in account, but fit well with the values obtained from the previous method which suggests that the FUV field at the expanding shells in the cavities is of the order of G$_{0}$ = 500-2000.\\
From this information it is possible to estimate a temperature in the PDR using the PDR Toolbox. The surface temperature of the PDR as a function of the density is presented in Fig. \ref{fig:temperaturePDRtoolbox} which indicates that a surface temperature of 100-500 K is expected for a FUV field strength between G$_{0}$ = 500-2000 and densities n = 10$^{2}$-10$^{5}$ cm$^{-3}$ \citep{Minier2013}. This surface temperature of the PDR corresponds to the temperatures traced by \CII. To verify this method, it is also found with the PDR Toolbox that the observed \CII\ intensity of $\sim$ 100 K km s$^{-1}$ towards the shells of RCW 36 fits with G$_{0} \sim$ 10$^{3}$ and n $\sim$ 10$^{3}$-10$^{4}$ cm$^{-3}$ which is expected for the expanding shells.

\section{Fitting the X-ray spectra}\label{sec:xpecFit}
\begin{figure*}
    \centering
    \includegraphics[width=\hsize]{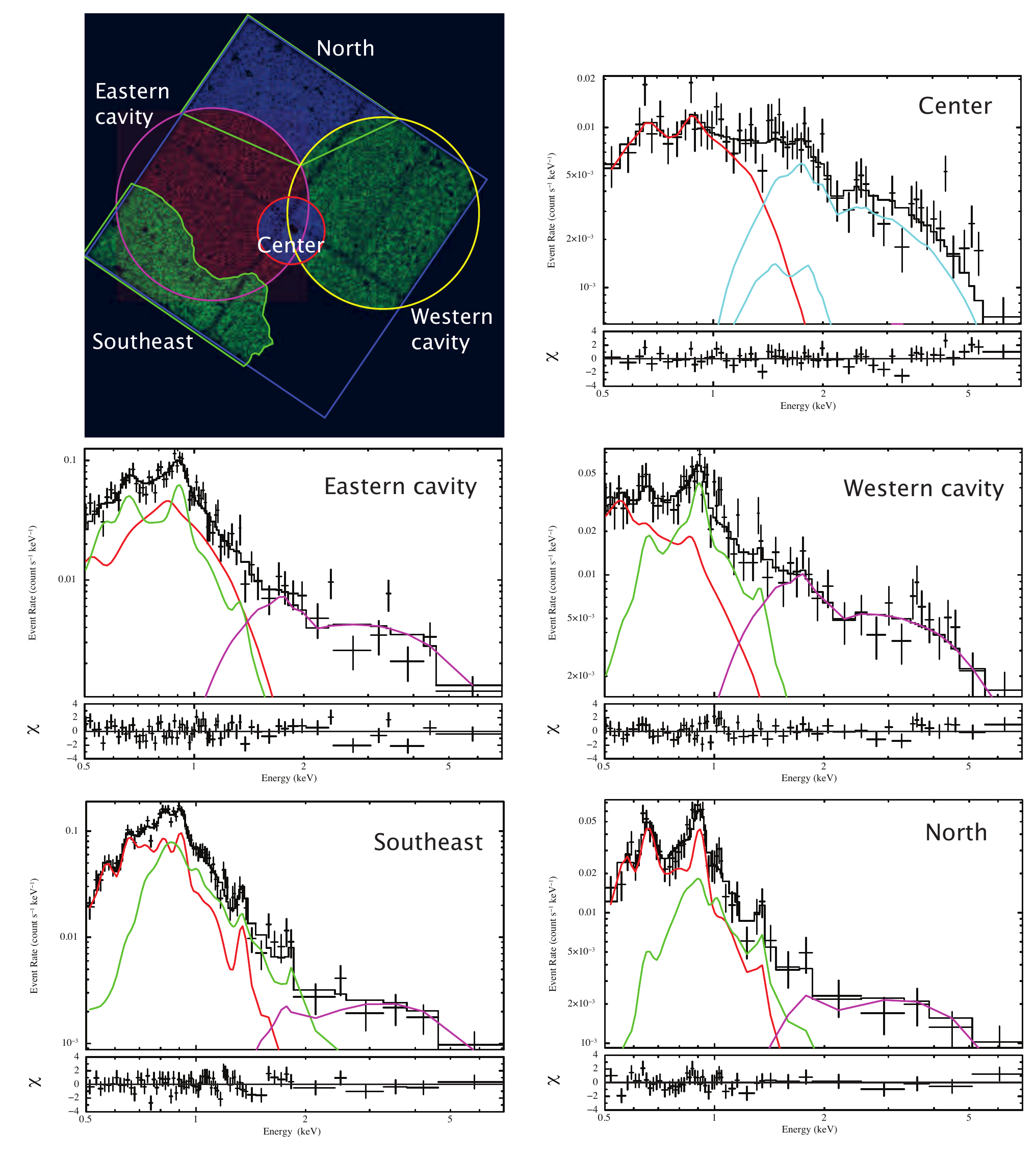}
    \caption{
    Chandra/ACIS spectra of diffuse X-ray emission in RCW~36, in the regions defined by Fig. \ref{fig:DiffuseXrayEmission} and repeated in the upper left panel here.  This panel also demonstrates that these regions are exclusive, with no X-ray event appearing in more than one spectrum:  Center and North events are color-coded in blue, Western Cavity and Southeast events are in green, and Eastern Cavity events are in red.  The spectra show the total model in black, the first diffuse thermal plasma component in red, the second diffuse thermal plasma component in green, and background emission in magenta.  For the Center region (upper right panel), cyan curves model the emission from unresolved pre-main sequence stars in the young cluster.  Lower panels in all spectra show the fit residuals.  Spectral fit parameters and model details are given in Tab. \ref{tab:PlasmaFit}.
    }
    \label{fig:xRaySpectra}
\end{figure*}

The fitting of the X-ray spectra, presented in Fig. \ref{fig:xRaySpectra}, uses the XSPEC package\footnote{https://heasarc.gsfc.nasa.gov/docs/xanadu/xspec/manual/manual.html} \citep{Arnaud1996}, and takes into account diffuse emission components from hot plasmas, unresolved stellar clusters, background emission and a constant absorption from foreground material over a considered field of view. When fitting a hot plasma the chemical abundance values from \citet{Wilms2000} are assumed. The fitting results are presented in Tab. \ref{tbl:diffuse_spectroscopy_style2} and Fig. \ref{fig:xRaySpectra}.\\
The spectrum of the central region around the OB cluster is best fitted by a hot plasma and an unresolved stellar cluster. The unresolved cluster has an X-ray luminosity of 0.17 L$_{\odot}$ between 0.5 and 7 keV. Converting this to a bolometric luminosity using log(L$_{x}$/L$_{bol}$) = -3.6 \citep{Bhatt2013}, gives a a bolometric luminosity of 6.8$\times$10$^{2}$ L$_{\odot}$. The fitted unresolved cluster can thus correspond to low-mass stars from the 1.3$\times$10$^{5}$ L$_{\odot}$ OB cluster \citep{Verma1994} that were not resolved with ACIS. For the fitted hot plasma, the ionization timescale $\tau$ can range between 10$^{8}$ and 5$\times$10$^{13}$ s cm$^{-3}$. The lower limit indicates a plasma with non-equilibrium ionization (NEI) and the upper limit indicates a plasma with collisional ionization equilibrium (CIE). For the central region, $\tau$ thus indicates a low to intermediate timescale.\\
The other four regions contain no unresolved cluster and are instead fitted by two hot plasma components because a single hot plasma component did not give a satisfactory fit to the spectra. During the fit of the two plasma components in the bipolar cavities, the first plasma component went to small ionization timescales and the second one to large ionization timescales, while temperatures of both hot plasma components converged to the same temperature. Therefore, in a second run, the temperatures for both plasmas were imposed to be the same. The presence of NEI and CIE with the same temperature in the fit indicates a recombining plasma that is in transition between from NEI to CIE. From the larger surface brightness in both cavities for the CIE components, it is found that the plasmas are dominated by CIE. In the region outside the \HII\ region, i.e. north and southeast, the temparatures of both plasma components did not converge. The first component has a temperature similar to the ones in the cavities, and is likely the hot plasma that is leaking from the bipolar \HII\ region. The second plasma component is slightly hotter and is probably fore- or background emission as RCW 36 is still relatively close to the galactic plane. The surface brightness indicates that the leaking plasma is dominating the emission.\\
The fitting also shows that the observed diffuse X-ray emission is significantly impacted by absorption from foreground layers with N$_{H} \sim$ 5$\times$10$^{21}$ cm$^{-2}$. This is particularly seen in column 23 of Tab. \ref{tbl:diffuse_spectroscopy_style2}, where it is found that the intrinsic surface brightness of both cavities, corrected for absorption, is higher than for the regions outside the \HII\ region. Interestingly, the fitted absorbing column density value for the eastern cavity is extremely similar to the typical column density found in the \textit{Herschel} maps. For the western cavity the fitting had more difficulties to constrain the typical column density, so we decided to supply the foreground column density based on the \textit{Herschel} data.

\setlength{\tabcolsep}{0.5mm}

%

\setlength{\tabcolsep}{0.5mm}
 \begin{deluxetable}{@{\hspace{1em}}lrrhrhhhhhh||llllc|llllc|hhhhh|llcc|lhlc|hhhh||cchh@{}}   
 
 \tabletypesize{\tiny}
\rotate

\tablecaption{Spectral Fits for Diffuse X-ray Regions \label{tbl:diffuse_spectroscopy_style2}}

\tablehead{
%
%
\multicolumn{4}{c}{Extraction Region} &
\colhead{Quality} &
\multicolumn{21}{c}{Diffuse Components} &
\multicolumn{4}{c}{Unresolved Stars} &                                                 
\multicolumn{4}{c}{Background} &
\multicolumn{4}{c}{} &
 \multicolumn{2}{c}{Total Diffuse} 
 \\[-10pt]
\multicolumn{4}{c}{} &
\multicolumn{1}{c}{} &
 \multicolumn{6}{c}{} &                  
\multicolumn{15}{c}{\hrulefill} &
\multicolumn{3}{c}{} &                                                 
\multicolumn{4}{c}{} \\[-3pt]
\multicolumn{4}{c}{} &
\multicolumn{1}{c}{} &
 \multicolumn{6}{c}{}           &                  
\multicolumn{5}{c}{pshock$^1$} &
\multicolumn{5}{c}{pshock$^2$} &
\multicolumn{5}{c}{} &
\multicolumn{4}{c}{apec$^4$+apec$^5$} &
\multicolumn{4}{c}{apec$^6$} &
\multicolumn{4}{c}{} &
 \multicolumn{2}{c}{Emission} \\[-7pt]
\multicolumn{4}{c}{\hrulefill} &
\multicolumn{1}{c}{\hrulefill} &
 \multicolumn{6}{c}{}           &                  
\multicolumn{5}{c}{\hrulefill} & 
\multicolumn{5}{c}{\hrulefill} & 
\multicolumn{5}{c}{} & 
\multicolumn{4}{c}{\hrulefill} & 
\multicolumn{4}{c}{\hrulefill} & 
\multicolumn{4}{c}{} & 
\multicolumn{2}{c}{\hrulefill}   \\[-5pt]
\colhead{Name} & \colhead{SNR} & \colhead{Area} & \nocolhead{} & 
\colhead{$\chi^2 \!/\! \rm{DOF}$} &
%
 \nocolhead{O} & \nocolhead{Ne} & \nocolhead{Mg} & \nocolhead{Si} & \nocolhead{S} & \nocolhead{Fe} & 
\colhead{$N_{H}^{1}$} & \colhead{$kT^1$} & \colhead{${\tau}^1$} & \colhead{$SEM^1$} & \colhead{$SB^{1}_{tc}$} &
\colhead{$N_{H}^{2}$} & \colhead{$kT^2$} & \colhead{${\tau}^2$} & \colhead{$SEM^2$} & \colhead{$SB^{2}_{tc}$} &
\nocolhead{$N_{H}^{3}$} & \nocolhead{$kT^3$} & \nocolhead{${\tau}^3$} & \nocolhead{$SEM^3$} & \nocolhead{$SB^{3}_{tc}$} &
\colhead{$N_{H}^{stars}$} &                                       \colhead{$SEM^5$} & \colhead{$SB^{stars}_{tc}$} & \colhead{$L^{stars}_{tc}$} &
\colhead{$N_{H}^{6}$} & \nocolhead{$kT^6$} &                      \colhead{$SEM^6$} & \colhead{$SB^{6}_{tc}$} &
\nocolhead{$N_{H}^{7}$} & \nocolhead{$\Gamma^{7}$} &                                      \nocolhead{$SB^{7}_{tc}$} & \nocolhead{$L^{7}_{tc}$} &
\colhead{$SB^{1+2}_{tc}$} & \colhead{$L^{1+2}_{tc}$}
%
%
\\
\colhead{} & \colhead{(cts)} & \colhead{(pc$^{2}$)} & \nocolhead{($\arcmin^{2}$)} &
\colhead{} & 
 \multicolumn{6}{c}{}              &        
\multicolumn{29}{c}{(\dotfill\ $N_H$: $10^{22}$cm$^{-2}$\ \dotfill\ $kT$:  keV\ \dotfill\ $\tau$: log cm$^{-3}$ s\ \dotfill\ $SEM$: log cm$^{-3}$ pc$^{-2}$\ \dotfill $SB$: log erg s$^{-1}$ pc$^{-2}$ \dotfill $L$: log erg s$^{-1}$ \dotfill)} 
\\[-10pt]
\colhead{(1)} & \colhead{(2)} & \colhead{(3)} & \nocolhead{} & \colhead{(4)} & 
\nocolhead{} & \nocolhead{} & \nocolhead{} & \nocolhead{} & \nocolhead{} & \nocolhead{} & 
\colhead{(5)} & \colhead{(6)} & \colhead{(7)} & \colhead{(8)} & \colhead{(9)} & 
\colhead{(10)} & \colhead{(11)} & \colhead{(12)} & \colhead{(13)} & \colhead{(14)} & 
\nocolhead{(15)} & \nocolhead{(16)} & \nocolhead{(17)} & \nocolhead{(18)} & \nocolhead{(19)} & 
\colhead{(15)} & \colhead{(16)} & \colhead{(17)} & \colhead{(18)} & 
\colhead{(19)} & \nocolhead{(20)} & \colhead{(20)} & \colhead{(21)} & 
\nocolhead{} & \nocolhead{} & \nocolhead{} & \nocolhead{} & 
\colhead{(22)} & \colhead{(23)} 
}
%
%
%
\startdata
                        Center      & 1650/61 &     0.9 &    12.6 &    54/ 53 & & & & & &  & $0.55\phd_{-0.3}^{+0.3}$   & $0.23\phd_{-0.09}^{+0.37}$  & ${\phn}8.98\phd_{-0.4}^{+0.3}$ & $ 55.9\phd_{\cdots}^{+1.5}$   &   32.26& \nodata                    & \nodata                      & \nodata                      & \nodata                     & \nodata& \nodata & \nodata & \nodata & \nodata & \nodata& $2.8\phd_{-0.8}^{+0.9}$           & $ 54.7\phd_{-0.09}^{+0.08}$ &   31.88 &   31.82& $4.8\phd\!*$ & $10\phd\!*$      & $ 53.9\phd\!*$ &   31.10& \nodata           & \nodata & \nodata & \nodata&   32.26 &   32.20&            grp3.0\_diffuse\_3pshock &                        HandFit \\[1ex]  
                    Eastern Cavity  & 4240/126 &     4.7 &    68.7 &    77/ 62 & & & & & &  & $0.51\phd_{-0.07}^{+0.11}$ & $0.15\phd_{-0.02}^{\cdots}$ & ${\phn}8.00\phd\!*$            & $ 56.5\phd_{\cdots}^{+0.4}$   &   31.69& $0.51\phd\!*$              & $0.15\phd\!*$                & $13.7\phd\!*$                & $ 55.3\phd_{-0.3}^{+0.3}$   &   32.00& \nodata & \nodata & \nodata & \nodata & \nodata& \nodata                           & \nodata                     &    \nodata &    \nodata& $2.4\phd\!*$ & $10\phd\!*$      & $ 53.9\phd\!*$ &   31.10& \nodata           & \nodata & \nodata & \nodata&   32.17 &   32.85&            grp3.0\_diffuse\_3pshock &                        HandFit \\[1ex]  
                    Western Cavity  & 3493/132 &     6.5 &    94.8 &    60/ 58 & & & & & &  & $0.11\phd_{\cdots}^{+0.3}$   & $0.17\phd_{-0.02}^{+0.03}$  & ${\phn}8.66\phd_{-0.5}^{+0.3}$ & $ 55.3\phd_{\cdots}^{+0.4}$ &   30.92& $0.80\phd\!*$  & $0.17\phd\!*$                & $13.7\phd\!*$                & $ 55.1\phd_{-0.4}^{+0.3}$ &   31.91& \nodata & \nodata & \nodata & \nodata & \nodata& \nodata                           & \nodata                     &    \nodata &    \nodata& $2.0\phd\!*$ & $10\phd\!*$      & $ 53.9\phd\!*$ &   31.10& \nodata           & \nodata & \nodata & \nodata&   31.95 &   32.76&            grp3.0\_diffuse\_3pshock &                        HandFit \\[1ex]  
                        Southeast   & 5269/116 &     3.6 &    52.5 &    64/ 59 & & & & & &  & $0.29\phd_{-0.1}^{+0.1}$   & $0.26\phd_{-0.08}^{+0.13}$  & $11.4\phd_{-0.4}^{+0.6}$       & $ 54.5\phd_{\cdots}^{+0.7}$   &   31.83& $0.14\phd_{\cdots}^{+0.1}$ & $0.68\phd_{\cdots}^{\cdots}$ & $12.9\phd_{\cdots}^{\cdots}$ & $ 53.7\phd_{-0.1}^{+0.1}$   &   31.03& \nodata & \nodata & \nodata & \nodata & \nodata& \nodata                           & \nodata                     &    \nodata &    \nodata& $4.8\phd\!*$ & $10\phd\!*$      & $ 53.9\phd\!*$ &   31.10& \nodata           & \nodata & \nodata & \nodata&   31.90 &   32.45&            grp3.0\_diffuse\_3pshock &                        HandFit \\[1ex]  
                        North       & 2207/97 &     3.1 &    45.1 &    39/ 40 & & & & & &  & $0.36\phd_{\cdots}^{+0.1}$ & $0.16\phd_{-0.02}^{+0.03}$  & $13.7\phd\!*$                  & $ 55.0\phd_{\cdots}^{+0.2}$   &   31.78& $0.36\phd\!*$              & $0.76\phd_{-0.2}^{\cdots}$   & $11.3\phd_{\cdots}^{\cdots}$ & $ 53.2\phd_{-0.3}^{+0.2}$   &   30.81& \nodata & \nodata & \nodata & \nodata & \nodata& \nodata                           & \nodata                     &    \nodata &    \nodata& $4.1\phd\!*$ & $10\phd\!*$      & $ 53.9\phd\!*$ &   31.10& \nodata           & \nodata & \nodata & \nodata&   31.83 &   32.32&            grp3.0\_diffuse\_3pshock &                        HandFit \\[1ex]  
\enddata
\begin{center}

\end{center}               
Properties of the diffuse regions:\\
Col.\ (1): Name of the diffuse extraction region shown in Figure~14.
\\Col.\ (2): Signal-to-noise ratio in the total energy band (0.5--7~keV), expressed as $NetCounts / \sigma_{NetCounts}$.
\\Col.\ (3): Geometric area of the region in square parsecs, irrespective of point source masking, assuming a distance of 900~pc.
\\\\Details of the spectral fits:
\\Col.\ (4): ``Reduced $\chi^2$'' = $\chi^2$ divided by ``degrees of freedom'' for the best-fit model.
\\All fits used the source model {\em ``tbabs$^1$*pshock$^1$ + tbabs$^2$*pshock$^2$  + tbabs$^{stars}$(apec$^4$ + apec$^5$) + tbabs$^6$*apec$^6$''} in \XSPEC. 
\\$N_{H}$ columns report the best-fit value for the extinction column density ({\it tbabs} components).
\\$kT$ columns report the best-fit value for the plasma temperature ({\it pshock} and {\it apec} components).
\\$\tau$ columns report the ionization timescale for the plasma ({\it pshock} components).
\\$SEM$ columns report the surface emission measure for the plasma ({\it pshock} and {\it apec} components), which is independent of distance.
\\Unresolved pre-main sequence stars are modeled with a pair of soft (0.86 keV) and hard (2.6 keV) thermal plasmas ({\em apec$^4$ + apec$^5$}) with a fixed normalization ratio ($norm4 = 0.5 \times norm5$).  Background emission ({\em apec$^6$}) is frozen at 10~keV; its absorption ({\em tbabs$^6$}) is always 
larger than the largest {\it tbabs} needed for other model components.  
%
\\\\
Quantities marked with an asterisk (*) were frozen in the fit, because the fit parameter ran to one of its limits or because it was poorly constrained.  
Uncertainties represent 90\% confidence intervals.
More significant digits are used for uncertainties $<$0.1 in order to avoid large rounding errors; for consistency, the same number of significant digits is used for both lower and upper uncertainties.
Uncertainties are missing when \XSPEC\ was unable to compute them or when their values were so large that the parameter is effectively unconstrained.  
$SB_{tc}$ and $L_{tc}$ columns report the absorption-corrected surface brightness and luminosity of model components in the total band (0.5--7~keV). 
$SB$ is independent of distance; $L = SB \times Area$.   
\\$SB^{stars}$ and $L^{stars}$ are the surface brightness and luminosity, respectively, of unresolved pre-main sequence stars, calculated as the sum from components {\it apec$^4$} and {\it apec$^5$}.
\\Col.\ (22): total surface brightness from both {\em pshock} components combined.
\\Col.\ (23): total luminosity from both {\em pshock} components combined.
%

\label{tab:PlasmaFit}
\end{deluxetable}

\section{Calculating the pressure terms in RCW 36.}\label{sec:linewidthShell}
\begin{figure}
    \centering
    \includegraphics[width=0.6\hsize]{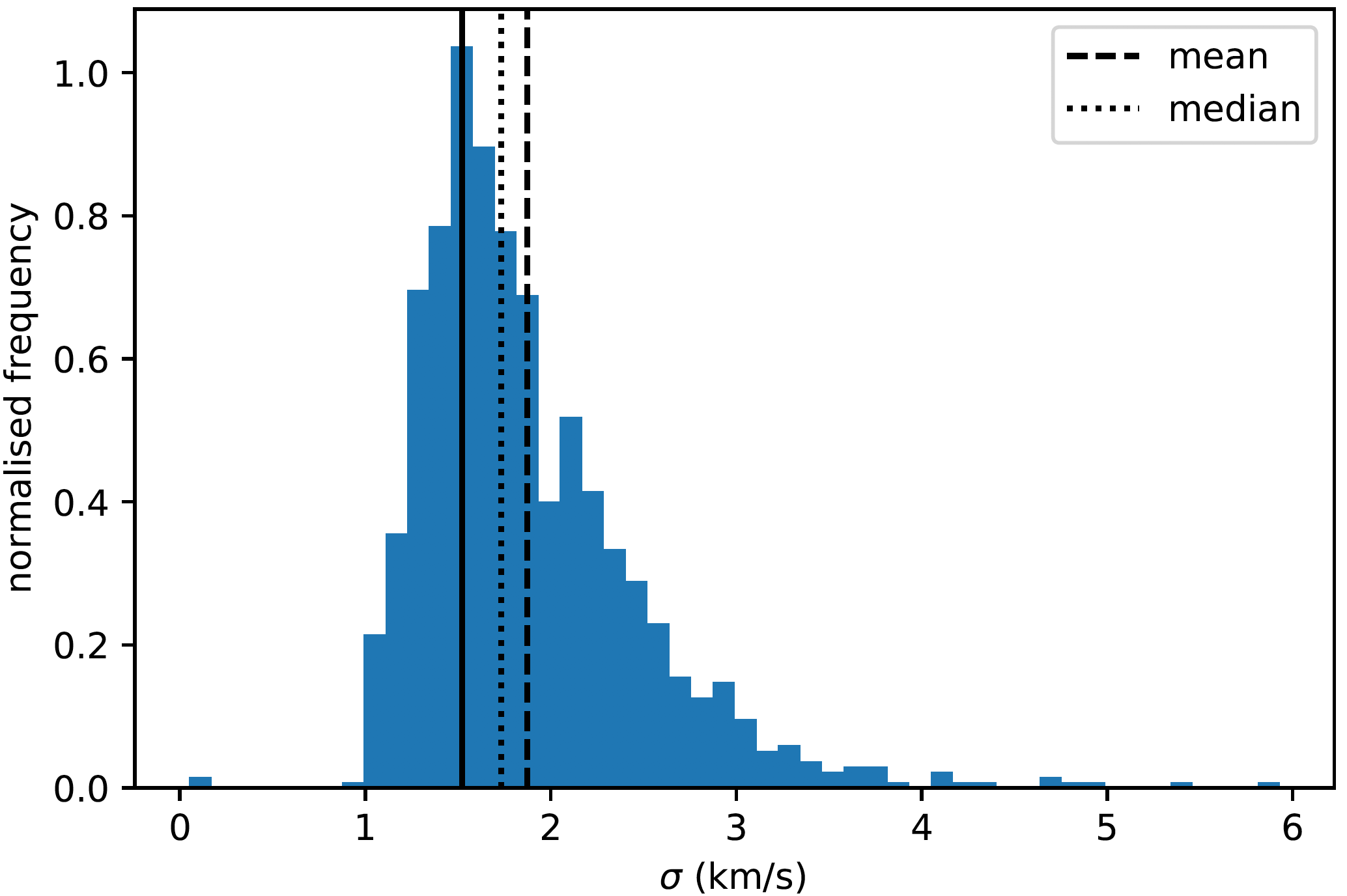}
    \caption{Histogram of the \CII\ linewidths in the expanding shells in the bipolar cavities. The dashed and dotted vertical lines indicate the mean and median linewidth, respectively.}
    \label{fig:shellHistogram}
\end{figure}

\begin{figure}
    \centering
    \includegraphics[width=0.6\hsize]{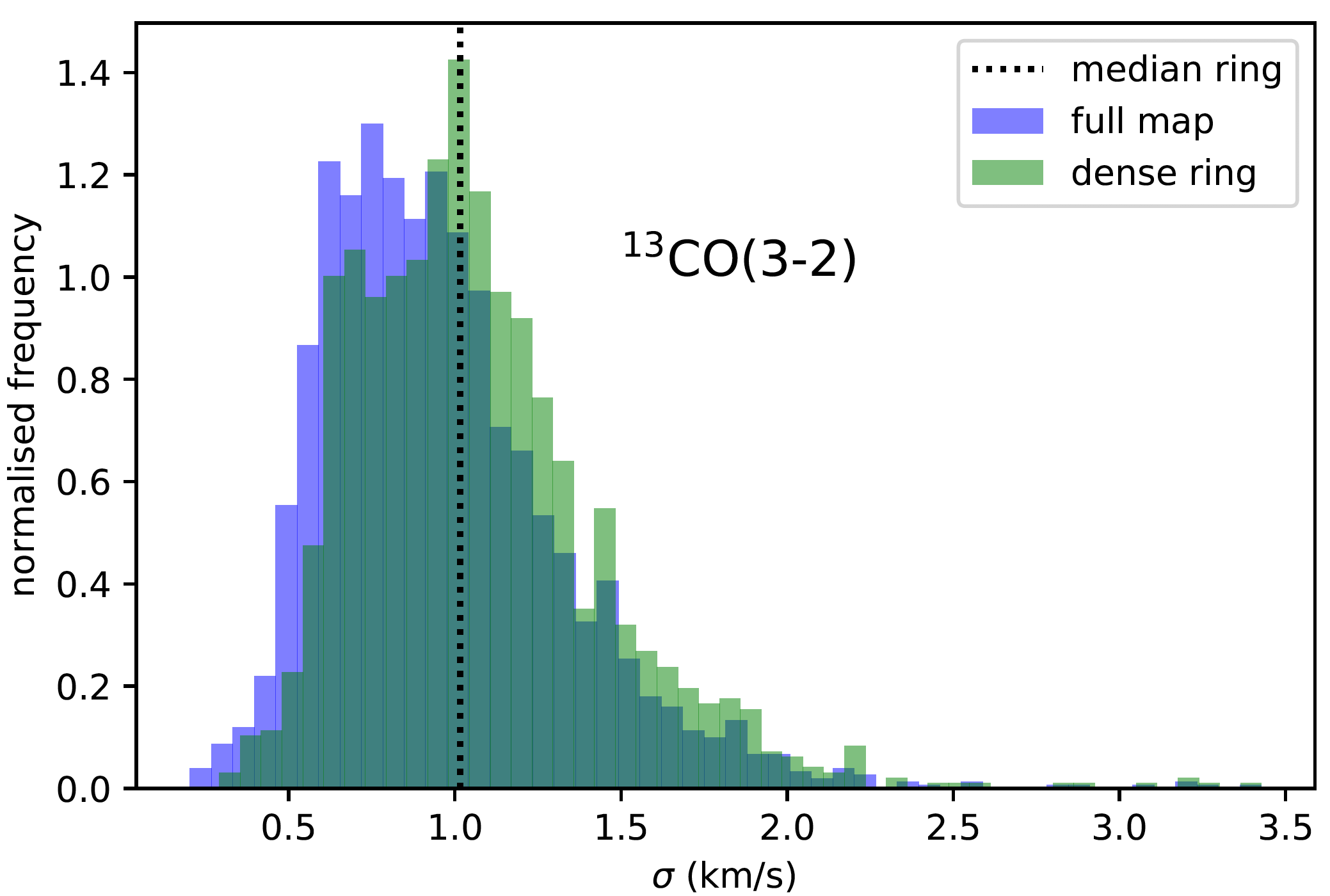}
    \caption{Histogram of the $^{13}$CO(3-2) linewidth over the full map of RCW 36 (blue). The green histogram indicates the linewidth over the swept-up ring, pointing to an increased linewidth with respect to the rest of the mapped region of Vela C.}
    \label{fig:ringHistogram}
\end{figure}

The hot plasma thermal pressure is obtained from fitting the X-ray spectra and the thermal pressure information for the ionized gas in the center is obtained from \citet{Minier2013}. The thermal pressure for the ionized gas in the cavities was derived from the SHASSA H$_{\alpha}$ data \citep{Gaustad2001}. From the SHASSA data it is possible to calculate the emission measure (EM) using EM = 4.2$\times$10$^{17}$ I$_{\alpha}$ after correction for the dust extinction, with I$_{\alpha}$ the intensity in the SHASSA region. The correction for the dust exctinction uses the curve from \citet{Calzetti2000} with k($\lambda$) = 2.38, A$_{V}$ = 3, based on the Herschel column density towards the cavities, and R$_{V}$ = 3.1 \citep{Draine2003}. Assuming a physical size of 1 pc for the cavities then gives an electron density of 58-62 cm$^{-3}$. Combining this with an electron temperature of 8000 K then gives the thermal pressure. This method based on the SHASSA data was not applied for the central region in the ring because the associated A$_{V}$ cannot be accurately constrained. To calculate the ionizing photon flux from the emission measure, we then use $\dot{N}_{ion}$ = EM$\times$A$\times \alpha_{B}$. Here, A is the surface area and $\alpha_{B}$ the recombination coefficient (= 2.7$\times$10$^{-13}$ cm$^{3}$ s$^{-1}$). The radiation pressure was directly estimated using I/c, where I is the radiation intensity and c the speed of light. The radiation intensity was determined from the luminosity of the central cluster (1.3$\times$10$^{5}$ L$_{\odot}$) and a distance from the cluster of 0.9 pc for the molecular ring and a distance of 1.5 pc for the cavity shells.\\
To calculate the ram pressure on the expanding shells, it was assumed that they are expanding in the ambient cloud of Vela C with a density n$_{H_{2}}$ = 10$^{2}$ - 5$\times$10$^{2}$ cm$^{-3}$ and a temperature of 50 K which is a typical temperature for the cold neutral medium. This density was estimated using a column density N$_{H_{2}}$ = 3$\times$10$^{21}$ cm$^{-2}$ in the ambient cloud, based on the Herschel data, and an estimated size in the line-of-sight of 2-10 pc which is poorly constrained. Inside the compressed shells of the cavities, we worked with n$_{H_{2}}$ = 10$^{3}$-3$\times$10$^{3}$ cm$^{-3}$  and a temperature of 250 K that was determined from the PDR modeling.  The density is estimated using a column density N$_{H_{2}}$ = 3$\times$10$^{21}$ cm$^{-2}$ and a depth of 0.3 to 1 pc. To calculate turbulent pressure, the \CII\ linewidth, associated with the expanding shell (i.e. the blueshifted velocity component varying between 1 and 6 km s$^{-1}$), was fitted. The details of this, both for the cavity shells and ring, are presented in the next paragraph. To calculate the ram pressure on the molecular ring, it is assumed that the ring is expanding into an ambient cloud/sheet with n$_{H_{2}}$ = 10$^{3}$-10$^{4}$ cm$^{-3}$ and a temperature of 20 K which is typical for these slightly denser molecular regions in the cloud. This large range of densities was assumed as it is difficult to constrain the density around the ridge observationally. For this estimate we assumed a typical column density around the ridge of 10$^{22}$ cm$^{-2}$ and a potential size range of 0.3 to 3 pc. To calculate the upper limit on the turbulent pressure in the ambient cloud, the velocity dispersion of 2.0 km s$^{-1}$ is taken 
\citep{Fissel2019}. The assumed internal density of the ring is n$_{H_{2}}$ = 2.4$\times$10$^{4}$ cm$^{-3}$ based on a typical N$_{H_{2}}$ = 1.5$\times$10$^{22}$ cm$^{-2}$ over the full ring and a diameter of 0.2 pc. The magnetic field strength in the ambient cloud was calculated using the assumed density range and the \citet{Crutcher2012} relation. The magnetic field in the shell of the cavities was estimated using equation 3 in \citet{vanMarle2015}, which assumes (initially) super Alfv\'enic expansion speed, which is uncertain leading to an upper limit. This assumption is likely not valid for the low expansion velocity of the ring, such that it is not calculated there. Future work using HAWC$^{+}$ data will further address the role of the magnetic field in the ring (Bij et al. in prep.).\\\\
To obtain the linewidth of the expanding shell in the cavities, the \CII\ velocity component between 0 and 6 km s$^{-1}$ was fitted with a single Gaussian. This fitting is complicated by contamination from the bright \CII\ emission at more redshifted velocities (the main velocity component associated with the ring). Therefore, a fitted Gaussian profile was only accepted if the the peak brightness is between 3.5 and 40 K. This peak brightness interval covers the full observed brightness range for the velocity component below 6 km s$^{-1}$, associated with the expanding shell, based on a visual inspection. As a result this constraint prevents that the Gaussian fitting tries to include additional emission that is associated with the bright component at 7-8 km s$^{-1}$ from the ring. The resulting \CII\ linewidth distribution for the expanding shells is presented in Fig. \ref{fig:shellHistogram}.\\
Fitting the linewidth was also done for $^{13}$CO(3-2). The resulting linewidth distribution for the detected emission over the full map is compared with the linewidth in the region that is identified as the dense ring in Fig. \ref{fig:ringHistogram}. This demonstrates that there is an increase in linewidth towards the dense ring. To determine the associated turbulent pressure and turbulent energy in the ring, it has to be taken into account that $^{13}$CO(3-2) is optically thick there which leads to opacity broadening given by \citep{Phillips1979}
\begin{equation}
    \frac{\Delta v}{\Delta v_{int}} = \frac{1}{\sqrt{\text{ln} 2}}\large[ \text{ln} \large( \frac{\tau_{0}}{\text{ln}(\frac{2}{\text{exp}(-\tau_{0})+1})} \large) \large]^{0.5}
\end{equation}
with $\Delta$v$_{int}$ the intrinsic velocity dispersion we are interested in, $\Delta$v the observed velocity dispersion and $\tau_{0}$ the opacity. As there is no optically thin tracer it is not straightforward to estimate the opacity. Therefore, RADEX \citep{vanderTak2007} was employed to estimate the optical depth. This was done using the observed linewidth, a typical column density N$_{H_{2}}$ = 1.5$\times$10$^{22}$ for the ring, [$^{12}$CO]/[H$_{2}$] = 10$^{-4}$, [$^{12}$CO]/[$^{13}$CO] = 60, T$_{kin}$ 15 K and n$_{H_{2}}$ = 2.4$\times$10$^{4}$ cm$^{-3}$. This gives an estimated opacity of $\sim$ 2 which results in a correction factor 1/1.35 for for the linewidth.

\section{Spitzer and Weaver solutions}\label{sec:SpitzerWeaverCalc}
In order to estimate the expansion timescale of the different components in RCW 36, we use the Spitzer \citep{Spitzer1978} and Weaver \citep{Weaver1977} solution. First we will discuss the Weaver solution which is derived from Eqs. 51 and 52 in \citet{Weaver1977}. This results in
\begin{equation}
    \text{t}_{exp} = \frac{16}{27}\frac{\text{R}}{\text{v}}
\end{equation}
with R the radius of the expanding structure and v the observed velocity. For the dense ring, we use R = 0.9 pc and v = 1.0-1.9 km~s$^{-1}$. For the shells in the cavities, we use R = 1.0 pc and v = 5.2 km~s$^{-1}$.\\
The Spitzer solution for the expansion timescale is obtained from
\begin{equation}
    \text{R} = \text{R}_{St}\left( 1 + \frac{7}{4}\sqrt{\frac{4}{3}}\frac{\text{c}_{i}\text{t}_{exp}}{\text{R}_{St}} \right)^{4/7}
\end{equation}
with c$_{i}$ the sound speed of the ionized gas (at a kinetic temperature T$_{k}$ = 8000 K) and R$_{St}$ the Str\"omgren radius which is given by
\begin{equation}
    \text{R}_{St} = \left( \frac{3\text{N}_{LyC}\text{m}_{p}^{2}}{4\pi \alpha_{B} \rho_{0}^{2}} \right)
\end{equation}
with N$_{LyC}$ the number of photons emitted that will ionize the Str\"omgren radius, m$_{p}$ the proton mass (= 1.67$\times$10$^{-27}$ kg), $\alpha_{B}$ the recombination coefficient (= 2.7$\times$10$^{-13}$ cm$^{3}$ s$^{-1}$) and $\rho_{0}$ the density of the neutral medium. For the number of ionizing photons we use 10$^{48}$ s$^{-1}$ inside the ring and 8.5$\times$10$^{46}$ s$^{-1}$ in the cavities. This was derived based on the radiocontinuum for the ring and the SHASSA H$\alpha$ flux for the cavities (see above). Based on the current mass in the dense ring, we can assume a density n$_{H_{2}}$ = 2.4$\times$10$^{4}$ cm$^{-3}$ for the neutral medium which gives t$_{exp}$ = 1.2 Myr. However, in Sec. \ref{sec:massEjection} we show that potentially half of the mass has already been removed from the central region, which would give a neutral medium density of n$_{H_{2}}\sim$5$\times$10$^{4}$ cm$^{-3}$ and a resulting t$_{exp}$ = 1.7 Myr. Considering that a significant amount of mass has been removed from the region might thus increase the Spitzer expansion timescale. Even though we have argued that the hot plasma from stellar winds might drive the expansion in the cavities, we also calculated the resulting Spitzer timescale for this regions. Using the above mentioned ionizing photon flux and a neutral medium density between n$_{H_{2}}$ = 10$^{2}$ cm$^{-3}$ and 5$\times$10$^{2}$ cm$^{-3}$ gives an expansion timescale between 0.2 and 0.4 Myr. These timescales are also consistent with the ones derived assuming a constant expansion velocity and confirm that, even if expansion in the cavities is driven by ionizing radiation, the expansion timescales of the cavities is short compared to the lifetime of the \HII\ region. All the obtained values for the expansion timescales using different methods are summarized in Tab. \ref{tab:tabExpansionTimescales}.
\begin{table}[]
    \centering
    \caption{Obtained expansion timescales for the dense molecular ring and cavities assuming (1) a constant expansion velocity, (2) the \citet{Weaver1977} solution and (3) the \citet{Spitzer1978} solution.}
    \begin{tabular}{cccc}
    \hline
    \hline
        region &  constant velocity & \citet{Weaver1977} & \citet{Spitzer1978}\\
        \hline
         dense ring & 0.5-0.9 Myr & 0.3-0.6 Myr & 1.2-1.7 Myr \\
         cavities & 0.2 Myr & 0.1 Myr & 0.2-0.4 Myr\\
         \hline
    \end{tabular}
    \label{tab:tabExpansionTimescales}
\end{table}






\end{document}